%% file: main.tex
\documentclass[preprint,12pt]{elsarticle}

\usepackage[margin=2.5cm]{geometry}

\usepackage{amssymb}

\input{configs/my-packages}
\input{configs/my-commands}

\journal{Journal of Computational Physics}

\begin{document}

\begin{frontmatter}



\title{A diffuse-interface method for compressible two-phase flows with seven- and six-equation models}

\author{Luis H. Hatashita}
\author{Suhas S. Jain}

\affiliation[]{organization={Flow Physics and Computational Science Lab,
Center for Multiphase flow Research, \\
Georgia Institute of Technology},
            postcode={30332}, 
            state={GA},
            country={USA}}

\begin{abstract}
In this work, a novel phase-field method is proposed for the six- and seven-equation non-equilibrium models for simulating compressible two-phase flows. Such formulations allow for monotonic mixture speed of sound, minimizing artificial wave delay during transmission across an interface. The proposed phase field formulation is constructed from the baseline seven-equation model, and interface-regularization terms are added in divergence form, while maintaining consistency between the partial differential equations (PDE) without introducing spurious source terms. It analytically admits conservative phasic and mixture entropy transport equations, thus facilitating the construction of discrete conservative schemes. The six-equation formulation is subsequently obtained under instantaneous velocity equilibrium. To avoid eigenvector degeneracy of the system of PDEs, the volumetric interface regularization flux is modified to account for a finite amount of conjugate phase, which improves on how phasic density is captured implicitly. Stability of compressible two-phase flow rely on the preservation of the interface-equilibrium conditions (IEC), and the preservation of discrete kinetic energy and entropy (KEEP). A detailed analysis of IEC demonstrates additional requirements on the consistency of flux splittings between the convective and interface-regularization terms for all quantities, as well as the effects on the phasic internal energy flux splittings. A KEEP-like discretization is proposed and evaluated over a suite of high-density ratio test cases, including interface advection, acoustic wave-induced bubble oscillation, oblique acoustic wave reflection and transmission, and two-phase Taylor-Green vortex flow. Results demonstrate accuracy, stability and robustness for very long time integrations, a desired feature for simulation of turbulent flows and acoustics, since the framework does not rely on the addition of numerical dissipation.
\end{abstract}


\end{frontmatter}

\input{chapters/intro}
\input{chapters/equations}
\input{chapters/numerics}
\input{chapters/results}
\input{chapters/conclusions}

\section*{Acknowledgments}

This work was supported by the Air Force Office of Scientific Research under award number FA9550-25-1-0137.
A preliminary version of this work has been published as conference proceedings of AIAA Scitech 2025~\cite{hatashita:2025}, 2026~\cite{hatashita:2026a}, and ILASS-Americas 2026~\cite{hatashita:2026b}.

The authors also acknowledge the generous computing resources from the DOE's 2024 and 2025 ALCC awards (TUR147 \& BubbleLaden, PI: Jain). This research used supporting resources at the Argonne and the Oak Ridge Leadership Computing Facilities. The Argonne Leadership Computing Facility at Argonne National Laboratory is supported by the Office of Science of the U.S. DOE under Contract No. DE-AC02-06CH11357. The Oak Ridge Leadership Computing Facility at the Oak Ridge National Laboratory is supported by the Office of Science of the U.S. DOE under Contract No. DE-AC05-00OR22725.

\appendix

\input{appendix/mcdi}

\bibliographystyle{elsarticle-num-names} 
\bibliography{cas-refs}





\end{document}

%% file: configs/my-packages.tex
\usepackage{todonotes}
\usepackage{amsmath}
\usepackage{amsthm}
\usepackage{mathtools}
\usepackage{yhmath}
\usepackage{multirow}
\usepackage{subcaption}
\usepackage{tikz}
\usepackage{cleveref}
\usepackage{comment}
\usetikzlibrary{arrows.meta}
\usetikzlibrary{positioning}
\usetikzlibrary{calc}

%% file: configs/my-commands.tex
\newcommand{\ppt}[1]{\frac{\partial}{\partial t} \left ( #1 \right )}

\newcommand{\ppxi}[1]{\frac{\partial}{\partial x_i} \left ( #1 \right )}
\newcommand{\ppxj}[1]{\frac{\partial}{\partial x_j} \left ( #1 \right )}

\newcommand{\ppxjsb}[1]{\frac{\partial}{\partial x_j} \left [ #1 \right ]}
\newcommand{\der}[1]{\text{d} \left ( #1 \right )}
\newcommand{\dldt}[1]{\frac{D_l}{Dt} \left ( #1 \right )}
\newcommand{\lrp}[1]{\left ( #1 \right )}
\newcommand{\lrsb}[1]{\left [ #1 \right ]}
\newcommand{\lrcb}[1]{\left \{ #1 \right \}}

\newcommand{\frb}[3]{\left . #1 \right |_{#2}^{#3}}
\newcommand{\highlight}[1]{\colorbox{lime}{$\displaystyle #1$}}

\makeatletter
\newsavebox\myboxA
\newsavebox\myboxB
\newlength\mylenA
\newcommand*\xoverline[2][0.75]{%
    \sbox{\myboxA}{$\m@th#2$}%
    \setbox\myboxB\null
    \ht\myboxB=\ht\myboxA%
    \dp\myboxB=\dp\myboxA%
    \wd\myboxB=#1\wd\myboxA
    \sbox\myboxB{$\m@th\overline{\copy\myboxB}$}
    \setlength\mylenA{\the\wd\myboxA}
    \addtolength\mylenA{-\the\wd\myboxB}%
    \ifdim\wd\myboxB<\wd\myboxA%
       \rlap{\hskip 0.5\mylenA\usebox\myboxB}{\usebox\myboxA}%
    \else
        \hskip -0.5\mylenA\rlap{\usebox\myboxA}{\hskip 0.5\mylenA\usebox\myboxB}%
    \fi}
\makeatother

\newtheorem{proposition}{Proposition}
\newtheorem{postulate}{Postulate}

\definecolor{mpl_blue}{RGB}{31, 119, 180}
\definecolor{mpl_orange}{RGB}{255, 127, 14}
\definecolor{mpl_green}{RGB}{44, 160, 44}
\definecolor{mpl_red}{RGB}{214, 39, 40}
\crefname{equation}{Eq.}{Eqs.}
\crefname{figure}{Fig.}{Figs.}

\crefname{section}{Section}{Sections}
\crefname{table}{Table}{Tables}
\crefname{postulate}{Postulate}{Postulates}
\crefname{proposition}{Proposition}{Propositions}

\newif\iffullmodel
\fullmodelfalse

\newif\ifsurfacetension
\surfacetensionfalse

\newif\ifkeepthm
\keepthmfalse

\newif\ifproofmcdi
\proofmcditrue

%% file: chapters/intro.tex
\section{Introduction}
\label{sec:intro}



Compressible two-phase flows are pervasive in both nature and industrial applications. For instance, bubble dynamics and collapse eventually rise as the governing phenomena in vascular injuries due to cavitation~\citep{coralic:2013} and as the source of structural damage in propulsion systems and turbomachinery~\citep{blake:1987,escaler:2006}. Moreover, it is of interest for the optimization of combustion systems in scramjets~\citep{benyakar:1998}, rotating detonation engines~\citep{tarey:2024}, liquid rocket propulsion~\citep{utturkar:2005}, and hybrid/solid rockets~\citep{carmicino:2015}. In addition to the presence of the interface and considerable compressibility effects in these flows, turbulence is expected to be observed for sufficiently high Reynolds numbers. Turbulence is characterized by chaotic fluctuations, which carry over to almost all quantities of interest, \textit{e.g.}, scalars such as temperature.

The primary challenge of compressible two-phase flows is the presence of discontinuities, \textit{e.g.} density jumps due to the interface, contact discontinuities, and shocks. Although sharp interface methods, such as level-set and ghost-fluid methods, exist, maintaining discrete conservation and being able to dynamically handle significant interface topology changes are more difficult to handle if compared to the diffuse interface method (DIM) approach~\citep{sethian:2003} (details on interface-capturing approaches can be found in~\citep{mirjalili:2017}). Furthermore, DIM is often preferred given that it requires only a single set of equations to describe motion for both phases without explicit tracking either, providing significant scale up to large problem sizes~\citep{saurel:2018}. In order to handle discontinuities, a standard approach for DIM is the use of upwind schemes coupled with Riemann solvers. The addition of numerical dissipation for stabilization further diffuses the initially ``thick'' interface, whereas they are inherently sharp~\citep{shukla:2010,jain:2023}, therefore, controlling interface thickness in DIM is a desirable feature. Earlier formulations focused on the construction of models for the mixture; however, a unique equation of state for multiphase flows is not trivial since each phase obeys its own equation of state~\citep{saurel:1999,saurel:2001,saurel:2009}. Hence, it is natural to adopt two-phase flow descriptions where the equations of state can be separate. One of the first seminal works using the two-phase formulation is the one from~\citet{baer:1986}, where the authors proposed a formulation for the detonation-to-deflagration transition in energetic materials. The model consists of the conservation equations for the phasic masses, momenta and energies appended by a volume fraction transport equation. The seven-equation model by~\citet{baer:1986} admits all non-equilibrium effects for velocity, pressure, temperature and chemical potential and is thus the most complete one. It has been reduced under the assumptions of infinitely fast relaxation process by~\citet{kapila:2001} to the six-~\citep{saurel:2009,zein:2010,pelanti:2014}, five-equation~\citep{perigaud:2005,murrone:2005,saurel:2008,tiwari:2013,abgrall:2014,rodio:2015} models, reducing to single mixture momentum; and, single mixture momentum and energy equations, respectively. Independently, \citet{allaire:2002} developed an alternative five-equation model~\citep{shukla:2010,coralic:2014,garrick:2017,jain:2020,huang2023consistent} without a non-conservative term in the volume transport equation obtained through the asymptotic analysis. Though robust, the model struggles with creation and destruction of interfaces by coalescence and breakup. Temperature equilibrium yields in the four-equation model~\citep{abgrall:1996,saurel:1999b,johnsen:2012,chiapolino:2017b,jain:2023,collis2022assessment}, with the transport of the mixture mass rather than individual phasic contributions. Lastly, chemical equilibrium yields in the three-equation model~\citep{clerc:2000}, recovering all mixture equations, namely a homogeneous equilibrium model. The suite of five-, four- and three-equation models allow for simpler extension of multiphysics effects, such as viscous, capillary~\citep{perigaud:2005}, and phase change~\citep{saurel:2008,demou:2022}. Since mixture equations are used for the momentum and energy equations, there is no need to define how much the contributions by each phase are.

Despite capturing the compressibility of both phases and producing accurate results~\citep{jain:2020} within an efficient computational cost, the original five-equation model formulation~\citep{kapila:2001,allaire:2002} and further reduced models fail to accurately capture wave transmission across interfaces under extreme conditions (\textit{e.g.} high density and pressure ratios), leading to variations in volume fraction and loss of positivity~\citep{petitpas:2007,saurel:2009,zein:2010,pelanti:2014}. Hence, the models which enable non-equilibrium effects for velocity and pressure are more suitable to the aforementioned applications. The six-equation model was shown to provide results comparable to the baseline seven-equation~\citep{zein:2010}, and is computationally simpler for its \textit{a priori} knowledge of the ordering of the hyperbolic operator eigenvalues~\citep{pelanti:2014}. Both six- and seven-equation models allow for distinct phasic pressures, eliminating the mixture closure which may not exist for any combination of arbitrary equations of state~\citep{saurel:2009}. Moreover, supporting distinct phasic pressure also permits the underlying models to sustain large pressure ratios. In the mixture zone across the interface, both models have a monotonic mixture speed of sound, this feature minimizes delays in wave transmission~\citep{pelanti:2014}. The velocity non-equilibrium effect is more subtle, the seven-equation model is the only one to allow for precise control of drag between the phases, which is of interest for modeling the dense-dilute transition~\citep{panchal:2023,viqueira:2024}. As a first baseline work, the more general formulations provide a foundational comparison to future reduced models.

Interface sharpening/regularization techniques circumvent the numerical diffusion of the interface using standard numerical schemes. 
\citet{shukla:2010} proposed a density and volume fraction correction in pseudo time to correct for the diffusion after time integration of a five-equation model~\citep{allaire:2002}. \citet{tiwari:2013} introduced regularization terms of a similar form to the operator proposed by~\citep{shukla:2010} in the five-equation model of~\citep{kapila:2001}, which was further extended to a seven-equation model by~\citet{panchal:2023}. However, these proposed formulations were not in divergence form, which limits conservation properties. Flux limiting has also been proposed by \citet{chiapolino:2017} to control interface thickness in a six-equation model~\citep{saurel:2009}; however, their approach still requires detection of the interface. More recently, \citet{jain:2020} consistently introduced interface regularization terms in all conservation equations in a baseline five-equation model of~\citep{allaire:2002} using a conservative form of a second-order phase field model. The phase field formulation was further extended to a four-equation model in \citet{jain:2023}. However, such phase field formulations have not been developed before for seven- and six-equation models.

Adding a phase field model as an interface regularization flux helps maintaining the interface thickness for long time integrations \citep{jain:2023}. However, one problem exists with all the current phase field models. Even if the volume fraction is initialized with a non-zero amount of other phase ($\delta$) everywhere, the phase field model will redistribute such that $\delta \rightarrow 0$. The practice of adding $\delta > 0$ is standard for accurate computation of phasic density, and it is even required to avoid degeneracy of eigenvectors for the seven-equation model~\citep{zein:2010}. Phasic density is an implicit quantity in most of the aforementioned models, and it is of significant importance in the model by~\citet{jain:2020} since it dictates regularization of mass, momentum and energy. In their work, \citet{jain:2020} introduced the approximation of $\rho_l \approx \rho_{0,l}$ in the limit of $\delta\rightarrow0$ imposed by the phase field model. Nevertheless, for accurate interface regularization and even pressure relaxation procedures~\citep{pelanti:2014} of six- and seven-equation models, a finite amount of volume fraction is required. \citet{jain:2023} proposed a modification of the Allen-Cahn based regularization flux to maintain the finite non-zero bound on the volume fraction field, constructed on a four-equation model. Therefore, it would be also of interest to further extend such formulation to more general seven- and six-equation models, and to develop more consistent modifications to phase field model than previously proposed.

The first stability test, models are evaluated under, is the requirement to satisfy the interface equilibrium conditions (IEC). Originally, \citet{abgrall:1996} proposed that a stable model should maintain uniform pressure and velocity fields if initialized as such. Given the initial assumption of uniform fields, IEC translates to conditions of how the volume fraction field has to be discretized and the form of certain non-conservative operators \citep{saurel:1999}. 
For formulations with interface regularization or other additional terms (\textit{e.g.}, artificial viscosity), IEC further imposes restrictions on flux splittings for face reconstructions~\citep{jain:2022a} and the form of such terms~\citep{tiwari:2013,jain:2020,jain:2024,collis2022assessment}. Interface regularization and a high-order Riemann solver that satisfies the IEC can produce accurate results. Nevertheless, as the Reynolds number increases, introducing turbulence, the small scale fluctuations are likely to be dissipated. Second-order central schemes are low-dissipative and can sustain even the smallest of the fluctuations. For such schemes, discretely conserving kinetic energy and entropy (KEEP schemes) is an additional stability test for compressible flows~\citep{honein:2004,chandrashekar:2013,kuya:2018,jain:2022a,tamaki:2022}. Specifically for two-phase flows, \citet{jain:2022a} developed consistency conditions between the fluxes to achieve discrete local and global kinetic energy preservation, therefore approximately preserving entropy. The authors were first to simulate compressible turbulent two-phase flows with the baseline five-equation model of~\citep{kapila:2001}, such as Taylor-Green vortex and homogeneous isotropic turbulence, at infinite Reynolds numbers demonstrating robustness. Since then, others have either extended such consistency conditions~\citep{hatashita:2025,hatashita:2026b} or constructed exact entropy preserving schemes~\citep{yoshida:2026} for the seven-equation model.

Thus, the main objective of this work is to develop and demonstrate a long-time accurate, robust and stable phase-field/diffuse interface method formulation for compressible two-phase flows suitable for capturing non-equilibrium effects. 
To this end, a phase field method is proposed for seven- and six-equation models; a consistent modification is proposed for phase field models to maintain finite non-zero bounds; and a robust KEEP-like scheme is proposed that ensures long time accuracy and stability.

The remainder of this paper is organized as follows: interface regularization terms in divergence form are first presented here for the six- and seven-equation models, enabling the construction of conservative schemes (such as the KEEP schemes) in \cref{sec:gov-eq}, along with the finite bound phase field model (it is formally derived and its effects are demonstrated in \ref{apx:modified-phase-field}); 
analytical entropy conservation properties are demonstrated for the seven-equation model in \cref{sec:mixture_entropy_eq}; in addition to IEC requirements on the non-conservative terms, flux splittings and consistency conditions among the discretization of all equations are thoroughly explored in \cref{sec:iec}; the numerical methods consisting of the hyperbolic integration and relaxation steps are presented in~\cref{sec:num_methods}; a suite of test cases is presented to evaluate accuracy, robustness and stability for the problems of interest and to demonstrate long-term stability in~\cref{sec:results}; and finally, the conclusions are presented in \cref{sec:conc}.

%% file: chapters/equations.tex
\section{Governing equations and phase-field formalism}
\label{sec:gov-eq}

We start with the fully-coupled two-phase flow equations, based on the formulation presented in the work of~\citet{saurel:1999}, which can be considered as a simplified version of the one from~\citet{baer:1986} (without considering solid-gas specific interactions). The system of equations is given by
\begin{equation}
\label{eq:7eq_baseline}
\begin{aligned}
& \frac{\partial \phi_1}{\partial t} + u_{I,j} \frac{\partial \phi_1}{\partial x_j} = \mu (p_1 - p_2), \\
& \ppt{\phi_l \rho_l} + \ppxj{\phi_l \rho_l u_{l,j}} = 0, \\
& \ppt{\phi_l \rho_l u_{l,i}} + \ppxj{\phi_l \rho_l u_{l,i}u_{l,j} + \phi_l p_l \delta_{ij}} = p_I \frac{\partial \phi_l}{\partial x_i} \pm \lambda(u_{2,i} - u_{1,i}), \\
& \begin{aligned}
\ppt{\phi_l \rho_1 e_{t,l}} & + \ppxjsb{\phi_l (\rho_l e_{t,l} + p_l) u_{l,j}} \\
& = p_I u_{I,j} \frac{\partial \phi_l}{\partial x_j} \mp \mu p_I (p_1 - p_2) \pm \lambda u_{I,j} (u_{2,j} - u_{1,j}), \\
\end{aligned}
\end{aligned}
\end{equation}
where the phasic conservation equations for mass, momentum and total energy for the phases 1 and 2 are supplemented by the transport equation for the volume fraction $\phi_1$ of the phase 1. $\phi_l \rho_l$, $\phi_l \rho_l u_l$ and $\phi_l \rho_l e_{t,l}$ are the phasic mass, momentum and total energy ($e_{t,l} = e_l + u_{l,i}u_{l,i}/2$, where $e_l$ is the phasic internal energy per unit mass). The $l$ index is restricted to represent the phase, $l=1,2$; $i$ and $j$ are Einstein indexes, which imply summation when repeated; and $\delta_{ij}$ is the Kronecker delta tensor. 

The phasic equations are obtained via the averaging procedure of~\citet{drew:1983}, which leads to the interface pressure $p_I$ force coupled with volume fraction gradient and the corresponding work term with the interface velocity $u_I$. The non-equilibrium effects are incorporated through the relaxation terms, the coefficient $\mu$ for the phasic pressures $p_l$ and the coefficient $\lambda$ for the phasic velocities $u_{l,i}$ dictating the rate at which these processes occur. Finite-rate relaxation may be incorporated following~\citep{pelanti:2022}, however in this study, both relaxation steps are taken to be instantaneous~\citep{saurel:1999}, \textit{i.e.} both $\mu, \lambda$ tend to $\infty$. The Euler equations are obtained for the mixture by summing the phasic equations, or in the single-phase limit. 

The system is closed by defining the expression for the interface quantities $p_I$ and $u_I$ and the equation of state to be used. Simpler expression as $p_I = p_g$ and $u_I = u_s$, where the subscripts $g$ and $s$ respectively refer to gas and solid phases, have been initially proposed by \citet{baer:1986}. \citet{saurel:1999} suggest the approximation of the interface quantities as the corresponding mixture values. \citet{saurel:2003} also propose closures function of the impedance of each phase. Herein, the interface quantities are defined as
\begin{equation}
\label{eq:p_u_I}
\begin{aligned}
& p_I = \sum_{l} \phi_l p_l, \\
& u_{I,i} = \frac{\sum_{l} \phi_l \rho_l u_{l,i}}{\sum_{l} \phi_l \rho_l}.
\end{aligned}
\end{equation}

In the context of phase change, \citet{zein:2010} alert to the use of cubic, Van der Waals equations of state, which may produce negative squared speed of sounds, and suggest a modified stiffened gas equation of state (SG-EoS). Although phase change is not considered here, the standard SG-EoS is sufficiently general \citep{saurel:1999,saurel:2003} and is therefore considered throughout this work. It is defined as
\begin{equation}
\label{eq:sg_eos}
p_l \left (\rho_l, e_l \right) = \left (\gamma_l - 1 \right) \rho_l e_l - \gamma_l \pi_l,
\end{equation}
where $\gamma_l$ and $\pi_l$ are phase specific constants.

Interface sharpening terms are consistently added in the phasic conservation equations and are highlighted in the system of equations in \cref{eq:7eq_proposed}, in addition to viscous terms. The full system of equations are given by
\begin{equation}
\label{eq:7eq_proposed}
\begin{aligned}
& \frac{\partial \phi_1}{\partial t} + \ppxj{\phi_1 u_{I,j}} = \phi_1 \frac{\partial u_{I,j}}{\partial x_j} + \highlight{\frac{\partial a_{1,j}}{\partial x_j}} + \mu (p_1 - p_2), \\
& \ppt{\phi_l \rho_l} + \ppxj{\phi_l \rho_l u_{l,j}} = \highlight{\frac{\partial R_{l,j}}{\partial x_j}}, \\
& \begin{aligned}
\ppt{\phi_l \rho_l u_{l,i}} & + \ppxjsb{\phi_l (\rho_l u_{l,i}u_{l,j} + p_l \delta_{ij} - \tau_{l, ij})} \\
& = \highlight{\ppxj{R_{l,j} u_{l,i}}} + p_I \frac{\partial \phi_l}{\partial x_i} \pm \lambda(u_{2,i} - u_{1,i}),
\end{aligned} \\
& \begin{aligned}
\ppt{\phi_l \rho_l e_{t,l}} & + \ppxjsb{\phi_l (\rho_l e_{t,l} + p_l) u_{l,j} - \phi_l \tau_{l, ij} u_{l,i}} \\
& = \highlight{\ppxj{R_{l,j}\frac{u_{l,i}u_{l,i}}{2}} + \ppxj{\rho_l e_l a_{l,j}}} \\ 
& + p_I u_{I,j} \frac{\partial \phi_l}{\partial x_j} \mp \mu p_I (p_1 - p_2) \pm \lambda u_{I,i} (u_{2,i} - u_{1,i}) ,
\end{aligned}
\end{aligned}
\end{equation}
where $a_{l,i}$ is the volumetric interface-regularization flux, which satisfies the condition $a_{2,i} = -a_{1,i}$. The phasic mass regularization flux is defined by $R_{l,i} = \rho_l a_{l,i}$, the phasic density $\rho_l$ is required to be explicitly computed, and if no explicit treatment is provided, standard calculations, \textit{e.g.}, $\rho_l = \phi_l \rho_l / \phi_l$, may yield inaccurate results in the limit of the pure other phase. A solution for this issue is further addressed in \cref{subsubsec:new_phase_field}. The phasic momentum, kinetic energy and internal energy are regularized based on the volumetric $a_{l,i}$ and the phasic mass $R_{l,i}$ regularization fluxes. There are several descriptions of the volumetric interface-regularization flux, such as conservative diffuse interface (CDI)~\citep{chiu:2011} and accurate conservative diffuse interface (ACDI)~\citep{jain:2022b} models.
However, as it will be further shown, the existing models asymptote to exact pure phases ($\phi_l = 0$) even if a small amount of the other phase is added ($\phi_l = \delta > 0$) initially. Therefore, a finite-bound modification is proposed to the phase field models in \cref{subsubsec:new_phase_field}. The viscous stresses are modeled as Newtonian fluids for both phases, where $\tau_{l,ij} = 2\mu_l S_{l,ij} + \beta_l \theta_l \delta_{ij}$ with $\mu_l$ as the phasic dynamic viscosity, $S_{l,ij} = (\partial u_{l,i}/\partial x_j + \partial u_{l,j}/\partial x_i)/2$ as the phasic strain rate tensor, $\beta_l$ as the phasic bulk viscosity, and $\theta_l = \partial u_{l,i}/\partial x_i$ as the dilatation rate. The bulk viscosity is closed under the Stokes' hypothesis as $\beta_l = -2\mu_l/3$.

The novel extension of interface regularization for the non-equilibrium models is obtained following the approach of \citet{jain:2020}. The interface regularization term $a_l$ is introduced in the volume fraction transport, and the corresponding terms are added to the phasic mass, momentum, and energy, all of which are in divergence form; hence, allowing for discrete conservation irrespective of the numerical scheme chosen. A systematic derivation is presented in the following subsections, alongside further proofs of entropy consistency and interface equilibrium conditions. 

The main difference between the newly proposed formulation and the one by \citet{jain:2020} is that non-equilibrium effects are allowed here, which changes the extra interface terms added to each phasic equation. For instance, the additional pressure terms $p_l a_l$ [see \cref{eq:phasic_internal_energy} in \cref{subsec:phasic_et_transp}] no longer cancel for the mixture energy equation resulting in a modified regularization term in the total energy equation in \cref{eq:7eq_proposed}. This seven-equation formulation has the additional advantage of admitting conservative phasic and mixture entropy transport equations (in the absence of relaxation steps), even in the presence of regularization terms [\cref{sec:mixture_entropy_eq}]. Mixture equations for the conservation equations for the proposed formulation also retain a conservative form. A six-equation formulation is also presented in the~\cref{subsec:pf-six}.

\subsection{Volume fraction transport}
\label{subsec:vol_frac_transp}

There are several descriptions of the volumetric interface-regularization flux based on the phase-field formulations. Herein, the scope is limited to second-order conservative Allen-Cahn-based formulations, given their lower derivative order, as opposed to fourth-order derivative in Cahn-Hilliard-based models.
For instance, the conservative diffuse-interface method (referred to as CDI)~\citep{chiu:2011,mirjalili:2020,jain:2020} and the more accurate model that is reformulated using the interface equilibrium profile analytically (the accurate conservative diffuse-interface method, ACDI~\citep{jain:2022b}) are respectively given by
\begin{equation}
\label{eq:phase_field_forms}
\begin{aligned}
 \frac{\partial \phi_1}{\partial t} + \ppxj{\phi_1 u_j} & = \phi_1 \frac{\partial u_j}{\partial x_j} + \frac{\partial}{\partial x_j} \lrcb{\Gamma \lrsb{\epsilon \frac{\partial \phi_1}{\partial x_j} - \phi_1(1-\phi_1) n_{1,j}^{\phi}}}, \\
  \frac{\partial \phi_1}{\partial t} + \ppxj{\phi_1 u_j} & = \phi_1 \frac{\partial u_j}{\partial x_j} + \frac{\partial}{\partial x_j} \lrcb{\Gamma \lrsb{\epsilon \frac{\partial \phi_1}{\partial x_j} - \frac{1}{4}\lrp{1-\tanh^2 \lrp{\frac{\psi_1}{2\epsilon}}} n_{1,j}^{\psi}}},
\end{aligned}
\end{equation}
where the interface regularization velocity, $\Gamma$, is set as the global maximum magnitude of the velocity field between the two phases $\Gamma = \max_{l} (|u_{l,i}|_{\infty})$, the interface thickness, $\epsilon$, is set as a constant times the grid spacing $\epsilon = c \max \Delta x_i$, chosen according to stability analysis and cost of performing simulations ($c\geq1$ for CDI, and $c>0.5$ for ACDI). In ACDI, the interface distance-like function is defined as $\psi_l = \epsilon \log((\phi_l + \varepsilon)/(1-\phi_l + \varepsilon))$, where the small number $\varepsilon = 10^{-100}$. As discussed in \citet{jain:2022b}, the interface normal $n_{l,i}$ is often computed directly from the volume fraction gradient in CDI as $n_{l,i}^{\phi} = \partial \phi_l/\partial x_i / |\partial \phi_l / \partial x_k |$. Nevertheless, one may show that it is analytically equivalent to compute it based on the interface distance-like function $\psi$, which is smoother than the volume fraction field. Hence, \citet{jain:2022b} proposed a more accurate normal computation, given by $n_{l,i}^{\psi} = \partial \psi_l/\partial x_i / |\partial \psi_l / \partial x_k |$. 

\subsubsection{Introducing finite non-zero bounds}
\label{subsubsec:new_phase_field}

It is a well-known issue that compressible two-phase diffuse interface formulations require division by $\phi$, \textit{e.g.}, in algebraic expressions for pressure relaxation \citep{pelanti:2014} and phasic density calculation \citep{jain:2020,jain:2022a}. Therefore, it is standard practice to add a small amount of the conjugate phase on the order of $10^{-8}$~\citep{saurel:1999}, which is not an issue in the absence of interface regularization terms. Nonetheless, standard phase field methods undesirably redistribute the small amount of the conjugate phase, initially introduced for the robustness of the numerical methods, and force the pure phases to take a value of $\phi=0$ or $\phi=1$.

In this work, we formally derive (see \ref{apx:modified-phase-field}) the modifications to the phase field models in such a way that $\phi$ asymptotes to $\delta$ and $1-\delta$. The final expression for the volume fraction transport equation, without the relaxation term, is
\begin{equation}
\label{eq:phase_field_proposed_7eq}
\begin{aligned}
\frac{\partial \phi_1}{\partial t} & + \ppxj{\phi_1 u_{I,j}} & \\
& = \phi_1 \frac{\partial u_{I,j}}{\partial x_j} + {\frac{\partial}{\partial x_j} \lrcb{\Gamma\lrsb{\epsilon(1 - 2\delta) \frac{\partial\phi_1}{\partial x_j} - (\phi_1 - \delta)(1 - \phi_1 - \delta)n_{1,j}^\phi}}}, \\
& = \phi_1 \frac{\partial u_{I,j}}{\partial x_j} + {\frac{\partial}{\partial x_j} \lrcb{\Gamma\lrsb{\epsilon(1 - 2\delta) \frac{\partial\phi_1}{\partial x_j} - \frac{(1-2\delta)^2}{4}\lrp{1 - \tanh^2 \lrp{\frac{\psi_1}{2\epsilon}}}n_{1,j}^{\psi}}}}.
\end{aligned}
\end{equation}

\subsection{Phasic mass transport equation}
\label{subsec:phasic_mass_transp}

The construction of the phasic mass transport equation is the same as the one proposed in~\citet{jain:2020} for the five-equation model, for which there are also two distinct phasic mass transport equations. For completeness, we report the major steps of the derivation and refer to \citet{jain:2020} for additional details.

Interface regularization terms are introduced in the conservation equation of each phasic mass such that in the limit of incompressible flow, \textit{i.e.}, $\rho_l \rightarrow \rho_{0,l}$ and $\partial u_{I,j}/\partial x_j = 0$, the volume fraction transport equation [\cref{eq:phase_field_proposed_7eq}] is recovered. Therefore, consider the transport equation for the mass of the phase $l$ with interface regularization as
\begin{equation}
\label{eq:phasic_mass} 
\ppt{\phi_l \rho_l} + \ppxj{\phi_l \rho_l u_{l,j}} = \highlight{\ppxj{\rho_l a_{l,j}}},
\end{equation}
where $a_{l,j}$ is any volumetric interface regularization term, nonetheless, we limit the scope to the one proposed in~\cref{subsubsec:new_phase_field} since $\rho_l$ is required to be explicitly computed. Given the assumptions on incompressible flow for both phases, the \cref{eq:phasic_mass} can be rearranged as
\begin{equation}
\label{eq:phasic_mass_incompressible} 
\rho_{0,l} \lrsb{\frac{\partial \phi_l}{\partial t} + \ppxj{\phi_l u_{l,j}}} = \rho_{0,l} \ppxj{ a_{l,j}},
\end{equation}
recovering the \cref{eq:phase_field_proposed_7eq} as
\begin{equation}
\label{eq:phasic_volume_fraction_incompressible} 
\frac{\partial \phi_l}{\partial t} + \ppxj{\phi_l u_{l,j}} = \phi_l \frac{\partial u_{l,j}}{\partial x_j} + \frac{\partial a_{l,j}}{\partial x_j}.
\end{equation}

The difference between the current approach and the one proposed by \citet{jain:2020} is that, given the modified phase field model with finite non-zero bound, the computation of $\rho_l$ is now accurate. Therefore, interface regularization may be computed directly based on $\rho_l$ and not on a reference state $\rho_{0,l}$ as originally proposed, which ultimately should improve accuracy for applications where compressibility effects are more significant, \textit{e.g.}, strong acoustics and shocks.

Similarly, the mixture mass transport equation follows from the sum of \cref{eq:phasic_mass} for each phase, yielding in
\begin{equation}
\label{eq:total_mass}
\frac{\partial \rho}{\partial t} + \ppxj{\rho u_j} = \highlight{\ppxj{R_{1,j} + R_{2,j}}},
\end{equation}
where the consistent mass regularization term $R_{l,j}$ is accurately computed based on the phasic density $\rho_l$ and the finite non-zero bound ACDI interface regularization as
\begin{equation}
\label{eq:mass_regularization} 
R_{l,j} = \rho_l a_{l,j} = \rho_l \Gamma\lrcb{\epsilon(1 - 2\delta) \frac{\partial\phi_l}{\partial x_j} - \frac{(1-2\delta)^2}{4}\lrsb{1 - \tanh^2 \lrp{\frac{\psi_l}{2\epsilon}}}n_{l,j}^\psi}.
\end{equation}

\subsection{Phasic momentum transport equation}
\label{subsec:phasic_mom_transp}

Given the distinct phasic momentum transport equations, the construction of the consistent momentum regularization term now requires that no spurious oscillations in the phasic kinetic energy are introduced, which is analogous to the approach proposed by \citet{jain:2020}. However, the constraint is now posed at the phasic level rather than on the mixture. Consider the consistent phasic momentum equation in the inviscid limit, without the source and relaxation terms, as
\begin{equation}
\label{eq:phasic_mom} 
\ppt{\phi_l \rho_l u_{l,i}} + \ppxj{\phi_l \rho_l u_{l,i} u_{l,j} + \phi_l p_l \delta_{ij}} = \highlight{\ppxj{\rho_l a_{l,j} u_{l,i}}}.
\end{equation}
Multiplying \cref{eq:phasic_mom} by $u_{l,i}/2$ and rearranging for the transport of $\phi_l \rho_l k_{l}$, where $k_l = u_{l,i}u_{l,i}/2$, yields in
\begin{equation}
\label{eq:phasic_kinetic_energy_int_1} 
\begin{aligned}
\ppt{\phi_l \rho_l k_{l}} & + \ppxj{\phi_l \rho_l k_{l} u_{l,j}} + u_{l,i}\ppxj{\phi_l p_l \delta_{ij}} \\
& + k_l \lrsb{\ppt{\phi_l \rho_l} + \ppxj{\phi_l \rho_l u_{l,j}} - \ppxj{\rho_l a_{l,j}}} \\
& - \frac{u_{l,i}}{2} \lrsb{\ppt{\phi_l \rho_l u_{l,i}} + \ppxj{\phi_l \rho_l u_{l,i} u_{l,j} + \phi_l p_l \delta_{ij}} - \ppxj{\rho_l a_{l,j} u_{l,i}}} \\
& + u_{l,i} \ppxj{\phi_l p_l \delta_{ij}} \\
& = {\ppxj{\rho_l a_{l,j} k_{l}}}.
\end{aligned}
\end{equation}
Using the identities from Eqs.~\eqref{eq:phasic_mass} and~\eqref{eq:phasic_mom} in Eq. \eqref{eq:phasic_kinetic_energy_int_1}, one may simplify both terms in the square brackets, which yields in
\begin{equation}
\label{eq:phasic_kinetic_energy} 
\ppt{\phi_l \rho_l k_{l}} + \ppxj{\phi_l \rho_l k_{l} u_{l,j}} + u_{l,i} \ppxj{\phi_l p_l \delta_{ij}} = {\ppxj{\rho_l a_{l,j} k_{l}}}.
\end{equation}
Therefore, it does not lead to the spurious generation of phasic kinetic energy. 
Furthermore, this form allows for the construction of a numerical scheme that also satisfies IEC.
This specific form of phasic momentum regularization due to interface preservation is essential particularly when coupled to low-dissipative numerical schemes, because otherwise, spurious generation of phasic kinetic energy would lead in nonphysical states and unstable simulations~\citep{jain:2020}. 

The momentum regularization in the seven-equation model [\cref{eq:phasic_mom_7eq}] is be included in the hyperbolic integration step as a part of the operator splitting described in~\cref{sec:num_methods}. The velocity relaxation step is performed separately.
\begin{equation}
\label{eq:phasic_mom_7eq} 
\ppt{\phi_l \rho_l u_{l,i}} + \ppxj{\phi_l \rho_l u_{l,i} u_{l,j} + \phi_l p_l \delta_{ij}} = {\ppxj{\rho_l a_{l,j} u_{l,i}}} + p_I \frac{\partial \phi_l}{\partial x_i} \pm \lambda (u_{2,i} - u_{1,i}).
\end{equation}
Furthermore, the mixture momentum transport equation may be shown to be in conservative form by summing~\cref{eq:phasic_mom_7eq} for each phase, as
\begin{equation}
\label{eq:7eq_wis_mix_mom}
\ppt{\rho u_i} + \ppxjsb{\rho u_i u_j + (\phi_1 p_1 + \phi_2 p_2)\delta_{ij}} = \highlight{\ppxj{R_{1,j} u_{1,i} + R_{2,j} u_{2,i}}},
\end{equation}
where both source and relaxation terms cancel out.

\subsection{Phasic total energy transport equation}
\label{subsec:phasic_et_transp}

The original five-equation model with interface preservation proposed by~\citet{jain:2020} could only require approximate entropy conservation. After the addition of the interface regularization terms, the sum of the phasic internal energy regularization terms was imposed to be approximate or equal to zero whilst maintaining IEC. Lastly, the regularization of the total energy was obtained by adding the internal energy to the consistent kinetic energy terms.

Given the distinct transport of phasic total energy in the seven-equation model, we now search for the consistent interface regularization of phasic internal energy that at least approximately conserves entropy. Define the phasic material derivative as
\begin{equation}
\label{eq:material_derivative_definition}
\frac{D_l}{Dt} (\cdot) = \ppt{\cdot} + u_{l,j} \ppxj{\cdot},  \quad l=1,2. 
\end{equation}
The phasic internal energy transport equation is obtained by subtracting the phasic kinetic energy equation from the phasic total energy equation. From the system [\cref{eq:7eq_baseline}], the construction of the baseline phasic kinetic energy equation follows from multiplying the phasic momentum equation by $u_{l,i}/2$ and using the identities for the conservation of phasic mass and momentum. It has the form of
\begin{equation}
\label{eq:phasic_kinetic_energy_baseline} 
\ppt{\phi_l \rho_l k_{l}} + \ppxj{\phi_l \rho_l k_{l} u_{l,j}} + u_{l,i} \ppxj{\phi_l p_l \delta_{ij}} = u_{l,j}p_I\frac{\partial \phi_l}{\partial x_j} \pm \lambda u_{l,j} (u_{2,j} - u_{1,j}),
\end{equation}
where the corresponding baseline phasic internal energy equation is as follows
\begin{equation}
\label{eq:phasic_internal_energy_baseline} 
\begin{aligned}
\ppt{\phi_l \rho_l e_{l}} & + \ppxj{\phi_l \rho_l e_{l} u_{l,j}} + \phi_l p_l \frac{u_{l,j}}{\partial x_j} \\
& = p_I(u_{I,j} - u_{l,j})\frac{\partial \phi_l}{\partial x_i} \pm \lambda (u_{I,i} - u_{l,i}) (u_{2,i} - u_{1,i}) \mp \mu p_I(p_1 - p_2).
\end{aligned}
\end{equation}
In the material derivative form [\cref{eq:material_derivative_definition}], it is
\begin{equation}
\label{eq:phasic_internal_energy_baseline_material} 
\begin{aligned}
\frac{D_l}{Dt}\lrp{\phi_l \rho_l e_{l}} & + \phi_l \rho_l h_l \frac{u_{l,j}}{\partial x_j} \\
& = p_I(u_{I,j} - u_{l,j})\frac{\partial \phi_l}{\partial x_i} \pm \lambda (u_{I,i} - u_{l,i}) (u_{2,i} - u_{1,i}) \mp \mu p_I(p_1 - p_2),
\end{aligned}
\end{equation}
where the phasic enthalpy is defined as $h_l = e_l + p_l/\rho_l$.

The goal is to obtain the consistent internal energy regularization term such that entropy is conserved in the absence of the source and relaxation terms; namely, solve for $X_l$ in
\begin{equation}
\label{eq:phasic_internal_energy_material_X} 
\dldt{\phi_l \rho_l e_{l}}  + \phi_l \rho_l h_l \frac{u_{l,j}}{\partial x_j} + X_l = 0.
\end{equation}
Let the Gibbs' relation for the phasic internal energy be
\begin{equation}
\label{eq:gibbs_phasic_internal_energy} 
\der{\rho_l e_l} = \rho_l \der{e_l} + e_l \der{\rho_l} = \rho_l T_l \der{s_l} + h_l \der{\rho_l}. 
\end{equation}
Distributing the material derivative in~\cref{eq:phasic_internal_energy_material_X} for $\phi_l$ and $\rho_l e_l$ and using~\cref{eq:gibbs_phasic_internal_energy}, it results in
\begin{equation}
\label{eq:phasic_internal_energy_material_X_int_1} 
\phi_l\lrp{\rho_l T_l \frac{D_l s_l}{Dt} + h_l \frac{D_l \rho_l}{Dt}} + \rho_l e_{l} \frac{D_l \phi_l}{Dt}  + \phi_l \rho_l h_l \frac{u_{l,j}}{\partial x_j} + X_l = 0,
\end{equation}
where the transport of the material derivative for $\rho_l$ follows from~\cref{eq:phasic_mass}, as
\begin{equation}
\label{eq:phasic_density_material} 
\phi_l \frac{D_l \rho_l}{Dt} = - \phi_l \rho_{l} \frac{\partial u_{l,j}}{\partial x_j} + a_{l,j}\frac{\partial \rho_{l}}{\partial x_j} - \rho_l \lrp{\frac{\partial \phi_l}{\partial t} + u_{l,j}\frac{\partial \phi_l}{\partial x_j} - \frac{\partial a_{l,j}}{\partial x_j}}.
\end{equation}
One may take the term in parentheses to be equal to 0 by the transport of volume fraction equation in the absence of the source and relaxation terms, which yields in
\begin{equation}
\label{eq:phasic_internal_energy_material_X_int_2} 
\phi_l\rho_l T_l \frac{D_l s_l}{Dt} + h_l \ppxj{\rho_l a_{l,j}} - p_l \frac{\partial a_{l,j}}{\partial x_j} + X_l = 0.
\end{equation}
Therefore, in order to satisfy phasic entropy conservation during internal energy regularization, the consistent phasic internal energy equation would have the form of
\begin{equation}
\label{eq:phasic_internal_energy_int_1} 
\ppt{\phi_l \rho_l e_l} + \ppxj{\phi_l \rho_l e_l u_{l,j}} + \phi_l p_l \frac{\partial u_{l,j}}{\partial x_j} = h_l \ppxj{\rho_l a_{l,j}} - p_l \frac{\partial a_{l,j}}{\partial x_j}.
\end{equation}
Nevertheless, one may show that~\cref{eq:phasic_internal_energy_int_1} does not satisfy IEC. Instead, we propose the IEC-compatible form, while maintaining entropy conservation through a conservative flux redistribution mechanism [\cref{eq:phasic_entropy}], as
\begin{equation}
\label{eq:phasic_internal_energy} 
\ppt{\phi_l \rho_l e_l} + \ppxj{\phi_l \rho_l e_l u_{l,j}} + \phi_l p_l \frac{\partial u_{l,j}}{\partial x_j} = \ppxj{\rho_l h_l a_{l,j}} - \ppxj{p_l a_{l,j}} = \highlight{\ppxj{\rho_l e_l a_{l,j}}},
\end{equation}
\noindent where both forms differ by
\begin{equation}
\label{eq:difference_phasic_internal_energy_regularization}
\lrsb{\ppxj{\rho_l h_l a_{l,j}} - \ppxj{p_l a_{l,j}}} - \lrsb{h_l \ppxj{\rho_l a_{l,j}} - p_l \frac{\partial a_{l,j}}{\partial x_j}} = a_{l,j} \lrp{\rho_l \frac{\partial h_l}{\partial x_j} - \frac{\partial p_l}{\partial x_j}}.
\end{equation}
This residual can be rearranged with the Gibbs relation, which further allows for conservative phasic [\cref{eq:phasic_entropy}] and mixture entropy [\cref{eq:mixture_entropy}] transport equations. Given the distinct phasic pressures, the proposed formulations differs from the original work based on the five-equation model of \citet{jain:2020}, and uses the interface regularization of internal energy per unit mass rather than enthalpy per unit mass. The consistent regularization of the phasic total energy equation follows from the addition of~\cref{eq:phasic_internal_energy} and~\cref{eq:phasic_kinetic_energy} as
\begin{equation}
\label{eq:phasic_total_energy} 
\ppt{\phi_l \rho_l e_{t,l}} + \ppxjsb{\phi_l (\rho_l e_{t,l} + p_l) u_{l,j}} = \highlight{\ppxj{\rho_l e_l a_{l,j}} + \ppxj{\rho_l k_l a_{l,j}}},
\end{equation}
where the regularizations of internal and kinetic energy are kept separately to construct a numerical scheme that discretely conserves kinetic energy and preserves entropy, which is described in~\cref{subsec:h}.

The final equation for the consistent transport of phasic total energy with all terms is given by
\begin{equation}
\label{eq:phasic_total_energy_7eq} 
\begin{aligned}
\ppt{\phi_l \rho_l e_{t,l}} & + \ppxjsb{\phi_l (\rho_l e_{t,l} + p_l) u_{l,j}} \\
& = {\ppxj{\rho_l e_l a_{l,j}} + \ppxj{\rho_l k_l a_{l,j}}} \\
& + p_I u_{I,j} \frac{\partial \phi_l}{\partial x_j} \pm \lambda u_{I,j} (u_{2,j} - u_{1,j}) \mp \mu p_I (p_1 - p_2).
\end{aligned}
\end{equation}
The mixture total energy transport equation can be further shown to be in conservative form after adding the contribution of each phases as
\begin{equation}
\label{eq:7eq_wis_mix_et}
\begin{aligned}
\ppt{\rho e_t} & + \ppxj{\rho e_t u_j} + \ppxj{\phi_1 p_1 u_{1,j} + \phi_2 p_2 u_{2,j}} \\
& = \highlight{\sum_{l=1}^{2} \left [ \ppxj{R_{l,j} \frac{u_{l,i}u_{l,i}}{2}} + \ppxj{\rho_l e_l a_{l,j}} \right ]}.
\end{aligned}
\end{equation}

\subsection{Phase field model for the six-equation formulation}
\label{subsec:pf-six}

The derivation of the six-equation model follows the regularization procedure developed for the seven-equation model in \cref{subsec:vol_frac_transp,subsec:phasic_mass_transp,subsec:phasic_mom_transp,subsec:phasic_et_transp}, together with the instantaneous velocity-relaxation limit. The main differences are the single mixture-momentum equation and the equilibrium of the velocity field, which requires momentum interface regularization in the form that does not contribute to any spurious oscillation in mixture kinetic energy. Following asymptotic analysis, a mass fraction $Y_l$ based splitting of the kinetic energy to each of the phases in the phasic total energy equation is used in the model. The phase field method for the six-equation model is given by
\begin{equation}
\label{eq:6eq_wis}
\begin{aligned}
& \frac{\partial \phi_1}{\partial t} + \ppxj{\phi_1 u_j} = \phi_1 \frac{\partial u_j}{\partial x_j} + \highlight{\frac{\partial a_{1,j}}{\partial x_j}} + \mu (p_1 - p_2) \\
& \ppt{\phi_l \rho_l} + \ppxj{\phi_l \rho_l u_j} = \highlight{\frac{\partial R_{l,j}}{\partial x_j}} \\
& \ppt{\rho u_i} + \ppxj{\rho u_i u_j} + \ppxi{\phi_1 p_1 + \phi_2 p_2} = \highlight{\ppxj{f_j u_i}} \\
& \begin{aligned}
\ppt{\phi_l \rho_l e_{t,l}} + \ppxjsb{\phi_l (\rho_l e_{t,l} + p_l) u_j} \pm \Lambda & = \highlight{Y_l \ppxj{f_j \frac{u_i u_i}{2}} + \ppxj{\rho_l e_l a_{l,j}}} \\
& \mp \mu p_I (p_1 - p_2), 
\end{aligned}
\end{aligned}
\end{equation}
where the mixture-mass regularization flux is $f_j = \sum_l R_{l,j}$, the mixture-velocity-pressure gradient term is ${\displaystyle \Lambda = -u_j \lrsb{Y_2 \ppxj{\phi_1 p_1} - Y_1 \ppxj{\phi_2 p_2}}}$, and the mass fraction is $Y_l = \phi_l\rho_l/\rho$.

\section{Entropy conservation properties}
\label{sec:mixture_entropy_eq}

In addition to satisfying IEC, the proposed interface regularization for the seven-equation formulation allows for the phasic internal energy to admit conservative phasic and mixture entropy transport equations. One may rearrange~\cref{eq:phasic_internal_energy_material_X_int_2} with the proposed internal energy regularization term as
\begin{equation}
\label{eq:phasic_entropy_int_1} 
\phi_l\rho_l T_l \frac{D_l s_l}{Dt} = \ppxj{\rho_l e_l a_{i,j}} - \lrsb{h_l \ppxj{\rho_l a_{l,j}} - p_l \frac{\partial a_{l,j}}{\partial x_j}} = \rho_l a_{i,j} \lrp{\frac{\partial e_j}{\partial x_j} - \frac{p_l}{\rho_l^2}\frac{\partial \rho_l}{\partial x_j}},
\end{equation}
where the last term in parentheses is in the form of the Gibbs' relation, \textit{i.e.},
\begin{equation}
\label{eq:gibbs_relation_internal_energy_density}
T_l\der{s_l} = \der{e_l} - \frac{p_l}{\rho_l^2} \der{\rho_l}.
\end{equation}
Moreover, given that the total derivative operator in~\cref{eq:gibbs_relation_internal_energy_density} is a geometrical representation of the change in entropy by the net flux in internal energy and the ratio of pressure to density, one may take the identity to also represent a gradient operator, which yields in
\begin{equation}
\label{eq:gibbs_relation_internal_energy_density_gradient}
T_l\frac{\partial s_l}{\partial x_j} = \frac{\partial e_l}{\partial x_j} - \frac{p_l}{\rho_l^2} \frac{\partial \rho_l}{\partial x_j}.
\end{equation}
Substituting the identity in~\cref{eq:phasic_entropy_int_1}, yields in
\begin{equation}
\label{eq:phasic_entropy_int_2} 
\phi_l\rho_l T_l \frac{D_l s_l}{Dt} = \rho_l a_{i,j} T_l\frac{\partial s_l}{\partial x_j}.
\end{equation}
Recasting~\cref{eq:phasic_entropy_int_2} in conservative form, leads to
\begin{equation}
\label{eq:phasic_entropy_int_3} 
T_l \lrsb{\ppt{\phi_l \rho_l s_l} + \ppxj{\phi_l \rho_l s_l u_{l,j}}} = \phi_l\rho_l T_l \frac{D_l s_l}{Dt} + T_l s_l \dldt{\phi_l \rho_l} + \phi_l \rho_l s_l \frac{\partial u_{l,j}}{\partial x_j}.
\end{equation}
Using~\cref{eq:phasic_mass} in ~\cref{eq:phasic_entropy_int_3} yields in
\begin{equation}
\label{eq:phasic_entropy} 
\ppt{\phi_l \rho_l s_l} + \ppxj{\phi_l \rho_l s_l u_{l,j}} = \ppxj{\rho_l s_l a_{l,j}}.
\end{equation}
Furthermore, interface regularization can be shown to be entropy conservative also at the mixture level, as
\begin{equation}
\label{eq:mixture_entropy} 
\ppt{\rho s} + \ppxj{\phi_1 \rho_1 s_1 u_{1,j} + \phi_2 \rho_2 s_2 u_{2,j}} = \ppxj{\rho_1 s_1 a_{1,j} + \rho_2 s_2 a_{2,j}},
\end{equation}
where $\rho s = \phi_1 \rho_1 s_1 + \phi_2 \rho_2 s_2$. In the absence of relaxation processes, entropy evolution is governed exclusively by conservative entropy fluxes associated with interface regularization.

Considering all non-equilibrium effects,~\cref{eq:phasic_entropy} would have the form of
\begin{equation}
\label{eq:phasic_entropy_7eq} 
\begin{aligned}
T_l \Bigg [\ppt{\phi_l \rho_l s_l} & + \ppxj{\phi_l \rho_l s_l u_{l,j}}   \Bigg ] & \\
& = T_l \ppxj{\rho_l s_l a_{l,j}} \\
& + (p_I-p_l)(u_{I,j} - u_{l,j})\frac{\partial \phi_l}{\partial x_j} \mp \mu (p_I - p_l)(p_1 - p_2) \\
& \pm \lambda (u_{I,j} - u_{l,j}) (u_{2,j} - u_{1,j}),
\end{aligned}
\end{equation}
and the corresponding mixture equation is
\begin{equation}
\label{eq:mixture_entropy_7eq} 
\begin{aligned}
\ppt{\rho s} & + \ppxj{\phi_1 \rho_1 s_1 u_{1,j} + \phi_2 \rho_2 s_2 u_{2,j}} & \\
& = \ppxj{\rho_1 s_1 a_{1,j} + \rho_2 s_2 a_{2,j}} \\
& + \sum_{l=1}^{2} \Bigg \{ \frac{1}{T_l} \Bigg [ (p_I-p_l)(u_{I,j} - u_{l,j})\frac{\partial \phi_l}{\partial x_j} \mp \mu (p_I - p_l)(p_1 - p_2) \\
& \hspace{4.65em} \pm \lambda (u_{I,j} - u_{l,j}) (u_{2,j} - u_{1,j}) \Bigg ] \Bigg \}.
\end{aligned}
\end{equation}

\section{Interface equilibrium condition}
\label{sec:iec}


The interface equilibrium conditions provide discretization consistency conditions in order to preserve uniform velocity and pressure fields, thus inhibiting the generation of spurious oscillations. It can be formally stated as the following postulate.
\begin{postulate}
\label{post:iec}
If $u|_i^n = u_0$ and $p|_i^n=p_0$ across the interface, where $i$ is the cell index and $n$ is the time integration index, then a model is said to satisfy the interface equilibrium conditions (IEC) if $u|_i^{n+1}=u_0$ and $p|_i^{n+1}=p_0$.
\end{postulate}
One may show that the proposed model in~\cref{eq:7eq_proposed} allows for the construction of a consistent discretization that satisfies IEC. One then searches for the necessary conditions such that IEC holds, \textit{i.e.}, the required discretization for the non-conservative terms related to $\partial \phi_l / \partial x_i$ and the discrete form of the volume fraction transport equation.

\subsection{Uniform velocity and consistency conditions}
\label{subsec:uniform_vel}

Let $n=0$, and assume an initial uniform velocity $u_{l,i}^n=u_0$ and pressure $p_{l,i}^n=p_0$ state for both phases on a uniform mesh for any cell $i$. For simplicity, limit the equations to one dimension. Discretize the phasic mass equation based on a first order Euler explicit time integration and an arbitrary spatial discretization for the fluxes as
\begin{equation}
\label{eq:phasic_mass_iec_int_1}
\begin{aligned}
\frb{\phi_l \rho_l}{i}{n+1} & = \frb{\phi_l \rho_l}{i}{n} \\
& - \frac{\Delta t}{\Delta x} \lrp{\frb{\widetilde{\phi_l \rho_l u_l}}{i+1/2}{n} - \frb{\widetilde{\phi_l \rho_l u_l}}{i-1/2}{n}} + \frac{\Delta t}{\Delta x} \lrp{\frb{\widetilde{\rho_l a_l}}{i+1/2}{n} - \frb{\widetilde{\rho_l a_l}}{i-1/2}{n}},
\end{aligned}
\end{equation}
where the ``$\widetilde{f} |_{i\pm1/2}$'' operator denotes the flux splitting of the quantity $f$ at the cell face $i\pm 1/2$. Given the uniform velocity assumption, it reduces to
\begin{equation}
\label{eq:phasic_mass_iec_int_2}
\begin{aligned}
\frb{\phi_l \rho_l}{i}{n+1} & = \frb{\phi_l \rho_l}{i}{n} \\
& - \frac{\Delta t}{\Delta x} \lrp{\frb{\widetilde{\phi_l \rho_l}}{i+1/2}{n} - \frb{\widetilde{\phi_l \rho_l}}{i-1/2}{n}}u_0 + \frac{\Delta t}{\Delta x} \lrp{\frb{\widetilde{\rho_l a_l}}{i+1/2}{n} - \frb{\widetilde{\rho_l a_l}}{i-1/2}{n}}.
\end{aligned}
\end{equation}
Discretizing the phasic momentum equation yields in
\begin{equation}
\label{eq:phasic_mom_iec_int_1}
\begin{aligned}
\frb{\phi_l \rho_l u_l}{i}{n+1} & = \frb{\phi_l \rho_l u_l}{i}{n} \\
& - \frac{\Delta t}{\Delta x} \lrp{\frb{\widetilde{\phi_l \rho_l u_l^2}}{i+1/2}{n} - \frb{\widetilde{\phi_l \rho_l u_l^2}}{i-1/2}{n}} \\
& - \frac{\Delta t}{\Delta x} \lrp{\frb{\widetilde{\phi_l p_l}}{i+1/2}{n} - \frb{\widetilde{\phi_l p_l}}{i-1/2}{n}} & \\
& + \frac{\Delta t}{\Delta x} \lrp{\frb{\widetilde{\rho_l a_l u_l}}{i+1/2}{n} - \frb{\widetilde{\rho_l a_l u_l}}{i-1/2}{n}} + \Delta t p_I \Delta,
\end{aligned}
\end{equation}
where $\Delta$ is the discrete form of $\partial \phi_l / \partial x$. Given the initial uniform velocity and pressure assumptions, and~\cref{eq:phasic_mass_iec_int_2}, \cref{eq:phasic_mom_iec_int_1} simplifies to
\begin{equation}
\label{eq:phasic_mom_iec_int_2}
\begin{aligned}
\frb{\phi_l \rho_l u_l}{i}{n+1} & = u_0 \Bigg [ \frb{\phi_l \rho_l}{i}{n+1} + \frac{\Delta t}{\Delta x} \lrp{\frb{\widetilde{\phi_l \rho_l}^{\phi\rho}}{i+1/2}{n} - \frb{\widetilde{\phi_l \rho_l}^{\phi\rho}}{i-1/2}{n}}u_0 \\
& \hspace{7.em} - \frac{\Delta t}{\Delta x} \lrp{\frb{\widetilde{\rho_l a_l}^{\phi\rho}}{i+1/2}{n} - \frb{\widetilde{\rho_l a_l}^{\phi\rho}}{i-1/2}{n}} \Bigg ] \\
& - \frac{\Delta t}{\Delta x} \lrp{\frb{\widetilde{\phi_l \rho_l}^{\phi\rho u}}{i+1/2}{n} - \frb{\widetilde{\phi_l \rho_l}^{\phi\rho u}}{i-1/2}{n}} u_0^2 \\
& - \frac{\Delta t}{\Delta x} \lrp{\frb{\widetilde{\phi_l}}{i+1/2}{n} - \frb{\widetilde{\phi_l}}{i-1/2}{n}}p_0 & \\
& + \frac{\Delta t}{\Delta x} \lrp{\frb{\widetilde{\rho_l a_l}^{\phi\rho u}}{i+1/2}{n} - \frb{\widetilde{\rho_l a_l}^{\phi\rho u}}{i-1/2}{n}} u_0 + \Delta t p_0 \Delta.
\end{aligned}
\end{equation}
Splitting the next iteration phasic momentum $\frb{\phi_l \rho_l u_l}{i}{n+1}$ as $\frb{\phi_l \rho_l}{i}{n+1} \frb{u_l}{i}{n+1}$ yields in
\begin{equation}
\label{eq:phasic_mom_iec_int_3}
\begin{aligned}
\frb{\phi_l \rho_l}{i}{n+1} \frb{u_l}{i}{n+1} & = \frb{\phi_l \rho_l}{i}{n+1}u_0 - {\Delta t} \lrp{\frac{\frb{\widetilde{\phi_l}}{i+1/2}{n} - \frb{\widetilde{\phi_l}}{i-1/2}{n}}{\Delta x} - \Delta}p_0 \\
& + \frac{\Delta t}{\Delta x} \lrsb{\lrp{\frb{\widetilde{\phi_l \rho_l}^{\phi\rho}}{i+1/2}{n} - \frb{\widetilde{\phi_l \rho_l}^{\phi\rho}}{i-1/2}{n}} - \lrp{\frb{\widetilde{\phi_l \rho_l}^{\phi\rho u}}{i+1/2}{n} - \frb{\widetilde{\phi_l \rho_l}^{\phi\rho u}}{i-1/2}{n}}} u_0^2 \\
& - \frac{\Delta t}{\Delta x} \lrsb{\lrp{\frb{\widetilde{\rho_l a_l}^{\phi\rho}}{i+1/2}{n} - \frb{\widetilde{\rho_l a_l}^{\phi\rho}}{i-1/2}{n}} - \lrp{\frb{\widetilde{\rho_l a_l}^{\phi\rho u}}{i+1/2}{n} - \frb{\widetilde{\rho_l a_l}^{\phi\rho u}}{i-1/2}{n}}} u_0.
\end{aligned}
\end{equation}
Therefore, the conditions for the preservation of uniform velocity, \textit{i.e.}, $\frb{u_l}{i}{n+1} = u_0 = \frb{u_l}{i}{n}$, are
\begin{equation}
\label{eq:iec_condition}
\highlight{\Delta = {\frac{\frb{\widetilde{\phi_l}}{i+1/2}{n} - \frb{\widetilde{\phi_l}}{i-1/2}{n}}{\Delta x}}},
\end{equation}
and the face reconstruction of the phasic mass in the convective and the interface regularization fluxes are consistent, given by
\begin{equation}
\label{eq:consistency_mass_mom}
\begin{aligned}
\frb{\widetilde{\phi_l \rho_l}^{\phi\rho}}{i\pm1/2}{n} & = \frb{\widetilde{\phi_l \rho_l}^{\phi\rho u}}{i\pm1/2}{n}, \\
\frb{\widetilde{\rho_l a_l}^{\phi\rho}}{i\pm1/2}{n} & = \frb{\widetilde{\rho_l a_l}^{\phi\rho u}}{i\pm1/2}{n},
\end{aligned}
\end{equation}
for the next time step $n+1 = 1$ for any cell $i$. By induction, uniform velocity is maintained for all future time steps and everywhere.

\subsection{Uniform pressure and consistency conditions}
\label{subsec:uniform_pres}

Furthermore, one may verify uniformity of pressure after time integration. Discretizing the phasic total energy equation yields in
\begin{equation}
\label{eq:phasic_total_energy_iec_int_1}
\begin{aligned}
\frb{\phi_l \rho_l e_{t,l}}{i}{n+1} & = \frb{\phi_l \rho_l e_{t,l}}{i}{n} \\
& - \frac{\Delta t}{\Delta x} \lrp{\frb{\widetilde{\phi_l \rho_l e_{t,l} u_l}}{i+1/2}{n} - \frb{\widetilde{\phi_l \rho_l e_{t,l} u_l}}{i-1/2}{n}} \\
& - \frac{\Delta t}{\Delta x} \lrp{\frb{\widetilde{\phi_l p_l u_l           }}{i+1/2}{n} - \frb{\widetilde{\phi_l p_l u_l           }}{i-1/2}{n}} & \\
& + \frac{\Delta t}{\Delta x} \lrp{\frb{\widetilde{\rho_l e_l a_l           }}{i+1/2}{n} - \frb{\widetilde{\rho_l e_l a_l           }}{i-1/2}{n}} \\
& + \frac{\Delta t}{\Delta x} \lrp{\frb{\widetilde{\rho_l k_l a_l           }}{i+1/2}{n} - \frb{\widetilde{\rho_l k_l a_l           }}{i-1/2}{n}} + \Delta t p_I u_I \Delta.
\end{aligned}
\end{equation}
Given the uniform velocity and pressure assumptions and the condition for the preservation of the uniform velocity in~\cref{eq:phasic_total_energy_iec_int_1}, the $\widetilde{\phi_l p_l u_l}$ and the $\Delta$ terms cancel out.
Splitting phasic total energy into internal and kinetic energy contributions gives
\begin{equation}
\label{eq:phasic_total_energy_iec_int_3}
\begin{aligned}
\frb{\phi_l \rho_l e_{l}}{i}{n+1} & + \frb{\phi_l \rho_l}{i}{n+1} \frac{u_0^2}{2} = \frb{\phi_l \rho_l e_{l}}{i}{n} + \frb{\phi_l \rho_l}{i}{n} \frac{u_0^2}{2} \\
& - \frac{\Delta t}{\Delta x} \lrsb{\lrp{\frb{\widetilde{\phi_l \rho_l e_{l}}}{i+1/2}{n} - \frb{\widetilde{\phi_l \rho_l e_{l}}}{i-1/2}{n}} + \lrp{\frb{\widetilde{\phi_l \rho_l}^{\phi\rho k}}{i+1/2}{n} - \frb{\widetilde{\phi_l \rho_l}^{\phi\rho k}}{i-1/2}{n}}\frac{u_0^2}{2}}u_0 \\
& + \frac{\Delta t}{\Delta x} \lrp{\frb{\widetilde{\rho_l e_l a_l}}{+1/2}{n} - \frb{\widetilde{\rho_l e_l a_l}}{i-1/2}{n}} \\
& + \frac{\Delta t}{\Delta x} \lrp{\frb{\widetilde{\rho_l a_l}^{\phi\rho k}}{+1/2}{n} - \frb{\widetilde{\rho_l a_l}^{\phi\rho k}}{i-1/2}{n}} \frac{u_0^2}{2}.
\end{aligned}
\end{equation}
Another consistency condition follows from~\cref{eq:phasic_total_energy_iec_int_3} upon substitution of~\cref{eq:phasic_mass_iec_int_1}. The face reconstruction of the phasic mass in the convective and the interface regularization fluxes between the phasic mass and phasic momentum discrete transport equations need to satisfy
\begin{equation}
\label{eq:consistency_mass_kinetic}
\begin{aligned}
\frb{\widetilde{\phi_l \rho_l}^{\phi\rho}}{i\pm1/2}{n} & = \frb{\widetilde{\phi_l \rho_l}^{\phi\rho k}}{i\pm1/2}{n}, \\
\frb{\widetilde{\rho_l a_l}^{\phi\rho}}{i\pm1/2}{n} & = \frb{\widetilde{\rho_l a_l}^{\phi\rho k}}{i\pm1/2}{n}.
\end{aligned}
\end{equation}
If both conditions are satisfied, then the discrete phasic internal energy equation is
\begin{equation}
\label{eq:phasic_internal_energy_iec_int_1}
\begin{aligned}
\frb{\phi_l \rho_l e_{l}}{i}{n+1} &  = \frb{\phi_l \rho_l e_{l}}{i}{n} \\
& - \frac{\Delta t}{\Delta x} \lrp{\frb{\widetilde{\phi_l \rho_l e_{l}}}{i+1/2}{n} - \frb{\widetilde{\phi_l \rho_l e_{l}}}{i-1/2}{n}}u_0 \\
& + \frac{\Delta t}{\Delta x} \lrp{\frb{\widetilde{\rho_l e_l a_l}}{i+1/2}{n} - \frb{\widetilde{\rho_l e_l a_l}}{i-1/2}{n}}.
\end{aligned}
\end{equation}
Following~\citet{saurel:1999}, nearly all equations of state (EoS) can be written under the Mie-Gruneisen form of
\begin{equation}
\label{eq:mie-gruneisen-eos-1}
p_l = \lrsb{\gamma_l (\rho_l) - 1} \rho_l e_l - \gamma_l(\rho_l) \pi_l(\rho_l).
\end{equation}
In the phasic internal energy form, the equation of state can be written as
\begin{equation}
\label{eq:mie-gruneisen-eos-2}
\begin{aligned}
\rho_l e_l & = \alpha(\rho_l)p_l - \eta(\rho_l), {\rm or,} \\
       e_l & = \alpha(\rho_l) \frac{p_l}{\rho_l} - \eta(\rho_l) \frac{1}{\rho_l}, \\
\end{aligned}
\end{equation}
where $\alpha(\rho_l) = 1/(\gamma(\rho_l)-1)$ and $\eta(\rho_l) = \gamma(\rho_l)\pi(\rho_l)/(\gamma(\rho_l)-1)$.

\subsubsection{Phasic internal energy convective and volumetric interface regularization flux splittings}
\label{subsubsec:pre-splittings}

The flux splittings of the phasic internal energy convection and volumetric interface regularization are crucial in obtaining the conditions to preserve uniform pressure. Therefore, we present the all possibilities for flux splitting. The phasic internal convective flux can be split into
\begin{align}
\frb{\widetilde{\phi_l \rho_l e_l u_l}}{i\pm1/2}{} & = \xoverline{\phi_l \rho_l e_l u_l}^{(i\pm1/2)} \label{eq:pre_conv_1}, \\
\frb{\widetilde{\phi_l \rho_l e_l u_l}}{i\pm1/2}{} & = \xoverline{\phi_l \rho_l e_l}^{(i\pm1/2)} \xoverline{u_l}^{(i\pm1/2)} \label{eq:pre_conv_2}, \\
\frb{\widetilde{\phi_l \rho_l e_l u_l}}{i\pm1/2}{} & = \xoverline{\phi_l \rho_l}^{(i\pm1/2)} \xoverline{e_l}^{(i\pm1/2)} \xoverline{u_l}^{(i\pm1/2)} \label{eq:pre_conv_3}, \\
\frb{\widetilde{\phi_l \rho_l e_l u_l}}{i\pm1/2}{} & = \xoverline{\phi_l}^{(i\pm1/2)} \xoverline{\rho_l}^{(i\pm1/2)} \xoverline{e_l}^{(i\pm1/2)} \xoverline{u_l}^{(i\pm1/2)} \label{eq:pre_conv_4},
\end{align}
where ``$\; \overline{(\cdot)}^{(i\pm1/2)} \;$'' operator denotes the face reconstruction scheme at the cell face $i\pm 1/2$, \textit{e.g.}, for second-order central $\xoverline{f}^{i\pm1/2} = (f|_i + f|_{i\pm1})/2$. The splittings are respectively referred to as the divergence form [\cref{eq:pre_conv_1}], the quadratic form [\cref{eq:pre_conv_2}], the cubic form [\cref{eq:pre_conv_3}], and the quartic form [\cref{eq:pre_conv_4}]. Analogously, the volumetric phasic internal energy interface regularization flux can be split into
\begin{align}
\frb{\widetilde{\rho_l e_l a_l}}{i\pm1/2}{} & = \xoverline{\rho_l e_l a_l}^{(i\pm1/2)} \label{eq:pre_int_1}, \\
\frb{\widetilde{\rho_l e_l a_l}}{i\pm1/2}{} & = \xoverline{\rho_l e_l}^{(i\pm1/2)} \xoverline{a_l}^{(i\pm1/2)} \label{eq:pre_int_2}, \\
\frb{\widetilde{\rho_l e_l a_l}}{i\pm1/2}{} & = \xoverline{\rho_l}^{(i\pm1/2)} \xoverline{e_l}^{(i\pm1/2)} \xoverline{a_l}^{(i\pm1/2)} \label{eq:pre_int_3},
\end{align}
where the splittings are respectively referred to as the divergence form [\cref{eq:pre_int_1}], the quadratic form [\cref{eq:pre_int_2}], and the cubic form [\cref{eq:pre_int_3}].

Upon substitution of equation of state [\cref{eq:mie-gruneisen-eos-2}] in the discrete transport equation for the phasic internal energy, it follows that, although there exist conditions for the preservation of uniform pressure for all combinations, only the quadratic and quartic phasic internal energy convective flux splittings paired with either quadratic or cubic volumetric phasic internal energy interface regularization flux splittings provide a condition consistent with the transport of the volume fraction equation. The divergence form for flux splittings is not evaluated here due to its increased aliasing errors~\citep{blaisdell:1996,jain:2022a}. 

\begin{proposition}
\label{prop:split_re_int}
Both quadratic and cubic splittings of the volumetric phasic internal energy interface regularization flux are at least approximately consistent with the discretization of the volume fraction equation.
\end{proposition}
\begin{proposition}
\label{prop:split_pre_conv}
Both quadratic and quartic splittings of the phasic internal energy convective flux are at least approximately consistent with the discretization of the volume fraction equation.
\end{proposition}
\begin{proof}
Consider first both quadratic splittings for the convective and interface regularization fluxes.
Splitting the cell-wise phasic internal energy $\phi_l \rho_l e_l |_i^n$ as $\phi_{l} |_i^n \rho_l e_l |_i^n$ and substituting~\cref{eq:mie-gruneisen-eos-2} in~\cref{eq:phasic_internal_energy_iec_int_1}, given the uniform pressure assumption, yields in
\begin{equation}
\label{eq:phasic_internal_energy_iec_int_2}
\begin{aligned}
\frb{\phi_l}{i}{n+1} \frb{(\alpha_l p_l + \eta_l)}{i}{n+1} & = \frb{\phi_l}{i}{n} \frb{(\alpha_l p_0 + \eta_l)}{i}{n} \\
& - \frac{\Delta t}{\Delta x} \lrsb{\lrp{\alpha_l \frb{\xoverline{\phi_l}}{i+1/2}{n} p_0 + \eta_l \frb{\xoverline{\phi_l}}{i+1/2}{n}} - \lrp{\alpha_l \frb{\xoverline{\phi_l}}{i-1/2}{n} p_0 + \eta_l \frb{\xoverline{\phi_l}}{i-1/2}{n}}}u_0 \\
& + \frac{\Delta t}{\Delta x} \lrsb{(\alpha_l p_0 + \eta_l) \frb{\xoverline{a_l}}{i+1/2}{n} - (\alpha_l p_0 + \eta_l) \frb{\xoverline{a_l}}{i+1/2}{n}}.
\end{aligned}
\end{equation}
In particular, for ideal and stiffened-gas EoS, the parameters $\alpha_l$ and $\eta_l$ are constants. Furthermore, Saurel and Abgrall~\citep{saurel:1999} argue that $\rho_l$ does not vary as significantly as $\phi_l \rho_l$ does across the interface, which suggests that $\alpha_l$ and $\eta_l$ can be locally approximated as constants. Rearranging~\cref{eq:phasic_internal_energy_iec_int_2} leads to
\begin{equation}
\label{eq:phasic_internal_energy_iec_int_3}
\begin{aligned}
\frb{\phi_l}{i}{n+1} \frb{(\alpha_l p_l + \eta_l)}{i}{n+1} = \Bigg [ \frb{\phi_l}{i}{n} \frb{(\alpha_l p_0 + \eta_l)}{i}{n} & - \frac{\Delta t}{\Delta x} \lrp{\frb{\xoverline{\phi_l}}{i+1/2}{n} - \frb{\xoverline{\phi_l}}{i-1/2}{n}}u_0 \\
& + \frac{\Delta t}{\Delta x} \lrp{\frb{\xoverline{a_l}}{i+1/2}{n} - \frb{\xoverline{a_l}}{i-1/2}{n}} \Bigg ] (\alpha p_0 + \eta).
\end{aligned}
\end{equation}
Hence, if the volume fraction transport equation is discretized as
\begin{equation}
\label{eq:volume_fraction_iec_int_1}
\begin{aligned}
\highlight{\frb{\phi_l}{i}{n+1} = \frb{\phi_l}{i}{n} - \frac{\Delta t}{\Delta x} \lrp{\frb{\xoverline{\phi_l}}{i+1/2}{n} - \frb{\xoverline{\phi_l}}{i-1/2}{n}}u_0 + \frac{\Delta t}{\Delta x} \lrp{\frb{\xoverline{a_l}}{i+1/2}{n} - \frb{\xoverline{a_l}}{i-1/2}{n}}},
\end{aligned}
\end{equation}
then, ~\cref{eq:phasic_internal_energy_iec_int_3} simplifies to
\begin{equation}
\label{eq:volume_fraction_iec_int_2}
\begin{aligned}
\alpha_l \frb{p_l}{i}{n+1} + \eta_l = \alpha_l p_0 + \eta_l \implies \frb{p_l}{i}{n_1} = p_0 = \frb{p_l}{i}{n},
\end{aligned}
\end{equation}
which provides the consistent discretization for the volume fraction transport equation.

With respect to~\cref{prop:split_re_int}, if one takes the cubic splitting for the volumetric phasic internal energy interface regularization flux, then the phasic internal energy transport equation would be in the form of
\begin{equation}
\label{eq:phasic_internal_energy_iec_int_4}
\begin{aligned}
\frb{\phi_l}{i}{n+1} \frb{(\alpha_l p_l + \eta_l)}{i}{n+1} & = \frb{\phi_l}{i}{n} \frb{(\alpha_l p_0 + \eta_l)}{i}{n} \\
& - \frac{\Delta t}{\Delta x} \lrsb{\lrp{\alpha_l \frb{\xoverline{\phi_l}}{i+1/2}{n} p_0 + \eta_l \frb{\xoverline{\phi_l}}{i+1/2}{n}} - \lrp{\alpha_l \frb{\xoverline{\phi_l}}{i-1/2}{n} p_0 + \eta_l \frb{\xoverline{\phi_l}}{i-1/2}{n}}}u_0 \\
& + \frac{\Delta t}{\Delta x} \Bigg [\frb{\xoverline{\rho_l}}{i+1/2}{n} \lrp{\alpha_l \frac{p_0}{\frb{\xoverline{\rho_l}}{i+1/2}{n}} + \eta_l \frac{1}{\frb{\xoverline{\rho_l}}{i+1/2}{n}}} \frb{\xoverline{a_l}}{i+1/2}{n} \\
& \hspace{2.5em} - \frb{\xoverline{\rho_l}}{i-1/2}{n} \lrp{\alpha_l \frac{p_0}{\frb{\xoverline{\rho_l}}{i-1/2}{n}} + \eta_l \frac{1}{\frb{\xoverline{\rho_l}}{i-1/2}{n}}} \frb{\xoverline{a_l}}{i-1/2}{n} \Bigg ] + O(\Delta \rho_l^2),
\end{aligned}
\end{equation}
where the last term in square brackets is the same as the one in~\cref{eq:phasic_internal_energy_iec_int_2}. The $O(\Delta \rho_l^2)$ error term\footnote{From approximating the arithmetic average of the inverse of phasic density at the cell face by the inverse of the arithmetic average for a second-order central scheme.} is negligible given the small variation in $\rho_l$ assumption, which is also used in taking $\alpha_l$ and $\eta_l$ as local constants~\citep{saurel:1999}. \cref{eq:phasic_internal_energy_iec_int_4} simplifies to ~\cref{eq:phasic_internal_energy_iec_int_3}, therefore requiring the same condition for the preservation of uniform pressure.

With respect to~\cref{prop:split_pre_conv}, if one takes the quartic splitting for the phasic internal energy convection flux, then the phasic internal energy transport equation would be in the form of
\begin{equation}
\label{eq:phasic_internal_energy_iec_int_5}
\begin{aligned}
\frb{\phi_l}{i}{n+1} \frb{(\alpha_l p_l + \eta_l)}{i}{n+1} & = \frb{\phi_l}{i}{n} \frb{(\alpha_l p_0 + \eta_l)}{i}{n} \\
& - \frac{\Delta t}{\Delta x} \Bigg [\frb{\xoverline{\phi_l}}{i+1/2}{n} \frb{\xoverline{\rho_l}}{i+1/2}{n} \lrp{\alpha_l \frac{p_0}{\frb{\xoverline{\rho_l}}{i+1/2}{n}} + \eta_l \frac{1}{\frb{\xoverline{\rho_l}}{i+1/2}{n}}} \\
& \hspace{2.5em}- \frb{\xoverline{\phi_l}}{i-1/2}{n} \frb{\xoverline{\rho_l}}{i-1/2}{n} \lrp{\alpha_l \frac{p_0}{\frb{\xoverline{\rho_l}}{i-1/2}{n}} + \eta_l \frac{1}{\frb{\xoverline{\rho_l}}{i-1/2}{n}}} \Bigg ]u_0 + O(\Delta \rho_l^2) \\
& + \frac{\Delta t}{\Delta x} \lrsb{(\alpha_l p_0 + \eta_l) \frb{\xoverline{a_l}}{i+1/2}{n} - (\alpha_l p_0 + \eta_l) \frb{\xoverline{a_l}}{i+1/2}{n}}.
\end{aligned}
\end{equation}
where the first term in square brackets is the same as the one in~\cref{eq:phasic_internal_energy_iec_int_2}, which again would simplify to ~\cref{eq:phasic_internal_energy_iec_int_3}, therefore requiring the same condition for the preservation of uniform pressure.

Nevertheless, if one takes the cubic splitting for the phasic internal energy convective flux, then the first term in the right-hand-side of ~\cref{eq:phasic_internal_energy_iec_int_2} would be distinct. The phasic internal energy transport equation would be in the form of
\begin{equation}
\label{eq:phasic_internal_energy_iec_int_6}
\begin{aligned}
\frb{\phi_l}{i}{n+1} \frb{(\alpha_l p_l + \eta_l)}{i}{n+1} & = \frb{\phi_l}{i}{n} \frb{(\alpha_l p_0 + \eta_l)}{i}{n} \\
& - \frac{\Delta t}{\Delta x} \Bigg [\frb{\xoverline{\phi_l \rho_l}}{i+1/2}{n} \lrp{\alpha_l \frac{p_0}{\frb{\xoverline{\rho_l}}{i+1/2}{n}} + \eta_l \frac{1}{\frb{\xoverline{\rho_l}}{i+1/2}{n}}} \\
& \hspace{2.5em}- \frb{\xoverline{\phi_l \rho_l}}{i-1/2}{n} \lrp{\alpha_l \frac{p_0}{\frb{\xoverline{\rho_l}}{i-1/2}{n}} + \eta_l \frac{1}{\frb{\xoverline{\rho_l}}{i-1/2}{n}}} \Bigg ]u_0 + O(\Delta \rho_l^2) \\
& + \frac{\Delta t}{\Delta x} \lrsb{(\alpha_l p_0 + \eta_l) \frb{\xoverline{a_l}}{i+1/2}{n} - (\alpha_l p_0 + \eta_l) \frb{\xoverline{a_l}}{i+1/2}{n}},
\end{aligned}
\end{equation}
which would then require the volume fraction transport equation, for the preservation of uniform pressure, to be in the form of
\begin{equation}
\label{eq:volume_fraction_iec_int_3}
\begin{aligned}
\frb{\phi_l}{i}{n+1} = \frb{\phi_l}{i}{n} - \frac{\Delta t}{\Delta x} \lrp{\frac{\frb{\xoverline{\phi_l\rho_l}}{i+1/2}{n}}{\frb{\xoverline{\rho_l}}{i+1/2}{n}} - \frac{\frb{\xoverline{\phi_l\rho_l}}{i-1/2}{n}}{\frb{\xoverline{\rho_l}}{i-1/2}{n}}}u_0 + \frac{\Delta t}{\Delta x} \lrp{\frb{\xoverline{a_l}}{i+1/2}{n} - \frb{\xoverline{a_l}}{i-1/2}{n}}.
\end{aligned}
\end{equation}
The same condition would follow if a cubic split were used for the interface regularization flux.
\end{proof}

In summary, the proposed model allows for the construction of a discretization that satisfies the IEC. Moreover, we demonstrated several combinations of splittings that satisfy these conditions. However, as it will be further discussed, in order to satisfy other consistency conditions for implicit discrete kinetic energy and entropy preservation, the quartic and cubic splittings will be chosen for the phasic internal energy convective and interface regularization fluxes, respectively.

%% file: chapters/numerics.tex
\section{Numerical implementation}
\label{sec:num_methods}

In this work, we perform operator splitting of the hyperbolic integration and the relaxation steps in the form of $\Vec{Q} |_i^{n+1} = L_{\rm p} L_{\rm u} L_{\rm h} (\Vec{Q} |_i^{n})$. The hyperbolic integration is detailed in~\cref{subsec:h}, and the velocity and pressure relaxations in~\cref{subsec:hu,subsec:hup}, respectively. In summary, the system of equations is presented in standard vector form. The phasic conservation equations in~\cref{eq:7eq_proposed} in one dimension may be arranged as
\begin{equation}
\label{eq:vec_form}
\frac{\partial \Vec{Q}}{\partial t} + \frac{\partial \Vec{E}}{\partial x} - \frac{\partial \Vec{E}_{\rm v}}{\partial x} - \frac{\partial \Vec{E}_{\rm s}}{\partial x} = \Vec{H} \frac{\partial \phi_1}{\partial x} + \Vec{\Psi}_\lambda + \Vec{\Psi}_\mu, 
\end{equation}
where the vectors are defined by
\begin{equation}
\label{eq:vec_defs}    
\begin{aligned}
&\Vec{Q} = \begin{bmatrix}
\phi_1 \rho_1 \\
\phi_1 \rho_1 u_1 \\
\phi_1 \rho_1 e_{t,1} \\
\phi_2 \rho_2 \\
\phi_2 \rho_2 u_2 \\
\phi_2 \rho_2 e_{t,2}
\end{bmatrix}, \;
\Vec{E} = \begin{bmatrix}
\phi_1 \rho_1 u_1 \\
\phi_1 \rho_1 u_1^2 + \phi_1 p_1 \\
\phi_1 \left (\rho_1 e_{t,1} + p_1 \right) u_1 \\
\phi_2 \rho_2 u_2 \\
\phi_2 \rho_2 u_2^2 + \phi_2 p_2 \\
\phi_2 \left (\rho_2 e_{t,2} + p_2 \right) u_2 
\end{bmatrix}, \\
& \Vec{E}_{\rm v} = \begin{bmatrix}
0 \\
\phi_1 \tau_1 \\
\phi_1 \tau_1 u_1 \\
0 \\
\phi_2 \tau_2 \\
\phi_2 \tau_2 u_2
\end{bmatrix}, \;
\Vec{E}_{\rm s} = \begin{bmatrix}
\rho_1 a_1 \\
\rho_1 u_1 a_1 \\
\rho_1  {\displaystyle \frac{u_1^2}{2}}  a_1+ \rho_1 e_1 a_1 \\
\rho_2 a_2 \\
\rho_2 u_2 a_2 \\
\rho_2 {\displaystyle \frac{u_2^2}{2}} a_2 + \rho_2 e_2 a_2
\end{bmatrix}, \\
& \Vec{H} = \begin{bmatrix}
0 \\
p_I \\
p_I u_I \\
0 \\
-p_I \\
-p_I u_I 
\end{bmatrix} , \;
\Vec{\Psi}_\lambda = \begin{bmatrix}
0 \\
\lambda \left (u_2 - u_1 \right) \\
u_I \lambda \left (u_2 - u_1 \right) \\
0 \\
-\lambda \left (u_2 - u_1 \right) \\
-u_I \lambda \left (u_2 - u_1 \right) \\
\end{bmatrix} , \;
\Vec{\Psi}_\mu = \begin{bmatrix}
0 \\
0 \\
-p_I \mu \left (p_1 - p_2 \right) \\
0 \\
0 \\
p_I \mu \left (p_1 - p_2 \right) \\
\end{bmatrix} . \\
\end{aligned}
\end{equation}
The~\cref{eq:vec_form} is appended by the volume fraction transport of the phase 1, which also contains one step to be solved first in the hyperbolic integration and second in the pressure relaxation steps.

\subsection{Hyperbolic integration with a KEEP scheme}
\label{subsec:h}

A finite volume approach is adopted to integrate~\cref{eq:vec_form}, which yields in
\begin{equation}
\label{eq:fv}
\begin{aligned}
\left . \Vec{Q} \right |_{i}^{n+1} & = \left . \Vec{Q} \right |_{i}^{n} - \frac{\Delta t}{\Delta x} \lrp{\left . \Vec{E} \right |_{i+1/2}^{n} - \left . \Vec{E} \right |_{i-1/2}^{n}} + \frac{\Delta t}{\Delta x} \lrp{\left . \Vec{E_{\rm v}} \right |_{i+1/2}^{n} - \left . \Vec{E_{\rm v}} \right |_{i-1/2}^{n}} \\
& + \frac{\Delta t}{\Delta x} \lrp{\left . \Vec{E_{\rm s}} \right |_{i+1/2}^{n} - \left . \Vec{E_{\rm s}} \right |_{i-1/2}^{n}} + \Delta t \left . \Vec{H} \right |_{i}^{n} \Delta,
\end{aligned}
\end{equation}
where the state variables for the $i^{\rm th}$ cell of the $n+1$ time step are updated based on the current time step, $n$, convective, $\Vec{E} |_{i+1/2}^n$, viscous, $\Vec{E_{\rm v}} |_{i+1/2}^n$, and interface regularization, $\Vec{E_{\rm s}} |_{i+1/2}^n$, fluxes constructed at the cell faces $i\pm 1/2$. In~\cref{eq:fv}, the $\Delta$ term multiplying the interface force and work term is the discrete form of $\partial \phi_1 / \partial x$. The fluxes are constructed based on a second-order central scheme, for which the splitting is performed as an analogous extension of the kinetic-energy and entropy-preserving (KEEP) scheme of~\citet{jain:2022a}. The convective, viscous and sharpening fluxes for phase 1 are given by
\begin{equation}
\label{eq:keepfluxes}
\begin{aligned}
\left . \Vec{E}_1 \right |_{i\pm1/2}^n 
& = \begin{bmatrix}
\xoverline{\phi_1}^{(i\pm1/2)} \xoverline{\rho_1}^{(i\pm1/2)} \xoverline{u_1}^{(i\pm1/2)} \\
\xoverline{\phi_1}^{(i\pm1/2)} \xoverline{\rho_1}^{(i\pm1/2)} \xoverline{u_1}^{(i\pm1/2)} \xoverline{u_1}^{(i\pm1/2)} + \xoverline{\phi_1 p_1}^{(i\pm1/2)} \\
\begin{aligned}
\xoverline{\phi_1}^{(i\pm1/2)} \xoverline{\rho_1}^{(i\pm1/2)} & \xoverline{e_1}^{(i\pm1/2)} \xoverline{u_1}^{(i\pm1/2)} + \xoverline{\phi_1}^{(i\pm1/2)} \xoverline{\rho_1}^{(i\pm1/2)} {\displaystyle \frac{\left. u_1 \right |_{i} \left. u_1 \right |_{i\pm1}}{2}} \xoverline{u_1}^{(i\pm1/2)} \\
& + {\displaystyle \frac{\left. u_1 \right |_{i}\left. \phi_1 p_1 \right |_{i\pm1} + \left. u_1 \right |_{i\pm1}\left. \phi_1 p_1 \right |_{i}}{2} }
\end{aligned}
\end{bmatrix}, \\
\left . \Vec{E}_{1,\rm v} \right |_{i\pm1/2}^n 
& = \begin{bmatrix}
0 \\
\xoverline{\phi_1}^{(i\pm1/2)} \xoverline{\tau_1}^{(i\pm1/2)} \\
\xoverline{\phi_1}^{(i\pm1/2)} \xoverline{\tau_1}^{(i\pm1/2)} \xoverline{u_1}^{(i\pm1/2)}
\end{bmatrix}, \\
\left . \Vec{E}_{1,\rm s} \right |_{i\pm1/2}^n 
& = \begin{bmatrix}
\xoverline{\rho_1}^{(i\pm1/2)} \xoverline{a_1}^{(i\pm1/2)} \\
\xoverline{\rho_1}^{(i\pm1/2)} \xoverline{u_1}^{(i\pm1/2)} \xoverline{a_1}^{(i\pm1/2)} \\
\xoverline{\rho_1}^{(i\pm1/2)} \xoverline{e_1}^{(i\pm1/2)} \xoverline{a_1}^{(i\pm1/2)} + \xoverline{\rho_1}^{(i\pm1/2)} {\displaystyle \frac{\left. u_1 \right |_{i} \left. u_1 \right |_{i\pm1}}{2}} \xoverline{a_1}^{(i\pm1/2)}
\end{bmatrix}.
\end{aligned}
\end{equation}
The fluxes for phase 2 are analogous to the ones defined in~\cref{eq:keepfluxes} with phasic index instead set to $l=2$.

The volume fraction transport in~\cref{eq:7eq_baseline} is not in conservative form. 
Stable discretization follows from the IEC as proposed by \citet{saurel:1999}. The preservation of uniform velocity requires $\Delta = \lrp{\overline{\phi_1}^{(i+1/2)} - \overline{\phi_1}^{(i-1/2)}}/ \Delta x$ in~\cref{eq:fv}. And the preservation of uniform pressure, considering the quartic splitting for the convection of phasic internal energy, requires the transport equation for the volume fraction field to be in the form of
\begin{equation}
\label{eq:vol_frac_int}
\begin{aligned}
\left .\phi_1 \right |_{i}^{n+1} & = \left .\phi_1 \right |_{i}^{n} - \frac{\Delta t}{\Delta x} \left [ u_{I,i}^{n} \lrp{\xoverline{\phi_1}^{(i+1/2)} - \xoverline{\phi_1}^{(i-1/2)}} \right ] + \frac{\Delta t}{\Delta x} \left [\lrp{\xoverline{a_1}^{(i+1/2)} - \xoverline{a_1}^{(i-1/2)}} \right ] .
\end{aligned}
\end{equation}

Time integration is performed using a fourth-order Runge-Kutta scheme for increased stability bounds, where $\Vec{Q} |_i^{n+1/2}$ are constructed based on the corresponding sub timestep fluxes and source terms.

\subsection{Velocity relaxation}
\label{subsec:hu}

After the hyperbolic integration step, the velocity relaxation is constructed and has the general form of
\begin{equation}
\label{eq:vel_relax}
\frac{\partial \Vec{Q}}{\partial t} = \Vec{\Psi}_\lambda,
\end{equation}
where both the volume fraction of phase 1 and each phasic mass are conserved. The approach of \citet{saurel:1999} is adopted herein; the reader is referred to their work for the complete derivation. One may rearrange the phasic momentum equation with the conservation of phasic mass to obtain the phasic velocity transport equations in the form of
\begin{equation}
\label{eq:vel_relax_sys_u}
\begin{aligned}
& \frac{\partial u_1}{\partial t} = \frac{\lambda \lrp{u_2 - u_1}}{\phi_1 \rho_1}, \\
& \frac{\partial u_2}{\partial t} = -\frac{\lambda \lrp{u_2 - u_1}}{\phi_2 \rho_2}.
\end{aligned}
\end{equation}
The steady state solution assuming $\lambda > 0$ is
\begin{equation}
\label{eq:vel_relax_sol}
u^{\rm hu} = u_1^{\rm hu} = u_2^{\rm hu} = \frac{\lrp{\phi_1 \rho_1 u_1}^{\rm h} + \lrp{\phi_2 \rho_2 u_2}^{\rm h}}{\lrp{\phi_1 \rho_1}^{\rm h} + \lrp{\phi_2 \rho_2}^{\rm h}},
\end{equation}
where the superscripts `h' and `hu' denote the values after hyperbolic integration and hyperbolic integration and velocity relaxation, respectively. One may then update the phasic momentum by multiplying the phasic mass $(\phi_l \rho_l)^{\rm hu} = (\phi_l \rho_l)^{\rm h}$ (conserved) with $u^{\rm hu}$. 
The phasic total energies must be updated accordingly given the energy transfer in the form of drag between the phases during the velocity relaxation process. The phasic internal energy transport equations are obtained by replacing the conservation of phasic mass and the phasic velocity transport equations [\cref{eq:vel_relax_sys_u}]. If $u_I$ is assumed to vary linearly during the velocity relaxation step, then the internal phasic energies per unit mass are updated by
\begin{equation}
\label{eq:vel_relax_sys_e3}
\begin{aligned}
& e_1^{\rm hu} = e_1^{\rm h} + \frac{1}{2} \lrp{u_1^{\rm hu} - u_1^{\rm h}}\lrp{u_I^{\rm h} - u_1^{\rm h}}, \\
& e_2^{\rm hu} = e_2^{\rm h} + \frac{1}{2} \lrp{u_2^{\rm hu} - u_2^{\rm h}}\lrp{u_I^{\rm h} - u_2^{\rm h}}.
\end{aligned}
\end{equation}
The phasic total energy is updated by multiplying the conserved phasic mass $(\phi_l \rho_l)^{\rm hu}$ by $e_l^{\rm hu} + {\displaystyle \frac{1}{2} \left ( u_l^{\rm hu} \right )^{2}}$. 

\subsection{Pressure relaxation}
\label{subsec:hup}

The general form of the pressure relaxation after the hyperbolic integration and velocity relaxation is
\begin{equation}
\label{eq:pres_relax}
\begin{aligned}
& \frac{\partial \phi_1}{\partial t} = \mu(p_1 - p_2), \\
& \frac{\partial \Vec{Q}}{\partial t} = \Vec{\Psi}_\mu,
\end{aligned}
\end{equation}
where both the phasic mass and momentum are conserved in the pressure relaxation step. One may replace $\mu (p_1 - p_2) = \partial \phi_1 / \partial t$ in the phasic total energy equations through $\Vec{\Psi}_\mu$ and obtain transport equations of the phasic internal energies per unit mass in the form of
\begin{equation}
\label{eq:pres_relax_sys_e1}
\begin{aligned}
& \frac{\partial e_1}{\partial t} = - \frac{p_I}{\phi_1 \rho_1} \frac{\partial \phi_1}{\partial t}, \\
& \frac{\partial e_2}{\partial t} =   \frac{p_I}{\phi_2 \rho_2} \frac{\partial \phi_1}{\partial t}.
\end{aligned}
\end{equation}
An approximate integration of~\cref{eq:pres_relax_sys_e1}, considering that $p_I$ varies linearly with $\phi_1$~\cite{pelanti:2014} or approximating $\overline{p_I} = {\displaystyle \frac{1}{2}} \lrp{p_I^{\rm hup} + p_I^{\rm hu}}$~\cite{saurel:1999,saurel:2009,zein:2010}, where the superscript `hup' denotes the stage after hyperbolic integration, velocity and pressure relaxations, is given by
\begin{equation}
\label{eq:pres_relax_sys_e2}
\begin{aligned}
& e_1^{\rm hup} = e_1^{\rm hu} - \frac{\overline{p_I}}{(\phi_1 \rho_1)^{\rm hu}} \lrp{\phi_1^{\rm hup} - \phi_1^{\rm hu}} = e_1^{\rm hu} - \frac{1}{(\phi_1 \rho_1)^{\rm hu}} \frac{p_I^{\rm hup} + p_I^{\rm hu}}{2} \lrp{\phi_1^{\rm hup} - \phi_1^{\rm hu}}, \\
& e_2^{\rm hup} = e_2^{\rm hu} + \frac{\overline{p_I}}{(\phi_2 \rho_2)^{\rm hu}} \lrp{\phi_1^{\rm hup} - \phi_1^{\rm hu}} = e_2^{\rm hu} + \frac{1}{(\phi_2 \rho_2)^{\rm hu}} \frac{p_I^{\rm hup} + p_I^{\rm hu}}{2} \lrp{\phi_1^{\rm hup} - \phi_1^{\rm hu}},
\end{aligned}
\end{equation}
which is a system of 2 equations and 2 unknowns, given that $p_I^{\rm hup} = p_1^{\rm hup} = p_2^{\rm hup}$. For arbitrary equations of state, the system of equations in~\cref{eq:pres_relax_sys_e2} must be solved iteratively, by replacing the $e_l$ in terms of $\rho_l$ and $p_l$ and solving for $\phi_1^{\rm hup}$ such that $p_1^{\rm hup} = p_2^{\rm hup}$~\cite{saurel:1999}. A more robust approach was proposed that analytically enforces $p_I^{\rm hup} = p_1^{\rm hup} = p_2^{\rm hup}$~\citep{lallemand:2000,pelanti:2014,furfaro:2015}, and is specific to the SG-EoS. It follows from using the equation of state to rearrange~\cref{eq:pres_relax_sys_e2}, as a function of only $p_I^{\rm hup}$ and $\phi_1^{\rm hup}$, to obtain an algebraic relation for $p_I^{\rm hup}$ in the form of
\begin{equation}
\label{eq:pres_relax_sol}
a\lrp{p_I^{\rm hup}}^2 + b\lrp{p_I^{\rm hup}} + c = 0,
\end{equation}
where the constants are given by
\begin{equation}
\label{eq:pres_relax_sol_consts}
\begin{aligned}
& a = 1 + \phi_1^{\rm hu} \gamma_2 + \phi_2^{\rm hu} \gamma_1, \\
& b = c_1 \phi_2^{\rm hu} + c_2 \phi_1^{\rm hu} - \lrp{1 + \gamma_2} \phi_1^{\rm hu} p_1^{\rm hu} - \lrp{1 + \gamma_1} \phi_2^{\rm hu} p_2^{\rm hu}, \\
& c = - \lrp{c_2 \phi_1^{\rm hu} p_1^{\rm hu} + c_1 \phi_2^{\rm hu} p_2^{\rm hu}}, \\
& c_1 = 2\gamma_1 \pi_1 + \lrp{\gamma_1 - 1} p_I^{\rm hu}, \\
& c_2 = 2\gamma_2 \pi_2 + \lrp{\gamma_2 - 1} p_I^{\rm hu}.
\end{aligned}
\end{equation}
The~\cref{eq:pres_relax_sol} has two analytical solutions, and only the physical positive one is retained. Rearranging~\cref{eq:pres_relax_sys_e2} yields in
\begin{equation}
\label{eq:pres_relax_sol_phi}
\phi_1^{\rm hup} = \phi_1^{\rm hu} \frac{\lrp{\gamma_1 - 1}p_I^{\rm hup} + 2 p_1^{\rm hu} + c_1}{\lrp{\gamma_1 + 1}p_I^{\rm hup} + c_1}.
\end{equation}
And lastly, $e_l^{\rm hup}$ is updated using~\cref{eq:pres_relax_sys_e2} and the phasic total energy by multiplying the conserved mass $\lrp{\phi_l \rho_l}^{\rm h} = \lrp{\phi_l \rho_l}^{\rm hu}$ with the relaxed internal energy $e_l^{\rm hup}$ plus the conserved phasic kinetic energy per unit mass ${\displaystyle \frac{1}{2}} \lrp{u_l^{\rm hu}}^2 = {\displaystyle \frac{1}{2}} \lrp{u_l^{\rm hup}}^2$.

%% file: chapters/results.tex
\section{Results}
\label{sec:results}

The following test cases are proposed to evaluate accuracy, stability and robustness of the proposed method under challenging and realistic conditions for moderately compressible two-phase flows. High-density ratio advection is evaluated in~\cref{subsec:int_advect}, acoustic problems in~\cref{subsec:comp}, and turbulence in~\cref{subsec:turb}.

All interfaces are initialized based on the finite-bound equilibrium profile [\cref{eq:equilibrium_profile_proposed}], where the interface thickness is set to $\epsilon=0.6\Delta x$; the interface regularization velocity to $\Gamma = \max_l (|u_{l,i}|_\infty)$; and the amount of the conjugate phase to $\delta = 10^{-8}$, unless stated otherwise. Phasic masses are initialized, considering initial uniform phasic density, with $\phi_l \rho_l = \phi_l \rho_{0,l}$. Phasic velocities and phasic pressure are problem dependent. The former close phasic momentum and the latter close phasic internal energies [\cref{eq:sg_eos}], and so does for phasic total energies. Herein, all initial phasic velocities and phasic pressures are assumed to be in equilibrium. Standard ambient ($p_l=p_{\rm atm} = 10^5$ Pa) stiffened gas equation of state parameters are used for the advection and acoustic cases, all presented in~\cref{tab:dim-sg-eos-par}, non-dimensional values are used for the compressible two-phase Taylor-Green vortex case, presented therein.
\begin{table}
\centering
\begin{tabular}{cccc}
\hline
                      & Air & Water            & Kerosene             \\ \hline
$\gamma_l \; [-]$     & 1.4 & 4.4              & 4.4                  \\
$\pi_l \; [{\rm Pa}]$ & 0   & $6\times 10^{8}$ & $3.266\times 10^{8}$ \\ \hline
\end{tabular}
\caption{Dimensional stiffened gas equation of state parameters.}
\label{tab:dim-sg-eos-par}
\end{table}

\subsection{Interface advection tests}
\label{subsec:int_advect}


The accuracy, stability and robustness of high-density ratio interface dynamics are evaluated by evolving a water drop in ambient air with uniform velocity and pressure. The preservation of uniform primitive (velocity and pressure) conditions is evidence of the IEC characteristics [\cref{post:iec}] of the scheme~\cite{saurel:1999}.

High density ratio is achieved by setting the air density to $1$ and the water density to $1000$, viscosity for both phases is set to $0$. Moderate compressibility is achieved by setting the phasic velocities to $u_l = 100$ m/s and the phasic pressures to $p_l = 10^{5}$ Pa, which approximately yields a $Ma_{\rm air} \approx 0.25$. The one-dimensional computational grid spans $[0, 1]$ m over 200 cells, and the water drop is initialized at $0.5$ with a radius $R=0.25$. Both boundaries are periodic for all quantities. The final time is at $t=2$ s, such that the drop advects 200 times over the domain (flow through times, $\tau_{\rm FT}$), to demonstrate long-time integration stability. The time step is set to a constant $\Delta t = 1.25\times10^{-6}$ s, which yields an acoustic CFL approximately equal to $0.5$. This setup is similar to the one proposed by~\citep{wong:2021}; nevertheless, the simulation is integrated to several $\tau_{\rm FT}$ here, whereas it was only advected for $1\tau_{\rm FT}$ in \citet{wong:2021}.

\begin{figure}
    \centering
    \begin{subfigure}{0.4\linewidth}
    \includegraphics[width=\linewidth]{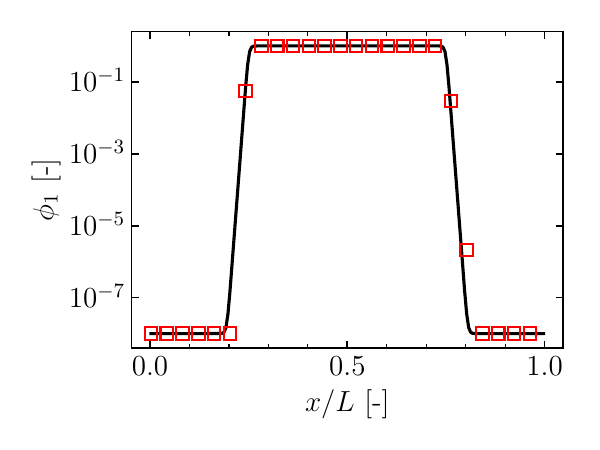}
    \caption{Volume fraction, $\phi_1$}
    \end{subfigure}%
    \begin{subfigure}{0.45\linewidth}
    \includegraphics[width=\linewidth]{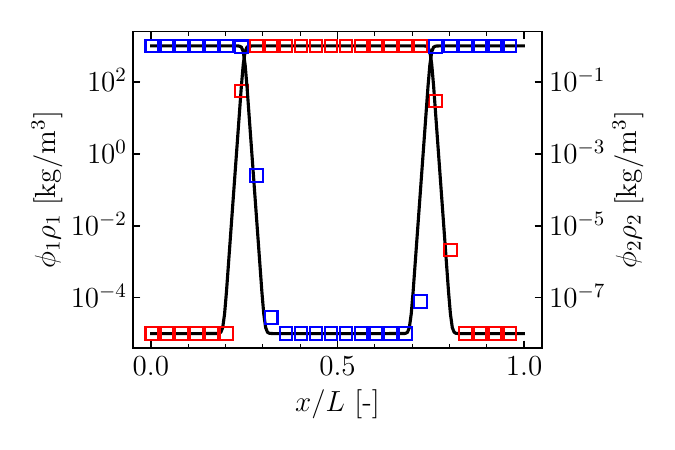}
    \caption{Phasic mass, $\phi_l \rho_l$}
    \end{subfigure}
    
    \begin{subfigure}{0.45\linewidth}
    \includegraphics[width=\linewidth]{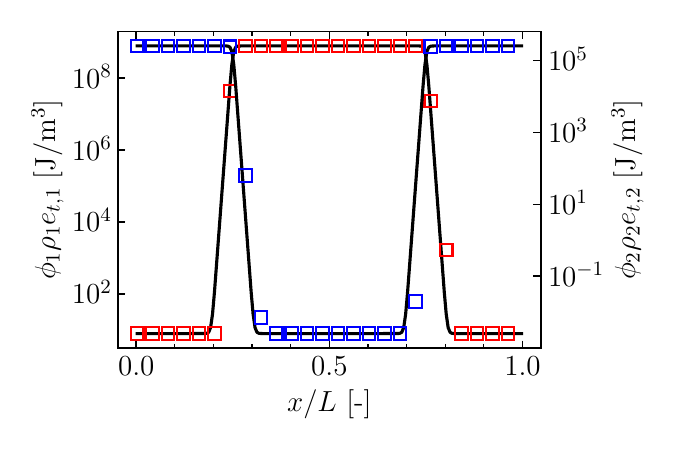}
    \caption{Phasic total energy, $\phi_l \rho_l e_{t,l}$}
    \end{subfigure}
    \caption{High-density ratio interface advection. Initial profile is depicted by solid black lines (\rule[2pt]{0.35cm}{1pt}), and the current simulation after advection by red ({\protect\tikz\protect\draw[line width = 0.75pt, red] (0,0) rectangle +(1ex,1ex) ;}, phase 1) and blue ({\protect\tikz\protect\draw[line width = 0.75pt, blue] (0,0) rectangle +(1ex,1ex) ;}, phase 2) squares. Profiles after advection in $\log$-scale of (a) volume fraction, (b) phase mass, and (c) phasic total energy, respectively.}
    \label{fig:advection-profiles}
\end{figure}

\Cref{fig:advection-profiles} depicts the conserved quantities compared to initial profiles after $200\tau_{\rm FT}$. Excellent agreement is observed even in $\log$-scale, which demonstrates the accuracy of the phase field model in preserving the interface thickness and finite bounds imposed for numerical stability. For instance, in the volume fraction profile [\cref{fig:advection-profiles} (a)], it is clear that the amount of conjugate phase is well captured, \textit{i.e.}, in this case $\phi_1 = \delta = 10^{-8}$ in the almost pure phase 2 regime. Profiles for individual phasic masses [\cref{fig:advection-profiles} (b)] and total energies [\cref{fig:advection-profiles} (c)] are also properly captured. Only a small deviation is observed toward the direction of advection; nonetheless, the deviation does not grow after a couple of $\tau_{\rm FT}$ and the profile in~\cref{fig:advection-profiles} is maintained.

\begin{figure}
    \centering
    \begin{subfigure}{0.425\linewidth}
    \includegraphics[width=\linewidth]{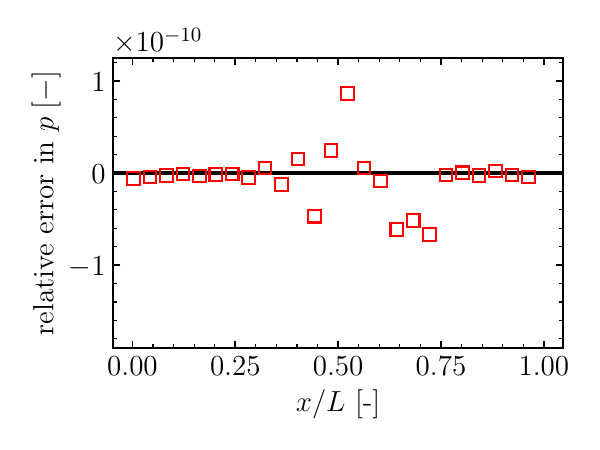}
    \caption{Pressure equilibrium}
    \end{subfigure}%
    \begin{subfigure}{0.425\linewidth}
    \includegraphics[width=\linewidth]{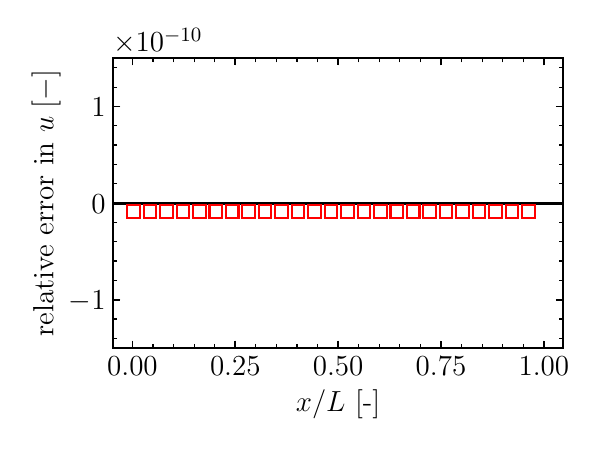}
    \caption{Velocity equilibrium}
    \end{subfigure}
    
    \begin{subfigure}{0.425\linewidth}
    \includegraphics[width=\linewidth]{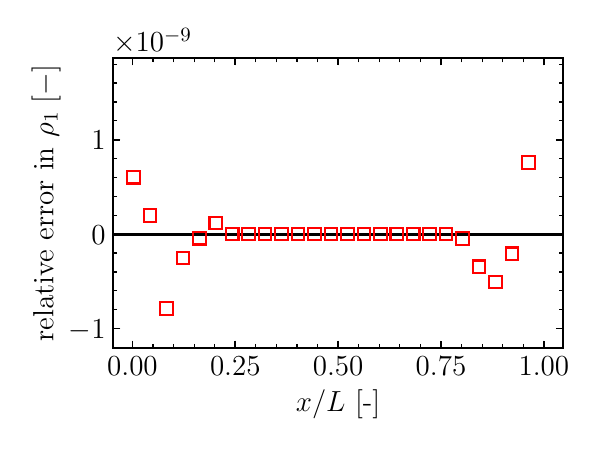}
    \caption{Error of density of phase 1, $\rho_1$}
    \end{subfigure}%
    \begin{subfigure}{0.425\linewidth}
    \includegraphics[width=\linewidth]{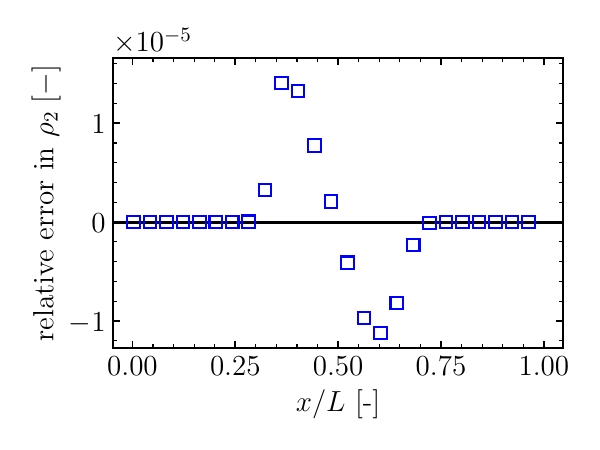}
    \caption{Error of density of phase 2, $\rho_2$}
    \end{subfigure}
    \caption{Interface equilibrium conditions of high-density ratio advection in terms relative error of (a) uniform pressure, and (b) uniform velocity. Relative errors of density of (c) phase 1, and (d) phase 2. Red ({\protect\tikz\protect\draw[line width = 0.75pt, red] (0,0) rectangle +(1ex,1ex) ;}) and blue ({\protect\tikz\protect\draw[line width = 0.75pt, blue] (0,0) rectangle +(1ex,1ex) ;}) squares denote values for phases 1 and 2, respectively.}
    \label{fig:advection-errors}
\end{figure}

In order to further quantify the accuracy and stability of the proposed formulation and discretization, \cref{fig:advection-errors} demonstrates relative errors compared to initial values of phasic pressure, phasic velocity and phasic densities. Characterization of relative errors of pressure and velocity is a measure of the accuracy with respect to the IEC. The test case imposes uniform phasic velocities and phasic pressures, which are expected to be preserved throughout the simulation, indicating a stable simulation. Indeed, \cref{fig:advection-errors} (a,b) highlight that the proposed method is capable of maintaining uniform phasic pressure and velocities with a normalized relative error up to an order of $10^{-10}$ for phase 1, similar results are observed for phase 2. 
Interestingly, there is no clear distinction of limitations in capturing accurate phasic pressure and velocities between the limits of almost pure and conjugate phases [\cref{fig:advection-errors} (a,b)], demonstrating that the method uniformly preserves pressure and velocity. Nevertheless, for phasic densities [\cref{fig:advection-errors} (c,d)], this distinction is observed. In spite of the greater error at the limit of conjugate phase, the errors remain bounded, where $\phi\rightarrow\delta$. Phasic density is challenging to capture accurately, given that it is an implicit quantity ($\rho_l = \phi_l \rho_l / \phi_l$). Furthermore, if the phasic density becomes negative or grows unbounded, stability is lost as it determines the rate of phasic mass interface regularization. \cref{fig:advection-errors} (c,d) depict the accuracy at which the formulation is able to capture each phasic density. In the limit of almost pure phase, both phasic densities are accurately captured. However, in the limit of conjugate phase, more noticeable errors are observed, up to $10^{-9}$ for phase 1 [\cref{fig:advection-errors} (c)] and up to $10^{-5}$ for phase 2 [\cref{fig:advection-errors}] (d). The increased relative error for phase 2 is due to $\rho_2 < \rho_1$, which allows phase 2 to deform more easily compared to phase 1. The specific error profile of phase is a representation of the small deviations observed in~\cref{fig:advection-profiles}. It is noteworthy that although a small deviation is observed, tied to the deformation of lighter phase, the formulation is able to keep the deformation limited, thus bounding phasic density, pivotal for long-time integration stability. Other approaches that explicitly track phasic density~\citep{brill:2024} have been proposed to a baseline six-equation model~\citep{saurel:2009}, which can improve phasic density error bounds; nevertheless, the method demands addition transport equation for all phasic densities increasing computation cost.

\begin{figure}
    \centering
    \begin{subfigure}{0.425\linewidth}
    \includegraphics[width=\linewidth]{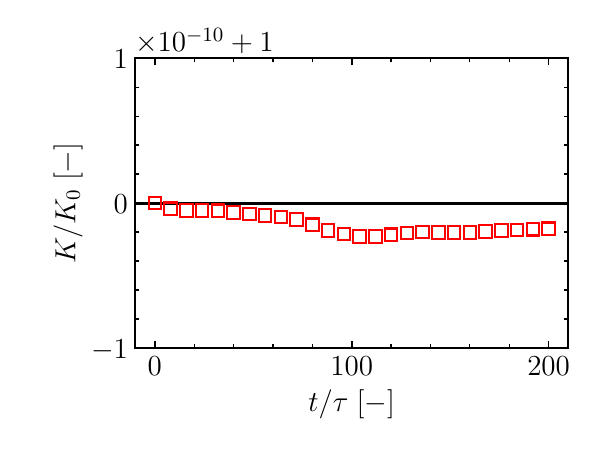}
    \caption{Mixture kinetic energy, $K$}
    \end{subfigure}%
    \begin{subfigure}{0.425\linewidth}
    \includegraphics[width=\linewidth]{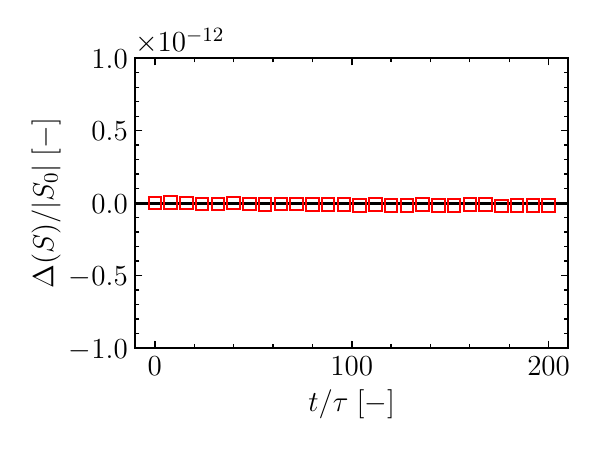}
    \caption{Change in mixture entropy, $\Delta(S)$}
    \end{subfigure}
    \caption{Mixture conservation properties of high-density ratio advection. Constant value depicted by solid black lines (\rule[2pt]{0.35cm}{1pt}), and the instantaneous values of the current simulation by red ({\protect\tikz\protect\draw[line width = 0.75pt, red] (0,0) rectangle +(1ex,1ex) ;}) squares. Evolution of normalized (a) total mixture kinetic energy $K = \sum_l K_l$, (b) change in mixture entropy $\Delta (S)$.}
    \label{fig:advection-mixture-conservation}
\end{figure}

In addition to the verified IEC, robust and stable simulations of compressible two-phase flows require preservation of kinetic energy and entropy. For this test case and the compressible two-phase Taylor-Green vortex case in \cref{subsec:turb}, phasic and mixture kinetic energy and entropy are computed throughout the simulations. Total phasic kinetic energy is defined as
\begin{equation}
\label{eq:phasic-ke}
K_l = \int_\Omega \phi_l \rho_l u_{l,i}u_{l,i} dV = \int_\Omega \phi_l \rho_l k_{l} dV,
\end{equation}
where the mixture kinetic energy follows from the sum of the contributions from each phase $K = \sum_l K_l$. Total phase entropy is defined as
\begin{equation}
\label{eq:phasic-ent}
S_l = \int_{\Omega} \phi_l \rho_l s_l dV,
\end{equation}
and analogously, the mixture entropy follows from the sum of the contributions from each phase $S = \sum_l S_l$. The phasic entropy per unit mass $s_l$ for the SG-EoS [\cref{eq:sg_eos}] is defined as
\begin{equation}
\label{eq:phasic-ent-pum} 
s_l = c_{v,l} \log \left (\frac{p_l + \pi_l}{\rho_l^{\gamma_l}} \right ) + s_{l}^{\rm o},
\end{equation}
where both the heat capacity at constant volume, $c_{v,l}$, and the reference state value for entropy, $s_l^{\rm o}$, remain to be defined. Realistic values for $c_{v,l}$ and $s_l^{\rm o}$ demand careful treatment, further details on computing these quantities is described in~\cite{lemetayer:2004}. For simplicity, we take $s_l^{\rm o} = 0$, which also does not contribute to phasic and mixture entropy change in a periodic domain where phasic mass is conserved, except for normalization. With this assumption, the normalized change in total phasic entropy becomes independent of $c_{v,l}$. The normalized change in total mixture entropy, on the other hand, still depends on $c_{v,l}$ given that each phase has its own value. Based on the phasic temperature, $c_{v,l}$ for a SG-EoS is obtained by
\begin{equation}
\label{eq:cv-def}
c_{v,l} = \frac{p_l + \pi_l}{T_l \rho_l (\gamma - 1)}.
\end{equation}
The phasic temperatures do not need to match in the proposed model [\cref{eq:7eq_proposed}], we estimate $c_{v,l}$ by assuming $T_l = 300$ K for simplicity, which yields in $c_{v,1} = 588.3$ and $c_{v,2} = 833.3$. \cref{fig:advection-mixture-conservation} depicts conservation properties at the mixture level with respect to total kinetic energy and change in total entropy. The proposed scheme [\cref{eq:keepfluxes}] is evaluated, and according to \cref{fig:advection-mixture-conservation}, both total mixture kinetic energy and the change in total mixture entropy ($\Delta(S) = S|^n - S|^0$) are well preserved, up to an error of order $10^{-10}$ for the former and $10^{-12}$ for the latter.

\begin{figure}
    \centering
    \begin{subfigure}{0.425\linewidth}
    \includegraphics[width=\linewidth]{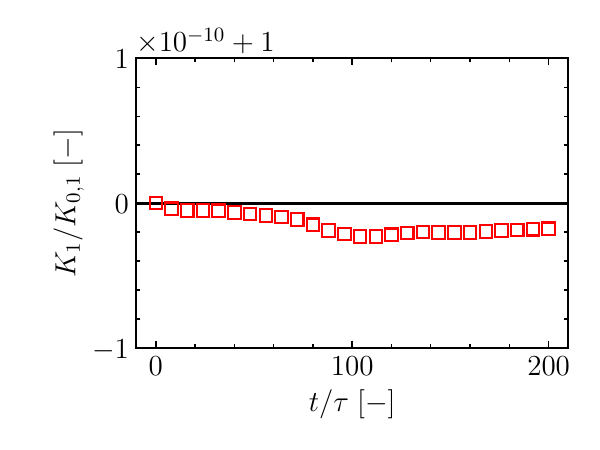}
    \caption{Kinetic energy of phase 1, $\phi_1 \rho_1 k_1$}
    \end{subfigure}%
    \begin{subfigure}{0.425\linewidth}
    \includegraphics[width=\linewidth]{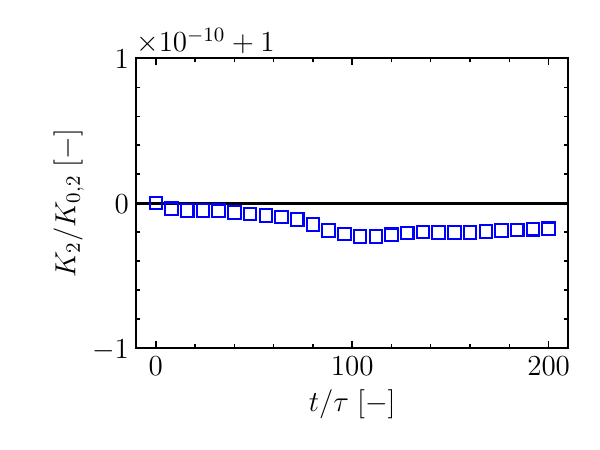}
    \caption{Kinetic energy of phase 2, $\phi_2 \rho_2 k_2$}
    \end{subfigure}
    
    \begin{subfigure}{0.425\linewidth}
    \includegraphics[width=\linewidth]{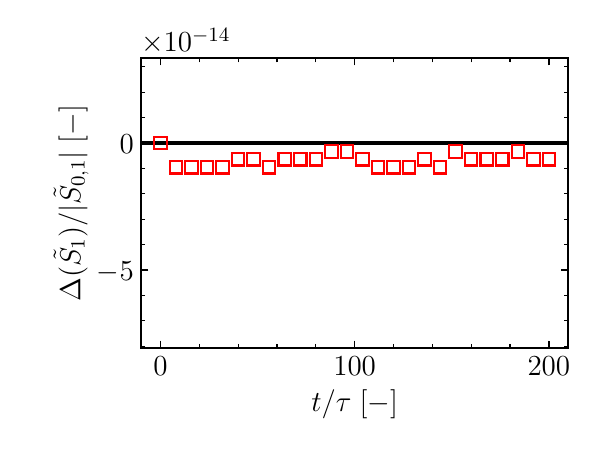}
    \caption{Change in entropy per unit mass of phase 1, $\phi_1 {s}_1$}
    \end{subfigure}%
    \begin{subfigure}{0.425\linewidth}
    \includegraphics[width=\linewidth]{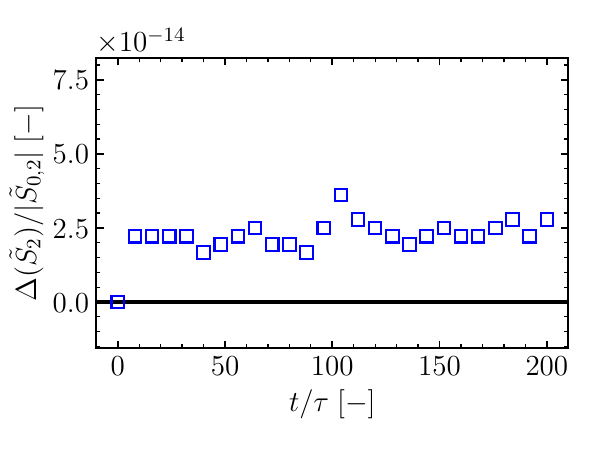}
    \caption{Change in entropy per unit mass of phase 2, $\phi_2 {s}_2$}
    \end{subfigure}
    \caption{Phasic conservation properties of high-density ratio advection. Constant value depicted by solid black lines (\rule[2pt]{0.35cm}{1pt}), and the instantaneous values of the current simulation by red ({\protect\tikz\protect\draw[line width = 0.75pt, red] (0,0) rectangle +(1ex,1ex) ;}, phase 1) and blue ({\protect\tikz\protect\draw[line width = 0.75pt, blue] (0,0) rectangle +(1ex,1ex) ;}, phase 2) squares. Evolution of normalized total phasic kinetic energy $K_l = \int \phi_l \rho_l k_l dV$ of (a) phase 1, and (b) phase 2. Evolution of normalized change in total phasic entropy per unit mass $\tilde{S}_l = \int \phi_l {s}_l dV$ of (c) phase 1, and (d) phase 2.}
    \label{fig:advection-phasic-conservation}
\end{figure}

Furthermore, \cref{fig:advection-phasic-conservation} demonstrates preservation properties at the phasic level, specially for the lighter phase, entropy changes are expected to be larger due to greater density fluctuations. Nevertheless, when weighed by mass, the larger changes in total phasic entropy are compensated at the mixture level. Similar to total mixture kinetic energy, the total phasic kinetic energy is also preserved up to an error of order $10^{-10}$ [\cref{fig:advection-phasic-conservation} (a,b)]. The change in the total phasic entropy by unit mass is bounded by an error of order $10^{-14}$ [\cref{fig:advection-phasic-conservation} (c,d)] for both phases, almost close to machine double precision, this is likely due to the approximate KEEP nature of the scheme. 

Qualitatively, the same test case is demonstrated in three dimensions. The domain spans $[0, 1]\times[0, 1]\times[0, 1]$ and is discretized with $64^3$ cells, the water drop is initialized at $(0.5, 0.5, 0.5)$ with a radius $R=0.25$. All boundaries are periodic. The final time is $t=0.05$ s, such that the drop is evolved for $5\tau_{\rm FT}$. The time step is the same as the 1D case, which reduces the CFL to $0.13$.

\begin{figure}
    \centering
    \begin{tikzpicture}
    \node (img12) {\includegraphics[width=0.24\textwidth]{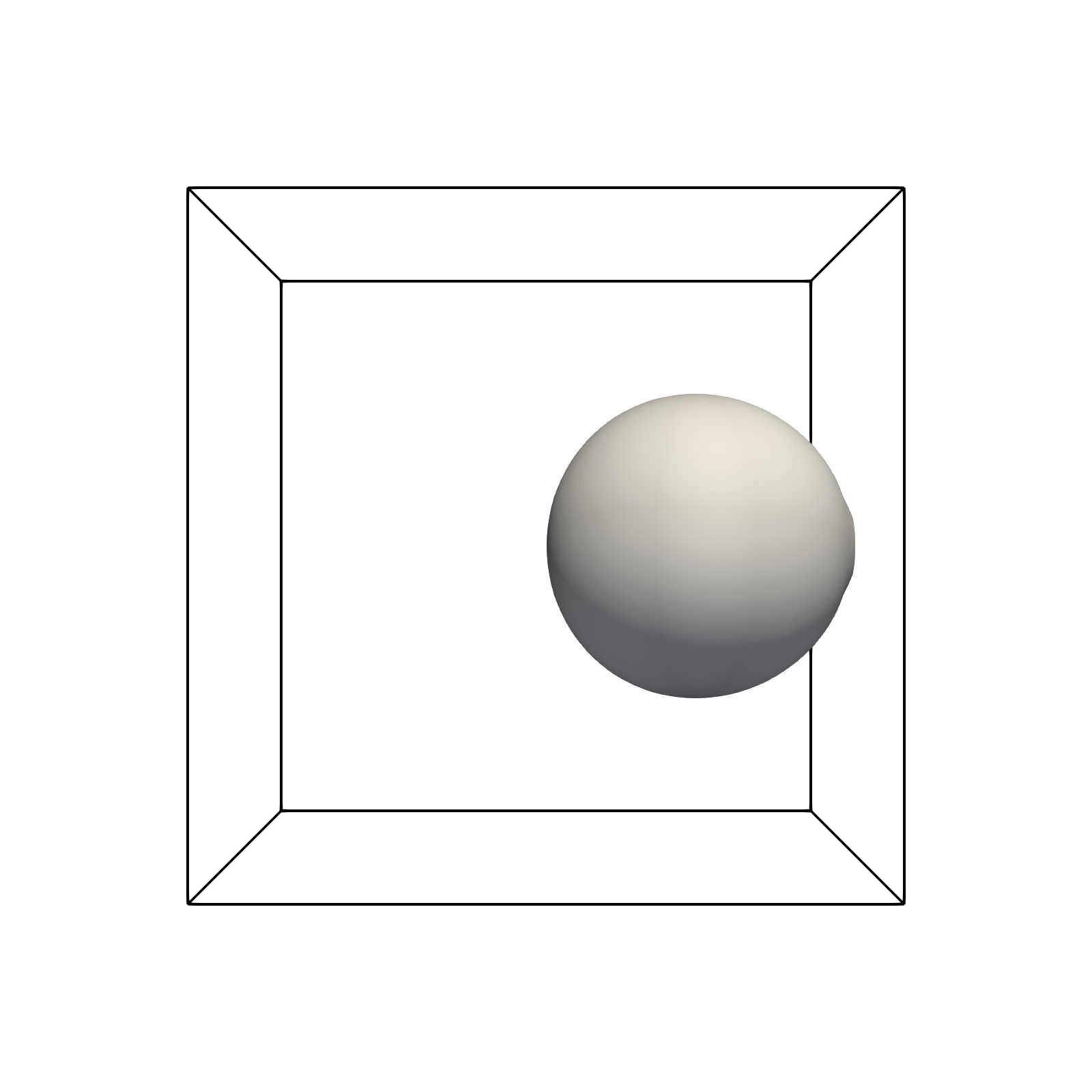}};
    \node (img13) [right=0pt of img12] {\includegraphics[width=0.24\textwidth]{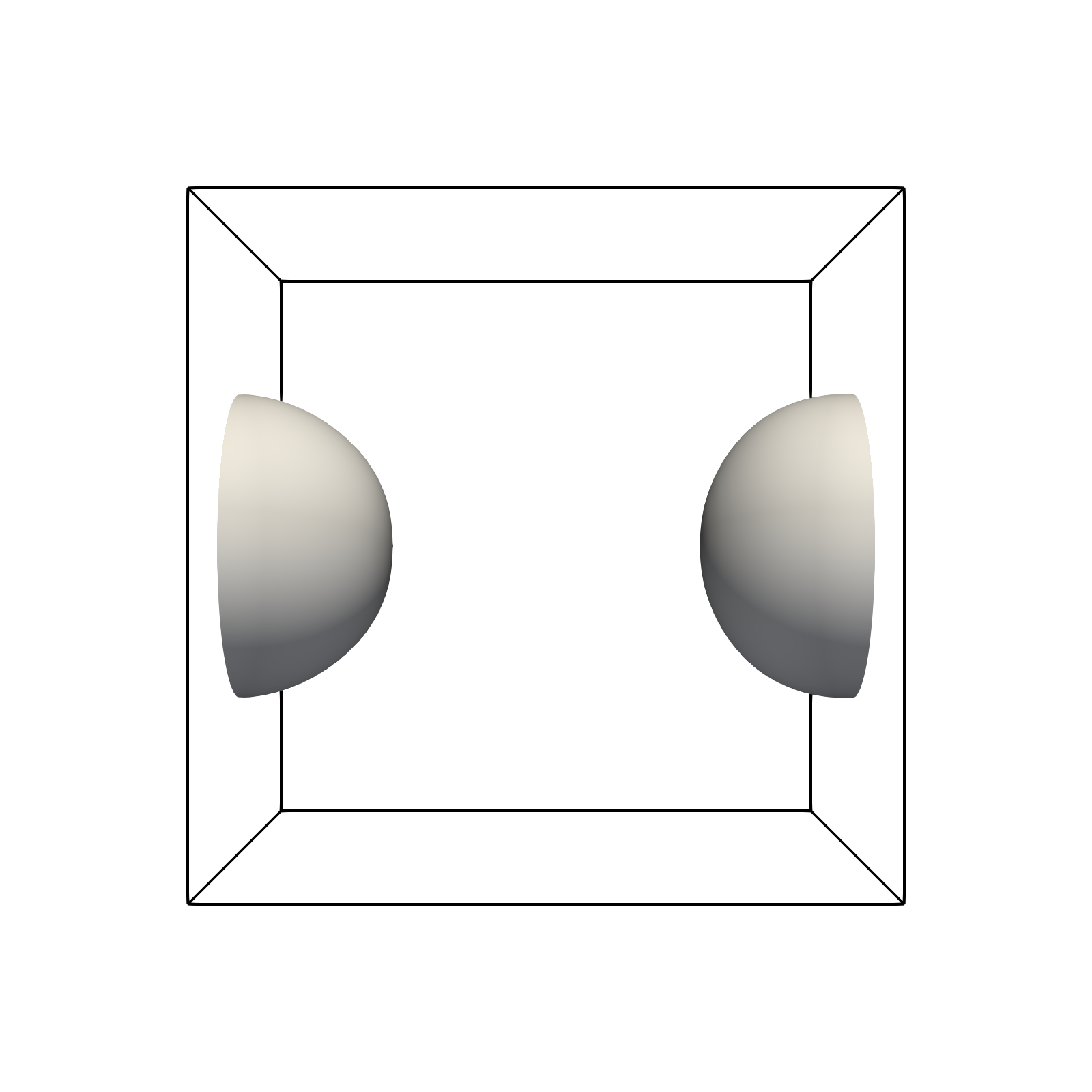}};
    \node (img14) [right=0pt of img13] {\includegraphics[width=0.24\textwidth]{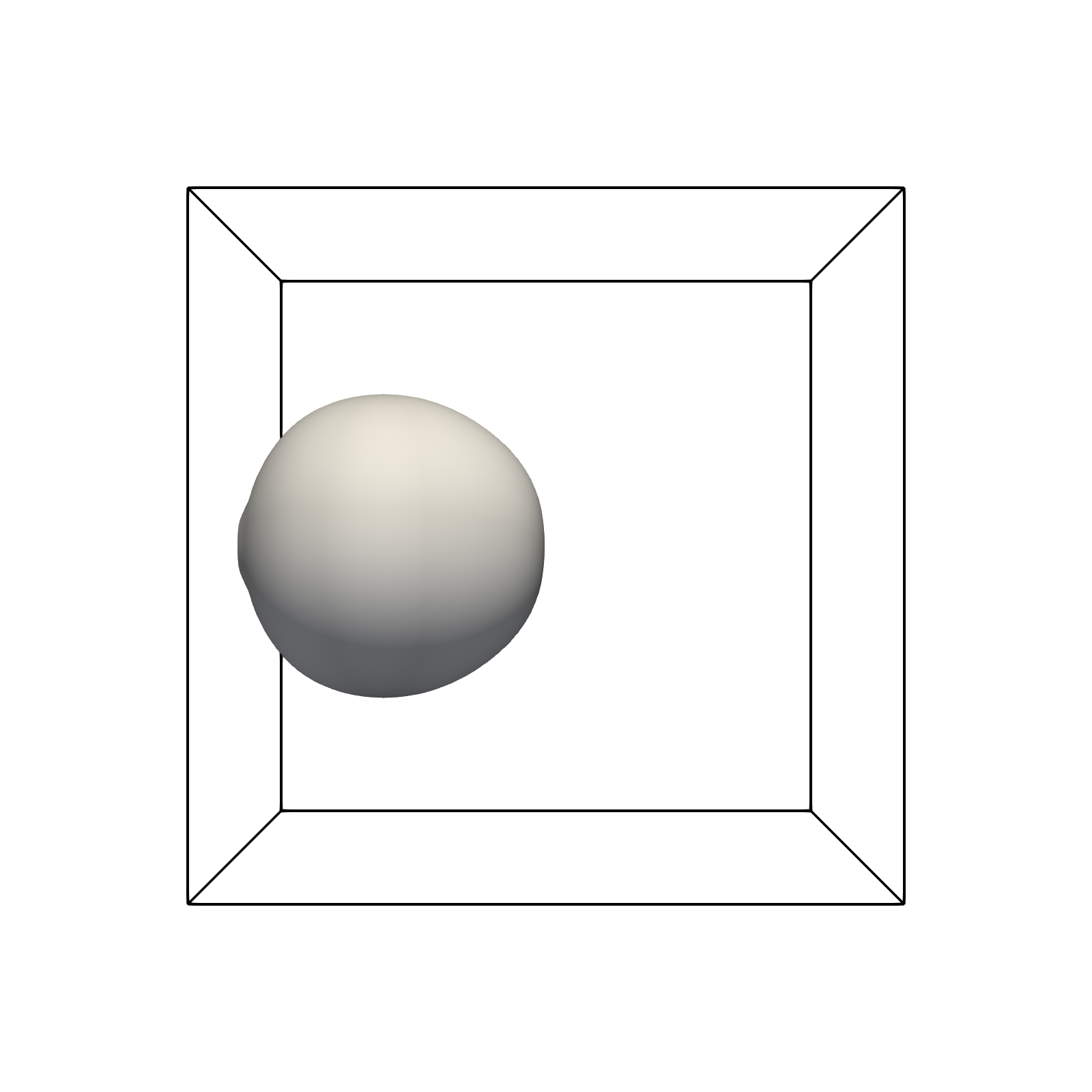}};
    \node (img15) [right=0pt of img14] {\includegraphics[width=0.24\textwidth]{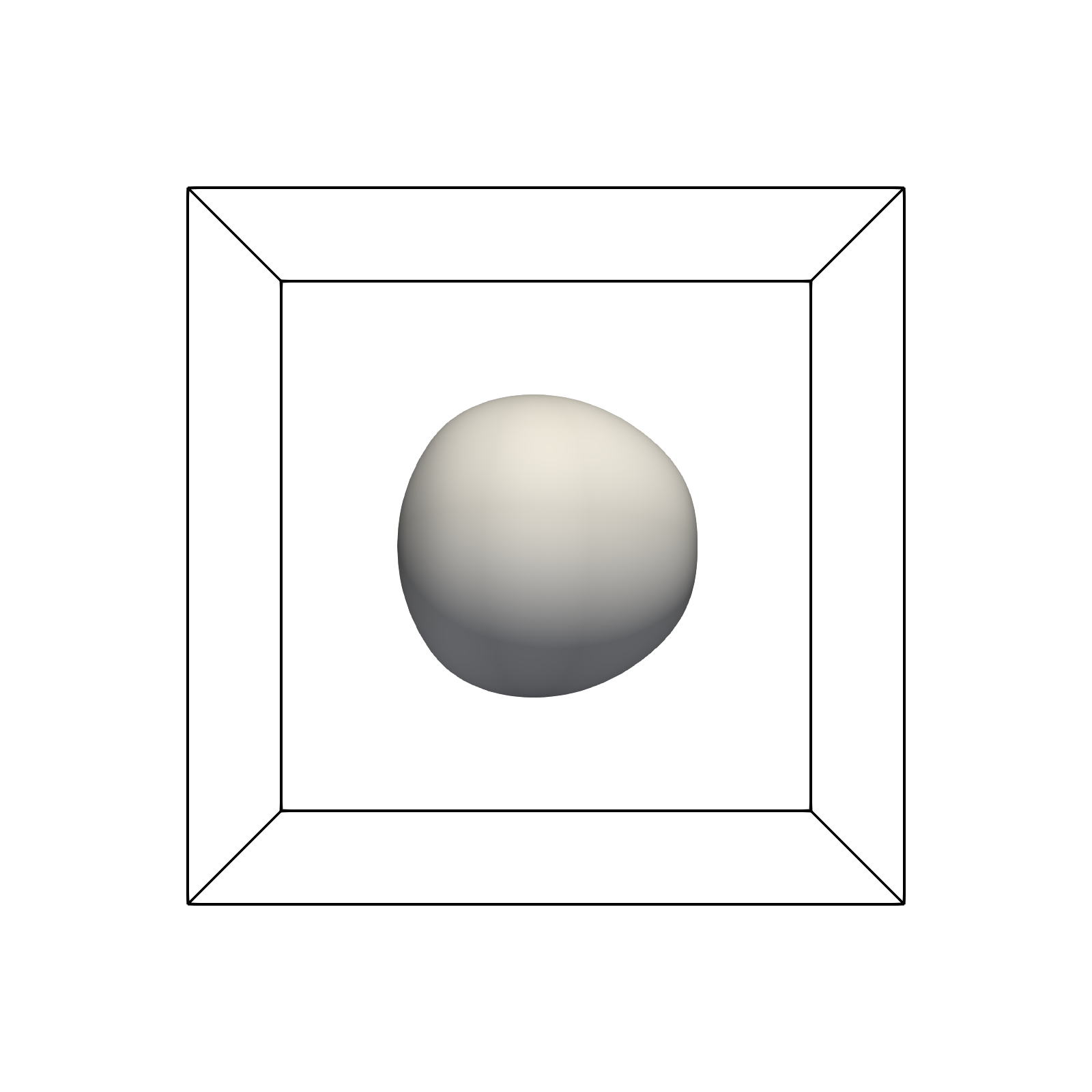}};
    \begin{scope}[shift={(-7em,-8.5em)}] 
    \node (x1) at (2.9  ,1.5) {1.25};
    \node (x2) at (7.25 ,1.5) {2.5};
    \node (x3) at (11.5 ,1.5) {3.75};
    \node (x4) at (15.75,1.5) {5};
    
    \node at (9.375,0.25) {Time, $t/\tau_{\rm FT} \; [-]$};
    
    \draw[->, thick] (x1.south west) ++(-1.5,-0.25) -- ($(x4.south east) + (1.7,-0.25)$);
    \end{scope}
    \end{tikzpicture}
    \caption{Three-dimensional high-density ratio interface advection. Evolution of interface isocontour ($\phi_1 = 0.5$) over several flow-through times ($\tau_{\rm FT}$).}
    \label{fig:advection-3d}
\end{figure}
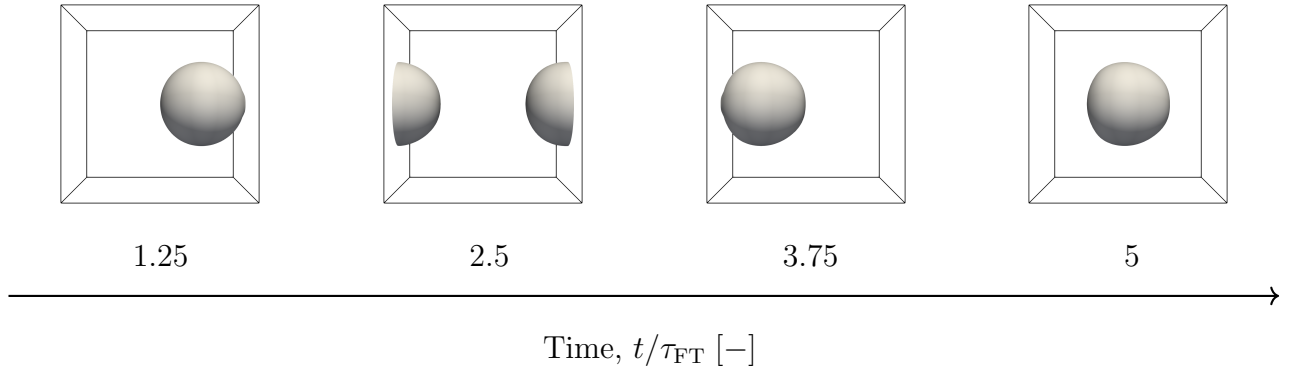

\Cref{fig:advection-3d} depicts the evolution of the isocontour ($\phi_1=0.5$) of the volume fraction field over $5\tau_{\rm FT}$. They simulation remains stable and bounded. The phase field method is able to retain the initial spherical shape of the interface, whilst maintaining conserved quantities bounded. Only a small deviation leads to a deformation of the spherical shape after a finite time, however, the deformation is not expected to increase once this ``new equilibrium'' state is achieved.

\subsection{Acoustic tests}
\label{subsec:comp}

Accurate propagation of acoustic waves require the underlying model to capture compressibility effects of both phases. The following two test cases were first proposed by~\citet{jain:2020} and are repeated here to evaluate the performance of the non-equilibrium model. 

The material properties for all phases are presented in~\cref{tab:dim-mat-par}. Temporally and spatially evolving boundary conditions are required for both cases, in order to control the timing of the pressure pulses for both, and radial focusing for the pressure-driven bubble oscillation case.
\begin{table}
\centering
\begin{tabular}{cccc}
\hline
                           & Air                 & Water              & Kerosene             \\ \hline
$\rho_l \; [{\rm kg/m^3}]$ & 1.225               & 997                & 820                  \\
$\mu_l \; [{\rm N/m^2}]$   & $1.81\times10^{-5}$ & $8.9\times10^{-4}$ & $1.64\times10^{-3}$ \\ \hline
\end{tabular}
\caption{Material properties for acoustic test cases.}
\label{tab:dim-mat-par}
\end{table}

\subsubsection{Pressure-driven bubble oscillation}
\label{subsubsec:pressure_osc_bubble}

The domain spans $[-5, 5] \times [-5, 5]$ $\mu$m and is discretized with $25^2$, $50^2$, and $100^2$ cells, evaluating the limit of coarse grid. An air bubble is initialized at the center $(0, 0)$ with a radius $R=2$ $\mu$m, surrounded by water, and the surface tension coefficient is set to $\sigma = 0$. The fluid properties follow from~\cref{tab:dim-mat-par} and the quiescent ambient conditions are set to $u_{l,i} = 0$ m/s and $p_l = 10^5$ Pa. Similar to the setup of~\citet{jain:2020}, the interface velocity regularization scale is set to $\Gamma=10\max_l (|u_{l,i}|_\infty)$. Temporally- and spatially-evolving boundary conditions are set for the phasic pressures to mimic a radially inward wave in the form of
\begin{equation}
\label{eq:bc_pres_osc_bubble} 
p_l((x, y, z), t) = 10^5 \left [1 + 0.1 \sin(10\omega_c \tilde{t}) \right ],
\end{equation}
where the phase shift is defined as $\tilde{t} = t - [r(x,y,z) - r_{\min}]/c$, with the local radial location of the wall as $r$, the minimum radius as $r_{\min} = 5$ $\mu$m, and $c$ as a reference wave speed. Although a mixture of both phases is defined everywhere, at the boundaries, the wave speed required for phase shifting is set as $c_{\rm water} = 1627.38$ m/s. The remaining quantities have zero-gradient type boundary conditions. The final time is $6.25\times10^{-7}\; {\rm s}$ such that $t/(2\pi/\omega_c) \approx 1$, where $\omega_c = 10208967.75 \;{\rm s^{-1}}$ is the characteristic resonance frequency of the bubble. The time window of interest is limited to the regime dominated by transient effects, and not to the long-time response. The time steps are set to $\Delta t = 5, 2.5,$ and $1.25\times10^{-11} \; {\rm s}$, respectively, such that the acoustic CFL is approximately equal to $0.2$ for the water phase.

The Rayleigh-Plesset equation for finite-size domains~\citep{jain:2020} is
\begin{equation}
\label{eq:rp_eq} 
\frac{p_{\rm B}(t) - p_{{\rm liquid}, S}(t)}{\rho_{\rm liquid}} = \log \left (\frac{S}{R} \right ) \left [ \left (\dot R \right)^2 + R \Ddot{R}\right ] + \left (\frac{R^2 - S^2}{2 S^2} \right ) \left (\dot R \right)^2 + \frac{2 \nu_{\rm liquid} \dot R}{R} + \frac{\sigma}{\rho_{\rm liquid} R},
\end{equation}
where $p_{\rm B}$ is the uniform pressure inside the bubble, and $p_{\rm liquid, S}$ is the liquid pressure at a finite distance from the center of the bubble at $r = S$. The reference fluid properties are the liquid density $\rho_{\rm liquid}$, the liquid kinematic viscosity $\nu_{\rm liquid}$, and the surface tension coefficient $\sigma$. Considering an isentropic compressible/expansion for the gas phase, the pressure variation can be measured based on the change in volume as
\begin{equation}
\label{eq:isentropic_process} 
\frac{p_{\rm B}(t)}{p_{\rm B}(0)} = \left [ \frac{V_{\rm B}(0)}{V_{\rm B}(t)} \right ]^{\gamma_{\rm B}} \xrightarrow{\text{in 2D}} \left [ \frac{A_{\rm B, 2D}(0)}{A_{\rm B,2D}(t)} \right ]^{\gamma_{\rm B}} = \left [ \frac{R_{\rm B, 2D}(0)}{R_{\rm B,2D}(t)} \right ]^{2\gamma_{\rm B}},
\end{equation}
which closes the ODE for $R$ and therefore for any derived quantity such as the perimeter, the area, and the volume of the bubble. Therefore, the analytical solution of the temporal evolution of the bubble's radius ($R_{\rm B} = R$ for conciseness) for the proposed setup follows from
\begin{equation}
\label{eq:rp_eq_final} 
\begin{aligned}
\frac{10^5}{\rho_{\rm liquid}} \left [ \frac{R(0)}{R(t)} \right ]^{2\gamma_{\rm B}} & -\frac{10^5 \left [1 + 0.1 \sin(10\omega_c t) \right ]}{\rho_{\rm liquid}} \\
& = \log \left (\frac{r_{\min}}{R} \right ) \left [ \left (\dot R \right)^2 + R \Ddot{R}\right ] + \left (\frac{R^2 - r_{\min}^2}{2 r_{\min}^2} \right ) \left (\dot R \right)^2 \\
& + \frac{2 \nu_{\rm liquid} \dot R}{R}.
\end{aligned}
\end{equation}

\begin{figure}
    \centering
    \includegraphics[width=0.85\linewidth]{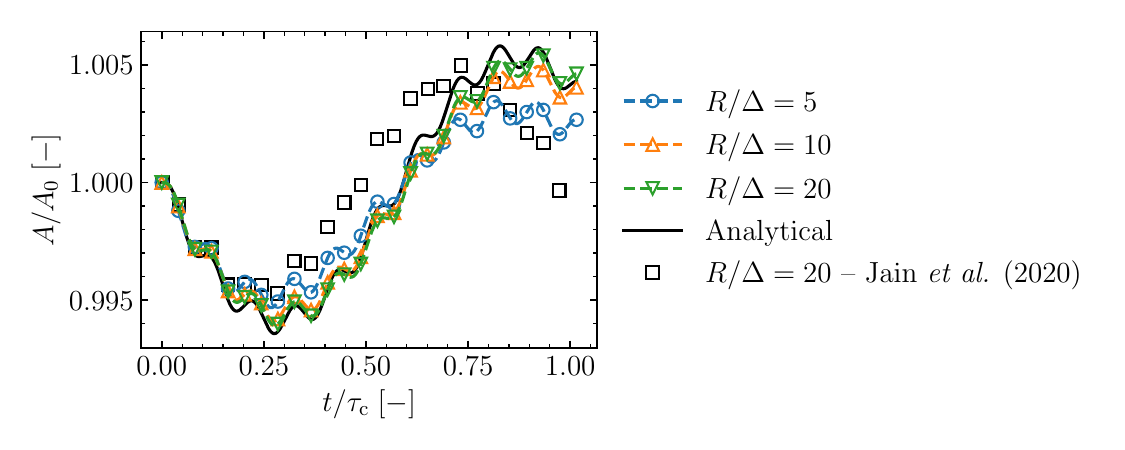}
    \caption{Pressure-driven bubble oscillation. Normalize bubble area evolution ($A/A_0$) over a span of one resonance period ($\tau_{\rm c}$). Colored dashed lines ({\protect\tikz[baseline=-0.5ex]\protect\draw[line width=1pt, black, dashed] (0,0) -- (1em,0);}) represent results from current simulation for 3 different grid resolution. Black squares ({\protect\tikz\protect\draw[line width = 0.75pt, black] (0,0) rectangle +(1ex,1ex) ;}) represent literature results~\cite{jain:2020} using a five-equation model and analogous grid resolution. Solid black line ({\rule[2pt]{0.35cm}{1pt}}) represent analytical solution of finite-domain Rayleigh-Plesset equation.}
    \label{fig:pressure_bubble_oscillation}
\end{figure}

\Cref{fig:pressure_bubble_oscillation} demonstrates the evolution of 2D area (equivalent to volume in 3D) of the bubble due to pressure fluctuations imposed by the boundary condition. A significant improvement with respect to the work of \citet{jain:2020} is observed. In their work~\cite{jain:2020}, a partial-equilibrium formulation is used, namely the five-equation model with a non-KEEP scheme. Given both improvements proposed in the current model [\cref{eq:7eq_proposed}] and the current discretization [\cref{eq:keepfluxes}], even with much coarser grids, the numerics converge better to the analytical solution. Furthermore, with the proposed formulation, a resolution of roughly 20 grid points over the diameter of the bubble ($R/\Delta=10$) is sufficient to capture the physics of the problem well, consistent with other works~\cite{dodd:2016,battistella:2020,crialesi:2022,hatashita:2026c}.

\subsubsection{Oblique wave incidence at high-impedance two-phase interface}
\label{subsubsec:waves}

The domain spans $[0, 10] \times [0, 10]$ $\mu$m where the water-kerosene interface is aligned with the north-west/south-east diagonal, discretized with $1000^2$ cells. The setup is consistent with the one proposed by~\citet{jain:2020}, further imposing a finite bound on $\phi$, where $\phi_1$ is initialized based on the equilibrium profile $1-0.5\left \{ 1 + (1-2\delta) \tanh \left [ (10^{-5} - x - y)/(2\epsilon) \right ] \right \}$, such that phase 1 (kerosene) is in the upper triangle and phase 2 (water) in the lower triangle. The fluid properties follow from~\cref{tab:dim-mat-par}, and the quiescent ambient conditions are set to $u_{l,i}=0$ m/s and $p_l = 10^5$ Pa. An improvement to~\citep{jain:2020} with respect to interface thickness is tested given that here $\epsilon = 0.6 \Delta x$. A leftward-leaning wave is imposed via the time varying phasic pressure boundary condition on the east face following $10^5 \left [ 1 - 0.5 \sin (\omega_{\rm ker}t)\right ]$ for $t<755.297$ ps, where $\omega_{\rm ker} = 2\pi c_{\rm ker}/\lambda$ and $\lambda=2$ $\mu$m. All other quantities support zero-gradient boundary conditions. Given the orientation of the interface and pressure pulse, the incident acoustic wave forms a $\theta_i=45^\circ$ angle with respect to the interface normal. The final time is $t=6$ ps to allow for the pressure pulse to traverse roughly half of the domain. The time step is $\Delta t=5$ ps such that the acoustic CFL are approximately 0.65 for kerosene and 0.8 for water.

For the proposed combination of fluids, where $c_{\rm ker} = 1324$ and $c_{\rm water} = 1627.4$, the angle of the transmitted wave, $\theta_t$, follows from the Snell's law of refraction, which is defined as
\begin{equation}
\label{eq:law-refrac}
\frac{\sin(\theta_i)}{c_i} = \frac{\sin(\theta_t)}{c_t}.
\end{equation}
Therefore, the expected angle of the transmitted wave is $\theta_t = 60.36^\circ$. Given that the incident angle is greater than the critical angle, there is no total internal reflection. The angle of reflection for an oblique wave is equal to the angle of incidence; thus, $\theta_r$ is expected to be equal to $45^\circ$. 

\begin{figure}
    \centering
    \includegraphics[width=0.6\linewidth]{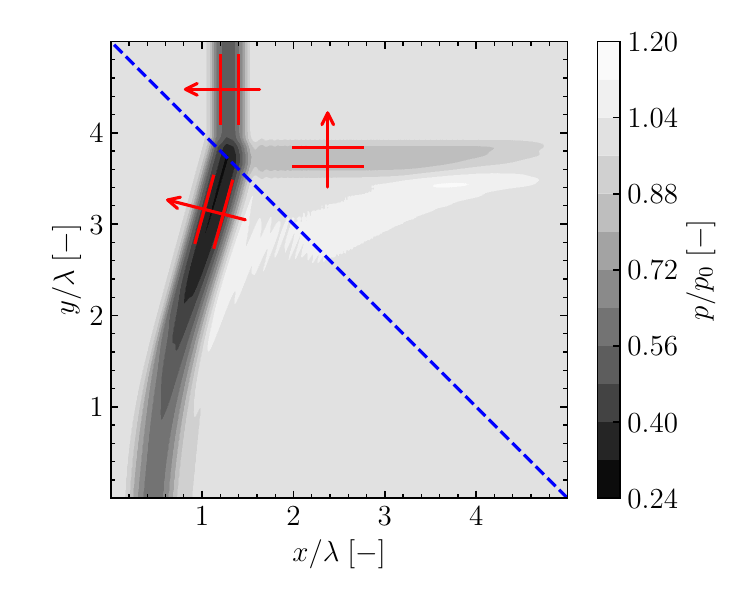}
    \caption{Mixture pressure contours of oblique acoustic wave incidence at high-impedance two-phase interface. $45^\circ$-angle interface represented by dashed blue line ({\protect\tikz[baseline=-0.5ex]\protect\draw[line width=1pt, blue, dashed] (0,0) -- (1em,0);}). Incident, reflected and transmitted waves (from top to bottom) depicted by solid red lines/arrows ({\color{red}\rule[2pt]{0.35cm}{1pt}}).}
    \label{fig:oblique_wave}
\end{figure}

\cref{fig:oblique_wave} depicts the mixture pressure contours after the incident wave has traversed roughly half of the domain. The red lines and arrows represent qualitatively the direction of propagation of the incident, transmitted, and reflected waves. The overlay of the wave schematics were constructed considering the theoretical predictions of $\theta_i = 45^\circ$, $\theta_t=60.36^\circ$, and $\theta_r=45^\circ$. Therefore, comparing the red lines and the pressure contours provides a qualitative description of the accuracy of the model in capturing acoustic wave propagation at interfaces. From which, we observe good agreement between the model and the theory. 

\subsection{Turbulent flow}
\label{subsec:turb}

Robustness and stability in compressible flows without any additional numerical dissipation, impeditive for simulations of turbulent flows, rely on discrete preservation of kinetic energy and entropy. The most challenging scenario is in the limit of $Re\rightarrow\infty$, without physical viscous dissipation, which is the target of the proposed test case. In this section, a two-phase Taylor-Green vortex case is simulated in the limit of infinite Reynolds number, which was proposed by \citet{jain:2022a} as a robustness test case for compressible two-phase flows.

The domain spans $[0, 2\pi]\times[0, 2\pi]\times[0, 2\pi]$ m and is discretized with $64^3$ cells to evaluate the limit of coarse grid without any numerical dissipation. A phase 1 slab of width $2L=2$ m is initialized in the middle of the domain, according to 
\begin{equation}
\label{eq:ic_phi_tgv} 
\phi_1 = 1 - 0.5\left \{1 + (1-2\delta)\tanh \left [\frac{\sqrt{(x-\pi)^2} - L}{2\epsilon} \right ] \right \}.
\end{equation}
The fluid properties of phase 1 are changed to achieve density ratios from 1 to 1000, while phase 2 is maintained as an ideal gas with unity phasic density. Two suites of test cases are evaluated: first maintaining both phases with ideal gas-like parameters and varying uniquely the density of phase 1 as in~\citep{jain:2022a}; and second replicating water-air with stiffened gas parameters as in~\citep{yoshida:2026}. For the former, all parameters are summarized in~\cref{tab:ndim-ig-eos-par}, and for the latter, all parameters are summarized in~\cref{tab:ndim-sg-eos-par}.
\begin{table}
\centering
\begin{tabular}{ccc}
\hline
                           & Phase 1       & Phase 2 \\ \hline
$\rho_l \; [{\rm kg/m^3}]$ & 1, 0.1, 0.01  & 1       \\
$\gamma_l \; [-]$          & 1.4           & 1.4     \\
$\pi_l \; [-]$             & 0             & 0       \\ \hline
\end{tabular}
\caption{Non-dimensional ($p_{\rm ref} = 1$) stiffened gas equation of state parameters for ideal gas compressible two-phase Taylor-Green vortex cases.}
\label{tab:ndim-ig-eos-par}
\end{table}
\begin{table}
\centering
\begin{tabular}{ccc}
\hline
                           & Phase 1       & Phase 2 \\ \hline
$\rho_l \; [{\rm kg/m^3}]$ & 1000          & 1       \\
$\gamma_l \; [-]$          & 4.4           & 1.4     \\
$\pi_l \; [-]$             & 6000          & 0       \\ \hline
\end{tabular}
\caption{Non-dimensional ($p_{\rm ref} = 1$) stiffened gas equation of state parameters for water-air compressible two-phase Taylor-Green vortex cases.}
\label{tab:ndim-sg-eos-par}
\end{table}

The domain is triply periodic for all quantities and the initial conditions are given by
\begin{equation}
\label{eq:ctgv_initial}
\begin{aligned}
u_l & =  Ma_0 \sin (x) \cos (y) \cos (z), \\
v_l & = -Ma_0 \cos (x) \sin (y) \cos (z), \\
w_l & = 0, \\
p_l & = p_0 + \frac{\rho Ma_0^2}{16} [\cos (2x) + \cos (2y)] [\cos (2z) + 2],
\end{aligned}
\end{equation}
where $Ma_0$ is the initial Mach number, $p_0$ is the initial baseline pressure, and $\rho = \phi_1\rho_{0,1} + (1 - \phi_1) \rho_{0,2}$ is the mixture density. The final non-dimensional time is $Ma_0 t = 40$ to evaluate long-term preservation properties of the scheme (qualitative 3D flow field depicted in \cref{fig:ctgv-stiffened-gas-3d}). The time steps for increasing density ratios are $\Delta t = 2.5, 1.25, 0.625,$ and $1.25\times10^{-3}$, respectively, which yields an acoustic CFL $\lessapprox 0.05$ to minimize the effects of time integration error.

For both test suites, the phasic dynamics viscosities are set to $\mu_l=0$ and the surface tension coefficient to $\sigma = 0$, to evaluate the limit of zero physical dissipation ($Re_l = \infty$) without capillary effects ($We_l = \infty$), thus characterizing robustness of the discretization in not generating any spurious entropy. Given the degree of difficulty of the problem and the approximate preservation nature of the scheme, the interface thickness is increased to $\epsilon = 1.25 \Delta x$ and the amount of the conjugate phase increased to $\delta=10^{-6}$, similar to what was used in~\citep{yoshida:2026}. In~\cref{subsubsec:tgv_1_cases}, the effect of density ratio on the preservation properties of the scheme are evaluated with an initial Mach number of $Ma_0 = 0.2$; and in~\cref{subsubsec:tgv_2_cases}, the preservation properties are evaluated for a realistic water-air density and equation of states with an initial Mach number of $Ma_0 = 0.05$.

\subsubsection{Density ratio sweep for $Ma_0=0.2$}
\label{subsubsec:tgv_1_cases}

In this section, the baseline pressure $p_0$ is set to $1/\gamma \approx 0.7$, which is sufficient for positive pressures, given that the maximum dynamic pressure $\rho_{0,2} Ma_0^2 \approx 0.04$. For sufficiently high $Ma$, kinetic energy is not fully conserved as in incompressible flows, primarily due to exchange to internal energy. The role of the density ratio in this exchange is not fully understood. However, it is expected that, due to the differences in inertia of each phase, for increasing density ratios, the rate of the exchange between kinetic energy and internal energy is likely to increase.

\begin{figure}
    \centering
    \begin{subfigure}{0.425\linewidth}
    \includegraphics[width=\linewidth]{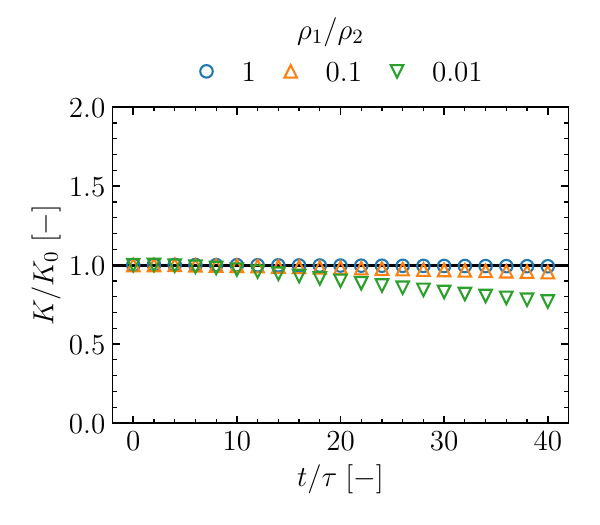}
    \caption{Mixture kinetic energy, $\rho k$}
    \end{subfigure}%
    \begin{subfigure}{0.425\linewidth}
    \includegraphics[width=\linewidth]{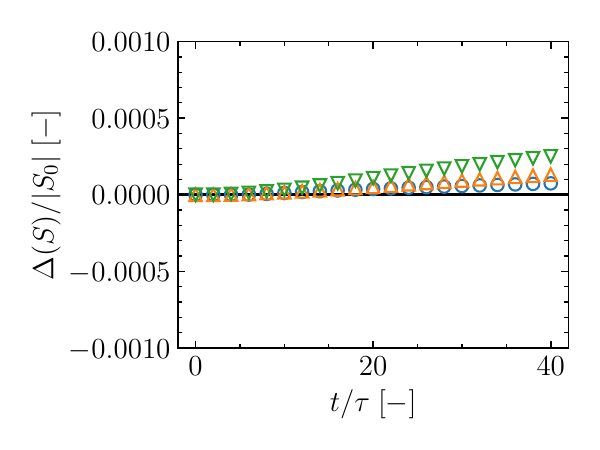}
    \caption{Mixture entropy, $\rho s$}
    \end{subfigure}
    \caption{Mixture conservation properties of compressible Taylor-Green vortex at $Ma=0.2$ for two ideal gases spanning density ratios from $1$ to $100$. Constant value depicted by solid black lines (\rule[2pt]{0.35cm}{1pt}), and the instantaneous values of the current simulation for different density ratios by blue ({\protect\tikz\protect\draw[line width = 0.75pt, mpl_blue] (0,0.5ex) circle (0.5ex);}, $\rho_1/\rho_2 = 1$), orange ({\protect\tikz\protect\draw[line width = 0.75pt, mpl_orange] (-0.5ex,0.1ex) -- (0.5ex,0.1ex) -- (0,0.966ex) -- cycle;}, $\rho_1/\rho_2 = 0.1$), and green ({\protect\tikz\protect\draw[line width = 0.75pt, mpl_green] (-0.5ex,0.866ex) -- (0.5ex,0.866ex) -- (0,0) -- cycle;}, $\rho_1/\rho_2 = 0.01$) symbols. Evolution of normalized (a) total mixture kinetic energy $K = \sum_l K_l$, (b) change in mixture entropy $S = \sum_l \int \phi_l \rho_l s_l dV = \int \rho s dV$.}
    \label{fig:ctgv-ideal-gas-mixture-conservation}
\end{figure}

\Cref{fig:ctgv-ideal-gas-mixture-conservation} depicts the kinetic energy and entropy preservation properties for the proposed scheme for varying density ratios. The unity density ratio does not present noticeable transfer of kinetic energy to internal energy for $Ma_0=0.2$. However, as the ratio increases in order of magnitude difference, this transfer is more significant (see for instance $\rho_1/\rho_2=0.01$). 

For the estimation of the change in total mixture entropy, one needs to determine the corresponding $c_{v,l}$. For the non-dimensional SG-EoS parameters with $p_{\rm ref} = 1$, we further assume $T_l = 1$, which yields in $c_{v,1} = 1.765$, and $c_{v,2} = 2.5$. \cref{fig:ctgv-ideal-gas-mixture-conservation} (b) presents the change in total mixture entropy normalized by its initial value. Overall, mixture entropy always increases, and it is ``preserved'' up to an order of $10^{-4}$, providing an evidence of robustness for increasing order of magnitude difference of density ratios. In addition to the faster exchange of kinetic and internal energy, the rate of entropy increase is also observed to be higher.

\begin{figure}
    \centering
    \begin{subfigure}{0.425\linewidth}
    \includegraphics[width=\linewidth]{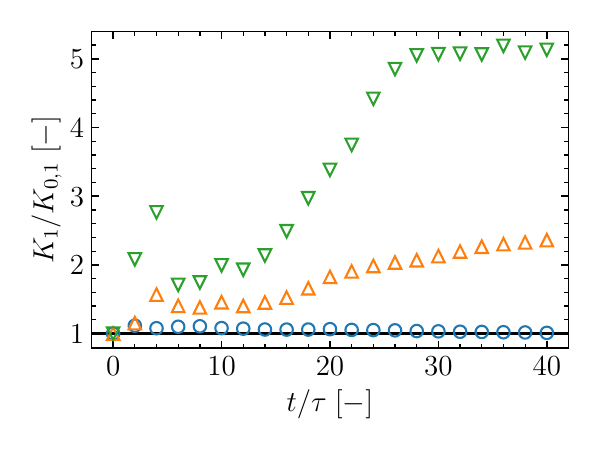}
    \caption{Kinetic energy of phase 1, $\phi_1 \rho_1 k_1$}
    \end{subfigure}%
    \begin{subfigure}{0.425\linewidth}
    \includegraphics[width=\linewidth]{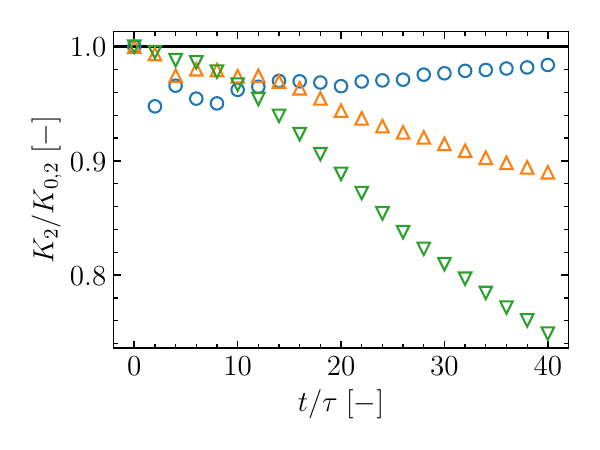}
    \caption{Kinetic energy of phase 2, $\phi_2 \rho_2 k_2$}
    \end{subfigure}
    
    \begin{subfigure}{0.425\linewidth}
    \includegraphics[width=\linewidth]{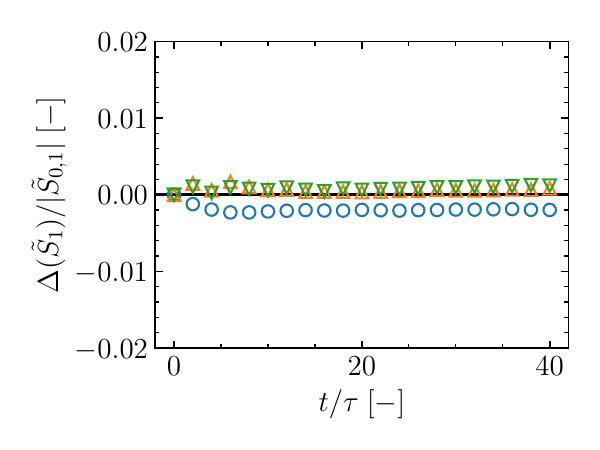}
    \caption{Change in entropy per unit mass of phase 1, $\phi_1 {s}_1$}
    \end{subfigure}%
    \begin{subfigure}{0.425\linewidth}
    \includegraphics[width=\linewidth]{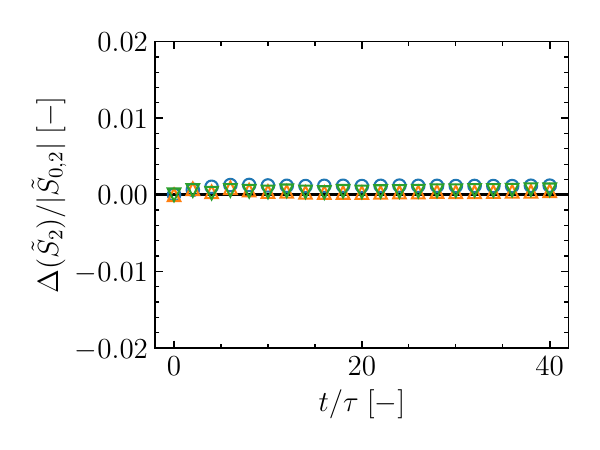}
    \caption{Change in entropy per unit mass of phase 2, $\phi_2 {s}_2$}
    \end{subfigure}
    \caption{Phasic conservation properties of compressible Taylor-Green vortex at $Ma=0.2$ for two ideal gases spanning density ratios from $1$ to $100$. Constant value depicted by solid black lines (\rule[2pt]{0.35cm}{1pt}), and the instantaneous values of the current simulation for different density ratios by blue ({\protect\tikz\protect\draw[line width = 0.75pt, mpl_blue] (0,0.5ex) circle (0.5ex);}, $\rho_1/\rho_2 = 1$), orange ({\protect\tikz\protect\draw[line width = 0.75pt, mpl_orange] (-0.5ex,0.1ex) -- (0.5ex,0.1ex) -- (0,0.966ex) -- cycle;}, $\rho_1/\rho_2 = 0.1$), and green ({\protect\tikz\protect\draw[line width = 0.75pt, mpl_green] (-0.5ex,0.866ex) -- (0.5ex,0.866ex) -- (0,0) -- cycle;}, $\rho_1/\rho_2 = 0.01$) symbols. Evolution of normalized total phasic kinetic energy $K_l = \int \phi_l \rho_l k_l dV$ of (a) phase 1, and (b) phase 2. Evolution of normalized change in total phasic entropy per unit mass $\tilde{S}_l = \int \phi_l {s}_l dV$ of (c) phase 1, and (d) phase 2.}
    \label{fig:ctgv-ideal-gas-phasic-conservation}
\end{figure}

Phasic preservation properties are presented in \cref{fig:ctgv-ideal-gas-phasic-conservation}. For the proposed test case, phase 2 is the heavier one and also the ``carrier'' phase since $\langle\phi_2\rangle > \langle \phi_1 \rangle$. Therefore, it is the phase to carry the majority of the mixture kinetic energy. As observed in \cref{fig:ctgv-ideal-gas-phasic-conservation} (a,b), the evolution of the total phasic kinetic energy of phase 2 is more representative of the behavior identified at the mixture level [\cref{fig:ctgv-ideal-gas-mixture-conservation}]. For larger differences in order of magnitude of density ratio, where the phasic kinetic energy also tends to decrease faster to the equilibrium value. For the lighter phase (1), the total phasic kinetic energy actually increases. With respect to the preservation characteristics of total phasic entropy, there are no noticeable differences for the density ratio sweep. Overall, the change in total phasic entropy normalized by its initial value is captured up to an error of $10^{-3}$.

\subsubsection{Water-air combination with $Ma_0=0.05$}
\label{subsubsec:tgv_2_cases}

\begin{figure}
    \centering
    \begin{tikzpicture}
    \node (img12) {\includegraphics[width=0.23\textwidth]{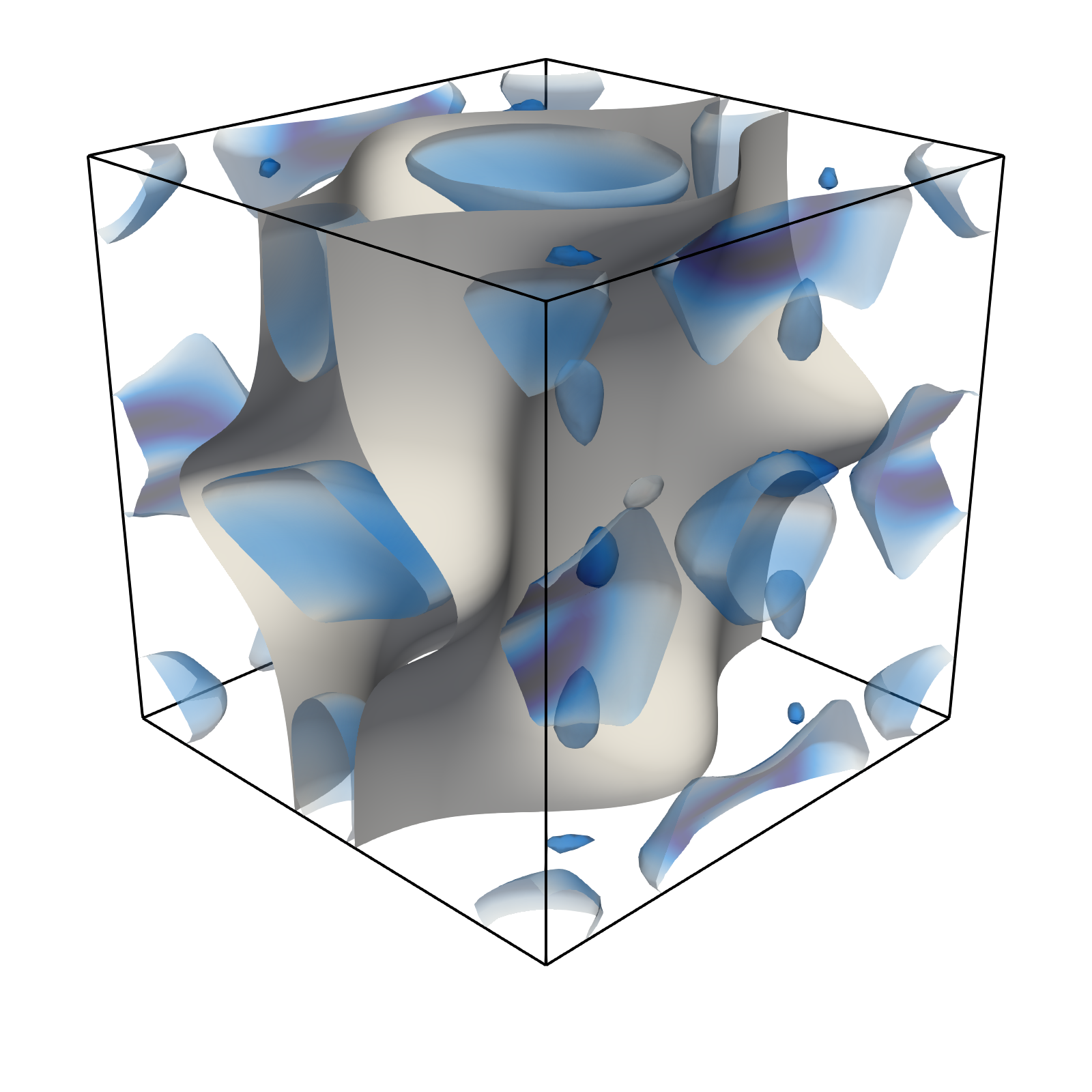}};
    \node (img13) [right=0pt of img12] {\includegraphics[width=0.23\textwidth]{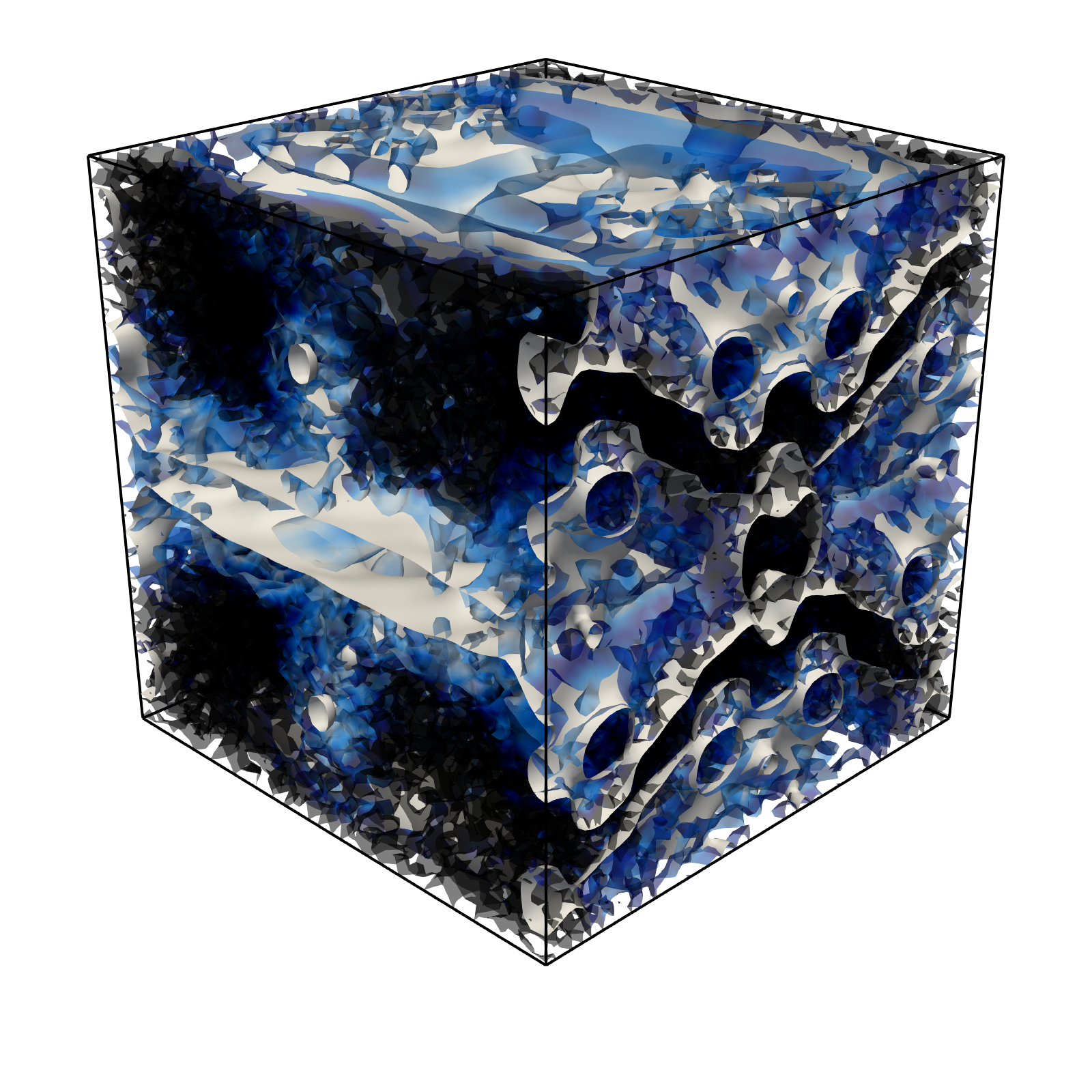}};
    \node (img14) [right=0pt of img13] {\includegraphics[width=0.23\textwidth]{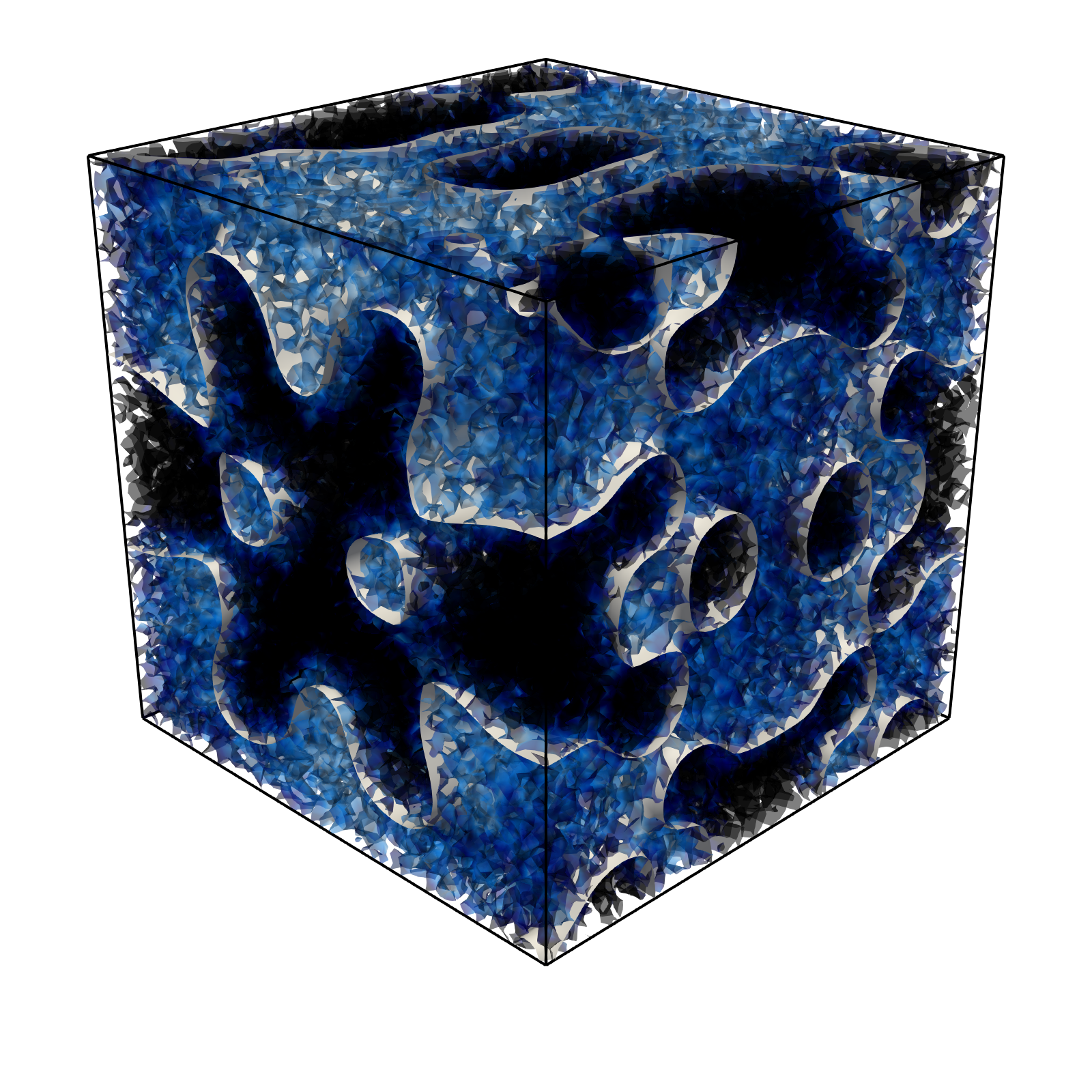}};
    \node (img15) [right=0pt of img14] {\includegraphics[width=0.27\textwidth, trim=10.5in 0in 3.5in 0in, clip]{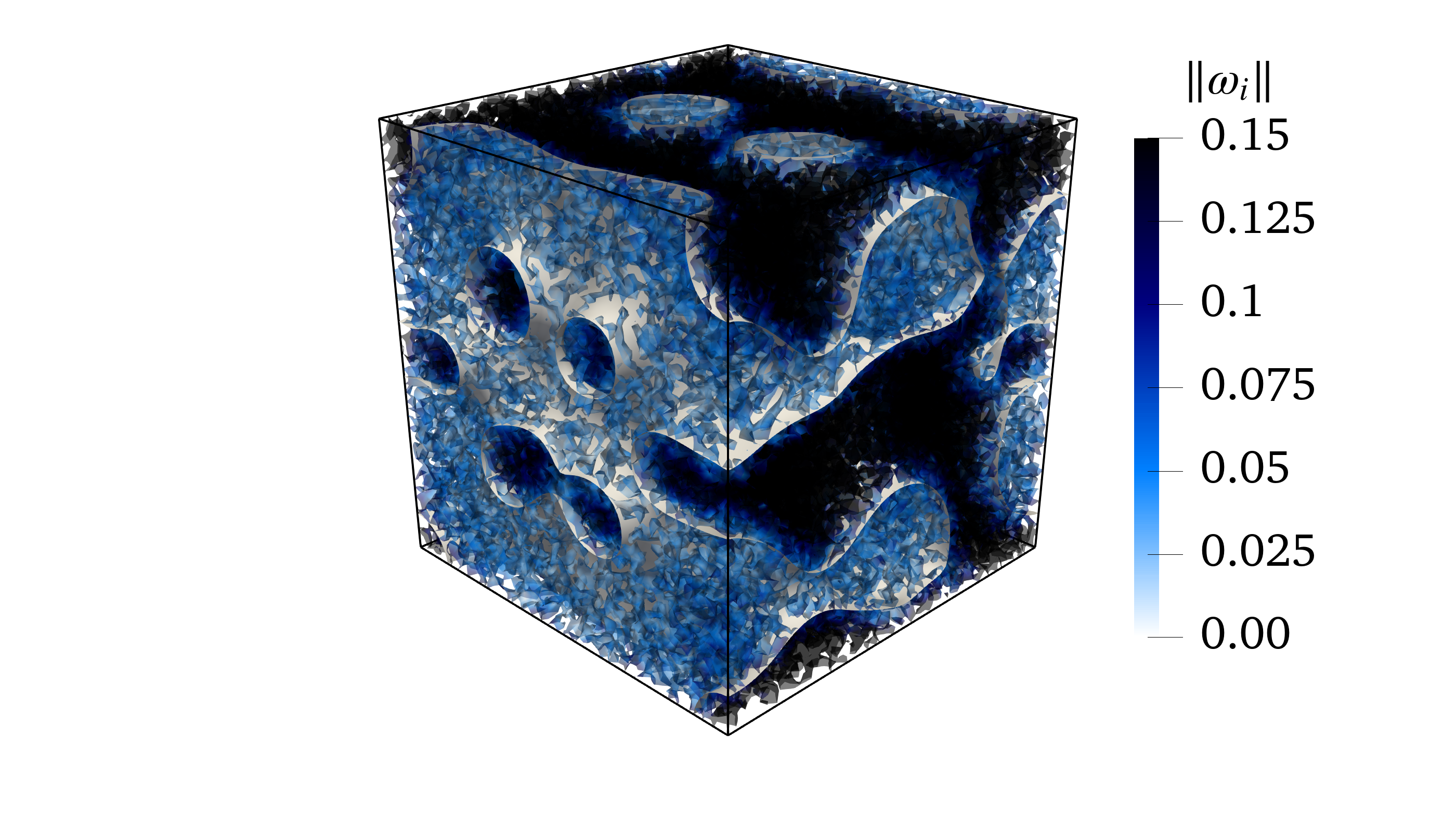}};
    \begin{scope}[shift={(-7em,-8.5em)}] 
    \node (x1) at (2.9  ,1.5) {1};
    \node (x2) at (7. ,1.5) {5};
    \node (x3) at (11.1 ,1.5) {20};
    \node (x4) at (14.9,1.5) {40};
    
    \node at (9.375,0.25) {Time, $Ma_0t \; [-]$};
    
    \draw[->, thick] (x1.south west) ++(-1.5,-0.25) -- ($(x4.south east) + (1.7,-0.25)$);
    \end{scope}
    \end{tikzpicture}
    \caption{
    Water-air compressible Taylor-Green vortex visualizations at different times. Interface iso-contour ($\phi_1 = 0.5$) as gray opaque surfaces, and Q-criterion iso-contour surfaces colored by vorticity magnitude.
    }
    \label{fig:ctgv-stiffened-gas-3d}
\end{figure}
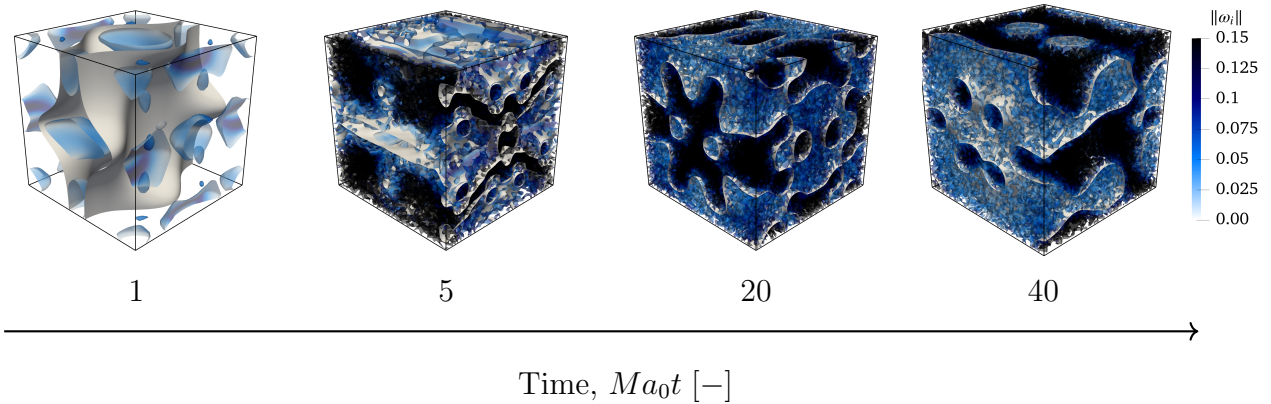

In this section, the baseline pressure is increased to $p_0 = 10$ to allow for positive pressures, given that the maximum dynamic pressure $\rho_{0,1} Ma_0^2 \approx 2.5 \gg 1/\gamma_1$. For this last test case, as the densities reach 3 orders of magnitude difference, the rate of exchange of kinetic and internal energy is likely to further increase compared to the previous cases [\cref{subsubsec:tgv_1_cases}]. On the other hand, the $Ma_0$ is decreased to compensate for the density ratio, which should decrease the rate of energy exchange due to compressibility effects. A qualitative flow field description is presented in~\cref{fig:ctgv-stiffened-gas-3d}, where the interface is contorted by the vortical flow.

\begin{figure}
    \centering
    \begin{subfigure}{0.425\linewidth}
    \includegraphics[width=\linewidth]{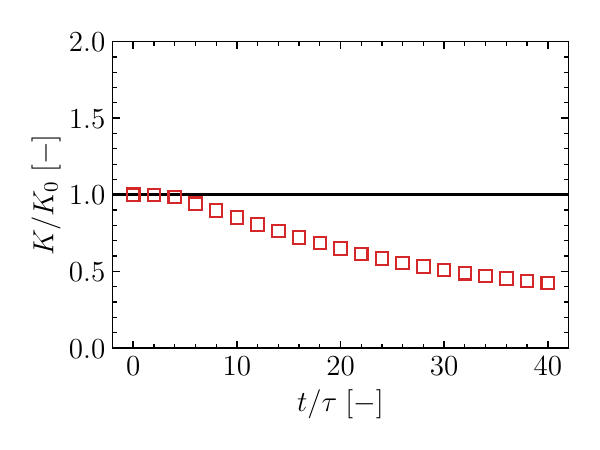}
    \caption{Mixture kinetic energy, $\rho k$}
    \end{subfigure}%
    \begin{subfigure}{0.425\linewidth}
    \includegraphics[width=\linewidth]{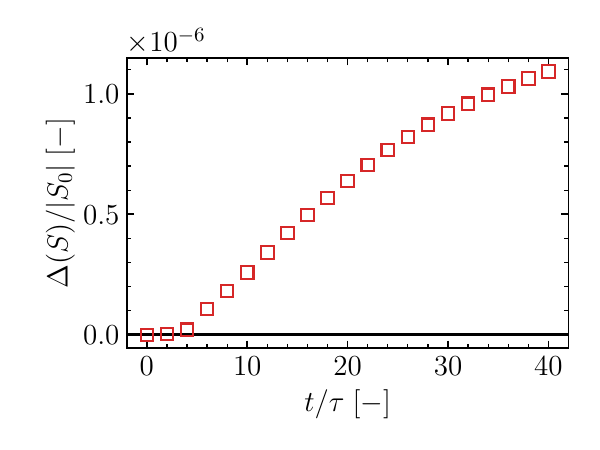}
    \caption{Mixture entropy, $\rho s$}
    \end{subfigure}
    \caption{Mixture conservation properties of compressible Taylor-Green vortex at $Ma=0.05$ for a water-air combination with $1000$ density ratio. Constant value depicted by solid black lines (\rule[2pt]{0.35cm}{1pt}), and the instantaneous values of the current simulation by red ({\protect\tikz\protect\draw[line width = 0.75pt, mpl_red] (0,0) rectangle +(1ex,1ex) ;}) squares. Evolution of normalized (a) total mixture kinetic energy $K = \sum_l K_l$, (b) change in mixture entropy $S = \sum_l \int \phi_l \rho_l s_l dV = \int \rho s dV$.}
    \label{fig:ctgv-stiffened-gas-mixture-conservation}
\end{figure}

\Cref{fig:ctgv-stiffened-gas-mixture-conservation} demonstrates the preservation properties of the scheme at the mixture level. Although the lower $Ma_0=0.05$, the rate of kinetic energy and internal energy exchange is higher compared to the previous cases. Conversely, the change in total mixture entropy is much lower, mostly influenced by $Ma_0$. It is noteworthy that the proposed scheme in this work, which is not an exact entropy preserving scheme as in~\cite{yoshida:2026}, is able to achieve comparable mixture entropy preservation levels up to order of $10^{-6}$. Possibly demonstrating an advantage of using the cheaper approximate scheme to achieve the same entropy preservation levels as an exact scheme.

\begin{figure}
    \centering
    \begin{subfigure}{0.425\linewidth}
    \includegraphics[width=\linewidth]{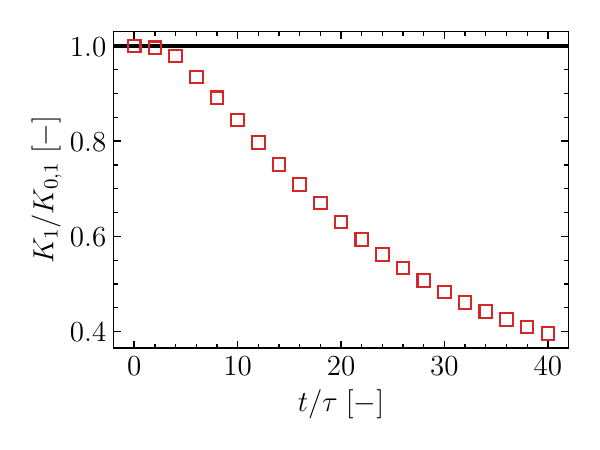}
    \caption{Kinetic energy of phase 1, $\phi_1 \rho_1 k_1$}
    \end{subfigure}%
    \begin{subfigure}{0.425\linewidth}
    \includegraphics[width=\linewidth]{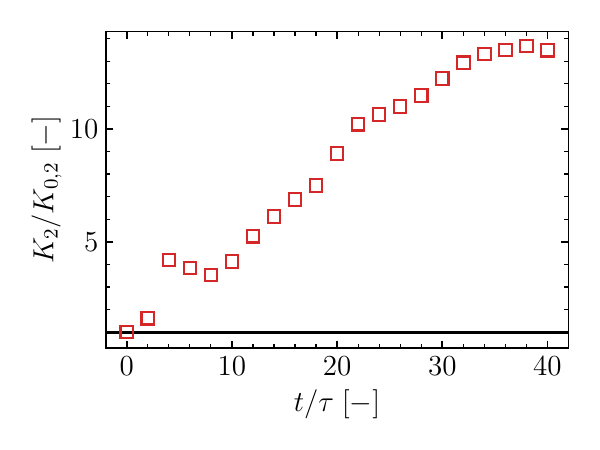}
    \caption{Kinetic energy of phase 2, $\phi_2 \rho_2 k_2$}
    \end{subfigure}
    
    \begin{subfigure}{0.425\linewidth}
    \includegraphics[width=\linewidth]{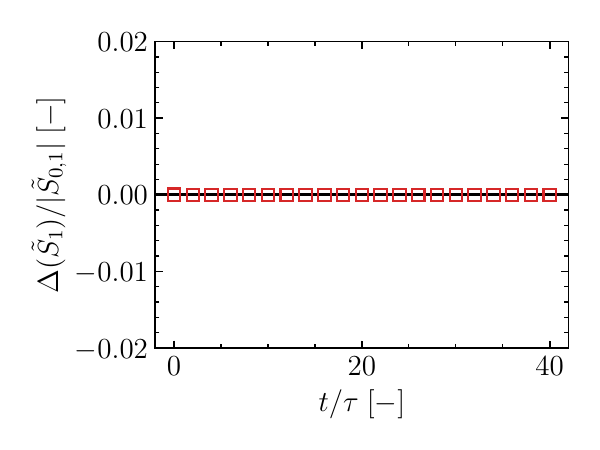}
    \caption{Change in entropy per unit mass of phase 1, $\phi_1 {s}_1$}
    \end{subfigure}%
    \begin{subfigure}{0.425\linewidth}
    \includegraphics[width=\linewidth]{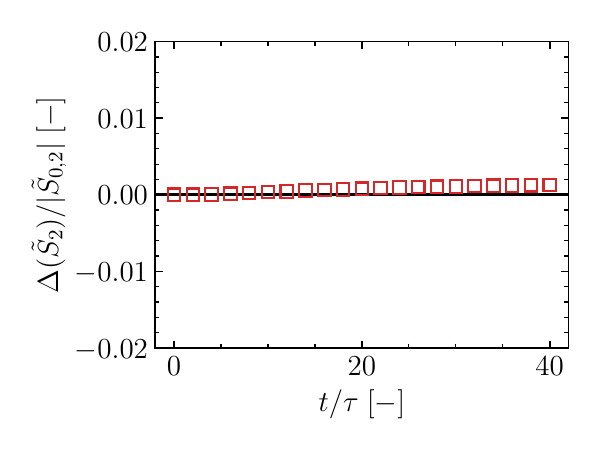}
    \caption{Change in entropy per unit mass of phase 2, $\phi_2 {s}_2$}
    \end{subfigure}
    \caption{Phasic conservation properties of compressible Taylor-Green vortex at $Ma=0.05$ for a water-air combination with $1000$ density ratio. Constant value depicted by solid black lines (\rule[2pt]{0.35cm}{1pt}), and the instantaneous values of the current simulation by red ({\protect\tikz\protect\draw[line width = 0.75pt, mpl_red] (0,0) rectangle +(1ex,1ex) ;}) squares. Evolution of normalized total phasic kinetic energy $K_l = \int \phi_l \rho_l k_l dV$ of (a) phase 1, and (b) phase 2. Evolution of normalized change in total phasic entropy per unit mass $\tilde{S}_l = \int \phi_l {s}_l dV$ of (c) phase 1, and (d) phase 2.}
    \label{fig:ctgv-stiffened-gas-phasic-conservation}
\end{figure}

Evaluating phasic preservation properties, for this test case, phase 1 (water) is the heavier one though still the ``dispersed'' one ($\langle \phi_1 \rangle \approx 0.3$). \cref{fig:ctgv-stiffened-gas-phasic-conservation} (a) demonstrates that the total phasic kinetic energy decay is dominated by the heavier phase (1), while the lighter one (2) increases. Even though the total kinetic energy of phase 2 increases by an order of magnitude, given that its phasic density is much lower, its initial value $K_{0,2}$ is also a small value, it does not yield in nonphysical kinetic energy generation the the mixture level [\cref{fig:ctgv-stiffened-gas-mixture-conservation} (a)]. The change in total phasic entropy is better captured compared to the previous cases as observed in \cref{fig:ctgv-stiffened-gas-phasic-conservation} (c,d). The heavier phase (1) presents a change in total phasic internal energy better than an order of $10^{-2}$, whereas phase 2 demonstrates an increase at an order of $10^{-2}$.

%% file: chapters/conclusions.tex
\section{Conclusions}
\label{sec:conc}


In this work, we proposed a new phase field model for the non-equilibrium six- and seven-equation formulations for compressible two-phase flows. Both models have a monotonic mixture speed of sound, which better captures wave transmission across interfaces. The phase field model for the seven-equation formulation specifically provides an IEC-compatible entropy-conservative regularization framework admitting conservative phasic and mixture entropy transport equations, a desirable feature for the construction of stable and robust numerical schemes. In contrast to equilibrium formulations, the non-equilibrium formulations require a finite amount of the conjugate phase to avoid degeneracy of eigenvectors. Therefore, we proposed a modification of the phase field model to account for the presence of a finite amount of the conjugate phase at interface equilibrium, with the analogous ACDI model also derived. These modifications allow for more accurate capturing of phasic densities, essential in regularization of phasic mass around the interface. 

Furthermore, we proved additional consistency conditions for flux splittings considering the interface-equilibrium conditions. Among the possible combinations of phasic internal energy convective and interface regularization flux splittings, we identified the ones to allow for long-time integration stability and consistent with discrete kinetic energy and entropy preservation requirements. The quartic flux splitting for the phasic internal energy convection, and the cubic flux splitting for the phasic internal energy interface regularization, both require the same discretization of the volume fraction transport equation as previous formulations.

Finally, to demonstrate accuracy, stability and robust of the underlying seven-equation formulation with phase field model that allows for finite amount of conjugate phase and the proposed KEEP-like flux splittings, we performed numerical simulations of high-density ratio interface advection, pressure-driven bubble oscillation, high-impedance interface oblique wave interaction, and compressible two-phase Taylor-Green vortex. The proposed framework is able to extend stability to previously non-reported long limits of time integration, and improved accuracy for acoustic test cases, while being nondissipative.

%% file: appendix/mcdi.tex
\section{Non-zero finite bound phase field model}
\label{apx:modified-phase-field}

\begin{figure}
    \centering
    \includegraphics[width=0.7\linewidth]{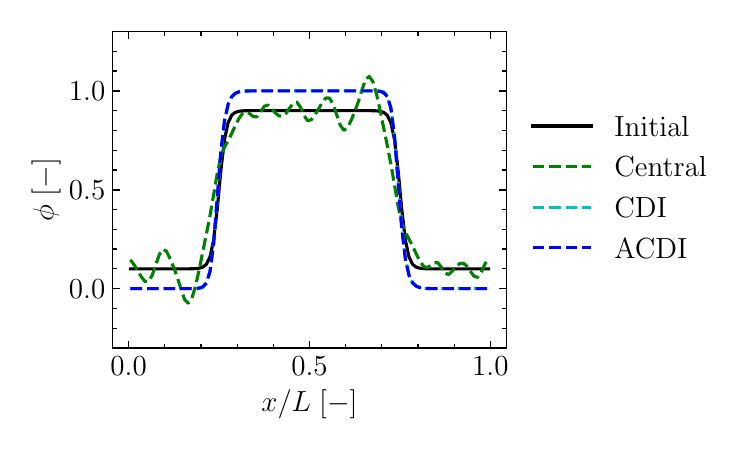}
    \caption{Redistribution of the other phase by standard phase field models in interface advection with small amount of other phase added, $\phi=10^{-1}$, in a periodic domain, compared to the initial profile, the standard central discrezation without phase field model and the standard central discretization with both CDI and ACDI as phase field models.}
    \label{fig:standard_phase_field}
\end{figure}

Consider, for instance, the standalone advection of an interface with a small amount of the other phase, $\phi_1=\delta=10^{-1}$ (exaggerated for illustrative purposes), in a periodic domain presented in the~\cref{fig:standard_phase_field}. Both approaches asymptote to 0 and 1, eliminating the small amount of the other phase added.

\begin{figure}
    \centering
    \includegraphics[width=0.55\linewidth]{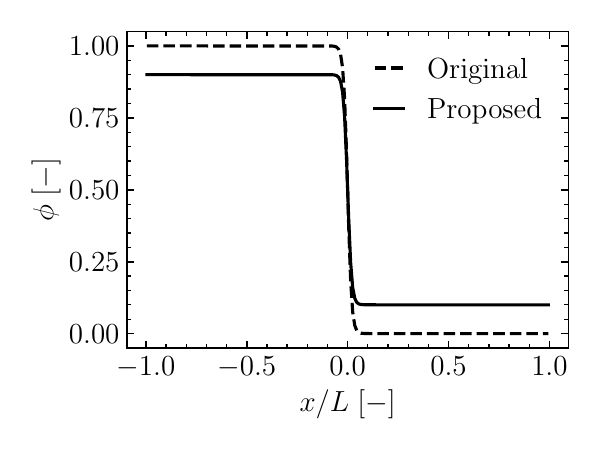}
    \caption{Comparison of interface equilibrium profiles with and without finite non-zero bounds.}
    \label{fig:comp_eq_profile}
\end{figure}

The equilibrium profile for both formulations is a $\tanh$ function described as
\begin{equation}
\label{eq:equilibrium_profile_original} 
\phi_{\rm original} = \frac{1}{2} \lrsb{1+\tanh\lrp{\frac{x}{2\epsilon}}},
\end{equation}
\noindent where $x$ indicates the distance from the interface and $\epsilon$ its finite interface thickness. It is clear that $\phi$ asymptotes respectively to 0 and 1 for large enough negative and positive $x$, \textit{i.e.}, the limit of pure phase. We propose a correction factor in the~\cref{eq:equilibrium_profile_original}, such that $\phi$ asymptotes to $\delta$ and $1-\delta$ (the small amount of the other phase), which is given by
\begin{equation}
\label{eq:equilibrium_profile_proposed} 
\phi_{\rm proposed} = \frac{1}{2} \lrsb{1+(1-2\delta)\tanh\lrp{\frac{x}{2\epsilon}}}.
\end{equation}
A comparison of both equilibrium profiles is presented in the~\cref{fig:comp_eq_profile}. The proposed equilibrium profile is constructed such that in the limit of the amount of other phase goes to zero, the original profile is recovered.

Let $x$ be the signed-distance-like function to the interface $\psi$, the~\cref{eq:equilibrium_profile_proposed} may be rearranged as follows
\begin{equation}
\label{eq:equilibrium_profile_proposed_v2} 
\phi_{\rm proposed} = \frac{1}{2} \lrsb{1+(1-2\delta)\tanh\lrp{\frac{x}{2\epsilon}}} = \frac{(1-\delta)e^{\psi/\epsilon} + \delta}{e^{\psi/\epsilon} + 1}.
\end{equation}
Differentiating the~\cref{eq:equilibrium_profile_proposed_v2} with respect to the signed-distance-like function yields in
\begin{equation}
\label{eq:dphi_dpsi_proposed}
\frac{d \phi}{d \psi} = \frac{1}{\epsilon}\frac{(1-2\delta)e^{\psi/\epsilon}}{(e^{\psi/\epsilon} + 1)^2} = \frac{1}{\epsilon} \frac{(\phi - \delta)(1 - \phi - \delta)}{(1 - 2\delta)},
\end{equation}
where the signed-distance function may be defined as $\psi(\phi) = s(\phi) - s(\phi = 0.5)$ with $s$ being the coordinate in the normal direction of the interface. Taking another derivative of the~\cref{eq:dphi_dpsi_proposed} with respect to $s$, where $d\phi/d\psi = d\phi/ds$, leads to.
\begin{equation}
\label{eq:d2phi_ds2_proposed}
\frac{d}{d s} \lrsb{\epsilon(1 - 2\delta) \frac{d\phi}{ds} - (\phi - \delta)(1 - \phi - \delta)} = 0.
\end{equation}
Recasting the~\cref{eq:d2phi_ds2_proposed} in Cartesian coordinates yields in (the modified Allen-Cahn for finite non-zero bounds formulation)
\begin{equation}
\label{eq:phase_field_proposed_cdi}
\begin{aligned}
\frac{\partial \phi_1}{\partial t} & + \ppxj{\phi_1 u_j} = \phi_1 \frac{\partial u_j}{\partial x_j} \\
& + \highlight{\frac{\partial}{\partial x_j} \lrcb{\Gamma\lrsb{\epsilon(1 - 2\delta) \frac{\partial\phi_1}{\partial x_j} - (\phi_1 - \delta)(1 - \phi_1 - \delta)n_{1,j}}}}.
\end{aligned}
\end{equation}
Substituting the proposed equilibrium profile from the~\cref{eq:equilibrium_profile_proposed} in the modified Allen-Cahn formulation in the~\cref{eq:phase_field_proposed_cdi} gives
\begin{equation}
\label{eq:phase_field_proposed_acdi}
\begin{aligned}
\frac{\partial \phi_1}{\partial t} & + \ppxj{\phi_1 u_j} = \phi_1 \frac{\partial u_j}{\partial x_j} \\
& + \highlight{\frac{\partial}{\partial x_j} \lrcb{\Gamma\lrsb{\epsilon(1 - 2\delta) \frac{\partial\phi_1}{\partial x_j} - \frac{(1-2\delta)^2}{4}\lrp{1 - \tanh^2 \lrp{\frac{\psi_1}{2\epsilon}}}n_{1,j}}}},
\end{aligned}
\end{equation}
where the signed distance-like function $\psi$ is adapted based on the proposed equilibrium profile. Solving for $\psi$ in the~\cref{eq:equilibrium_profile_proposed_v2} yields in
\begin{equation}
\label{eq:proposed_psi}
\psi = \epsilon \log \lrp{\frac{\phi - \delta}{1 - \phi - \delta}}. 
\end{equation}
In order to control the limit of $\psi$ when $\phi$ or $1-\phi \rightarrow \delta$, one may add a small number $\varepsilon = 10^{-100}$ in both numerator and denominator inside of the $\log$ of the Eq.~\eqref{eq:proposed_psi}.

\begin{figure}
    \centering
    \includegraphics[width=0.7\linewidth]{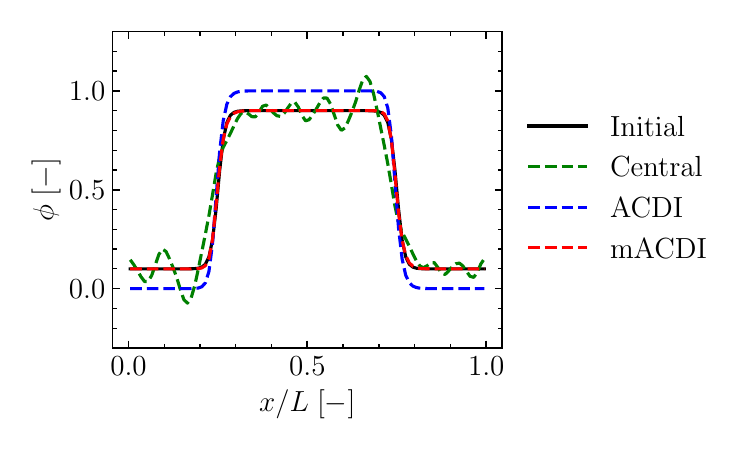}
    \caption{Interface advection with small amount of other phase, $\phi=10^{-1}$, in a periodic domain, with the proposed modified interface equilibrium profile compared to the initial profile, standard central discrezation without phase field model and standard central discretization with ACDI as the phase field model.}
    \label{fig:proposed_phase_field}
\end{figure}

One may solve the standalone advection of an interface with a small amount of the other phase in a periodic domain with the Eq.~\eqref{eq:phase_field_proposed_acdi}. The Fig.~\ref{fig:proposed_phase_field} highlights how the proposed phase field model with finite non-zero bounds is able to retain the modified equilibrium interface profile and the amount of the other phase added, contrary to standard phase field models. This proposed model has potential to more accurately capture phasic density $\rho_l$ and allow for more robust numerical schemes that require division by $\phi$, such as the step for pressure relaxation with an algebraic expression for stiffened-gas equation of state. Furthermore, this model is not restricted to the seven-equation model for compressible two-phase formulation, it can be used whenever division by $\phi$ is a concern.

In the context of the seven-equation model, instead of $\phi_1$ being advected by the relaxed velocity $u$, it is advected by the interface velocity $u_I$. Thus, the final expression for volume fraction transport equation without the relaxation term is
\begin{equation}
\label{eq:phase_field_proposed_7eq_apx}
\begin{aligned}
\frac{\partial \phi_1}{\partial t} & + \ppxj{\phi_1 u_{I,j}} & \\
& = \phi_1 \frac{\partial u_{I,j}}{\partial x_j} + {\frac{\partial}{\partial x_j} \lrcb{\Gamma\lrsb{\epsilon(1 - 2\delta) \frac{\partial\phi_1}{\partial x_j} - (\phi_1 - \delta)(1 - \phi_1 - \delta)n_{1,j}}}}, \\
& = \phi_1 \frac{\partial u_{I,j}}{\partial x_j} + {\frac{\partial}{\partial x_j} \lrcb{\Gamma\lrsb{\epsilon(1 - 2\delta) \frac{\partial\phi_1}{\partial x_j} - \frac{(1-2\delta)^2}{4}\lrp{1 - \tanh^2 \lrp{\frac{\psi_1}{2\epsilon}}}n_{1,j}}}}.
\end{aligned}
\end{equation}

%% file: cas-refs.bib
@article{mirjalili:2020,
title = {A conservative diffuse interface method for two-phase flows with provable boundedness properties},
journal = {Journal of Computational Physics},
volume = {401},
pages = {109006},
year = {2020},
issn = {0021-9991},
doi = {10.1016/j.jcp.2019.109006},
author = {Shahab Mirjalili and Christopher B. Ivey and Ali Mani},
}

@article{chiu:2011,
title = {A conservative phase field method for solving incompressible two-phase flows},
journal = {Journal of Computational Physics},
volume = {230},
number = {1},
pages = {185-204},
year = {2011},
issn = {0021-9991},
doi = {10.1016/j.jcp.2010.09.021},
author = {Pao-Hsiung Chiu and Yan-Ting Lin},
}

@article{coralic:2014,
title = "{Finite-volume WENO scheme for viscous compressible multicomponent flows}",
journal = {Journal of Computational Physics},
volume = {274},
pages = {95-121},
year = {2014},
issn = {0021-9991},
doi = {10.1016/j.jcp.2014.06.003},
author = {Vedran Coralic and Tim Colonius},
}

@article{tiwari:2013,
    title = {A diffuse interface model with immiscibility preservation},
    journal = {Journal of Computational Physics},
    volume = {252},
    pages = {290-309},
    year = {2013},
    issn = {0021-9991},
    author = {Arpit Tiwari and Jonathan B. Freund and Carlos Pantano},}

@article{coralic:2013,
    title = {Shock-induced collapse of a bubble inside a deformable vessel},
    journal = {European Journal of Mechanics - B/Fluids},
    volume = {40},
    pages = {64-74},
    year = {2013},
    issn = {0997-7546},
    author = {Vedran Coralic and Tim Colonius},}

@article{blake:1987,
    author = "Blake, J. R. and Gibson, D. C.",
    title = "Cavitation Bubbles Near Boundaries", 
    journal= "Annual Review of Fluid Mechanics",
    year = "1987",
    volume = "19",
    number = "Volume 19, 1987",
    pages = "99-123",
    publisher = "Annual Reviews",
    issn = "1545-4479",
    type = "Journal Article",}

@article{escaler:2006,
    title = {Detection of cavitation in hydraulic turbines},
    journal = {Mechanical Systems and Signal Processing},
    volume = {20},
    number = {4},
    pages = {983-1007},
    year = {2006},
    issn = {0888-3270},
    author = {Xavier Escaler and Eduard Egusquiza and Mohamed Farhat and François Avellan and Miguel Coussirat},}

@article{benyakar:1998,
    author = {Ben-Yakar, Adela and Natan, Benveniste and Gany, Alon},
    title = {Investigation of a Solid Fuel Scramjet Combustor},
    journal = {Journal of Propulsion and Power},
    volume = {14},
    number = {4},
    pages = {447-455},
    year = {1998},}

@article{tarey:2024,
    title = {Evolution of a shock-impacted reactive liquid fuel droplet with evaporation effects: A numerical study},
    journal = {International Journal of Multiphase Flow},
    volume = {174},
    pages = {104744},
    year = {2024},
    issn = {0301-9322},
    author = {Prashant Tarey and Praveen Ramaprabhu and Jacob A. McFarland},}

@article{utturkar:2005,
    title = {Recent progress in modeling of cryogenic cavitation for liquid rocket propulsion},
    journal = {Progress in Aerospace Sciences},
    volume = {41},
    number = {7},
    pages = {558-608},
    year = {2005},
    issn = {0376-0421},
    author = {Yogen Utturkar and Jiongyang Wu and Guoyo Wang and Wei Shyy},}

@article{carmicino:2015,
    author = {Carmicino, C. and Russo Sorge, A.},
    title = {Experimental Investigation into the Effect of Solid-Fuel Additives on Hybrid Rocket Performance},
    journal = {Journal of Propulsion and Power},
    volume = {31},
    number = {2},
    pages = {699-713},
    year = {2015},}

@article{shukla:2010,
    title = {An interface capturing method for the simulation of multi-phase compressible flows},
    journal = {Journal of Computational Physics},
    volume = {229},
    number = {19},
    pages = {7411-7439},
    year = {2010},
    issn = {0021-9991},
    author = {Ratnesh K. Shukla and Carlos Pantano and Jonathan B. Freund},}

@article{jain:2023,
    title = {Assessment of diffuse-interface methods for compressible multiphase fluid flows and elastic-plastic deformation in solids},
    journal = {Journal of Computational Physics},
    volume = {475},
    pages = {111866},
    year = {2023},
    issn = {0021-9991},
    author = {Suhas S. Jain and Michael C. Adler and Jacob R. West and Ali Mani and Parviz Moin and Sanjiva K. Lele},}

@article{saurel:1999,
    title = {A Multiphase {G}odunov Method for Compressible Multifluid and Multiphase Flows},
    journal = {Journal of Computational Physics},
    volume = {150},
    number = {2},
    pages = {425-467},
    year = {1999},
    issn = {0021-9991},
    author = {Richard Saurel and R\'emi Abgrall},}

@article{saurel:2001, 
    title={A multiphase model for compressible flows with interfaces, shocks, detonation waves and cavitation}, 
    volume={431}, 
    journal={Journal of Fluid Mechanics}, 
    author={Saurel, Richard and Le M\'etayer, Olivier}, 
    year={2001}, 
    pages={239–271}}

@article{saurel:2009,
    title = {Simple and efficient relaxation methods for interfaces separating compressible fluids, cavitating flows and shocks in multiphase mixtures},
    journal = {Journal of Computational Physics},
    volume = {228},
    number = {5},
    pages = {1678-1712},
    year = {2009},
    issn = {0021-9991},
    author = {Richard Saurel and Fabien Petitpas and Ray A. Berry},}

@article{baer:1986,
    title = {A two-phase mixture theory for the deflagration-to-detonation transition ({DDT}) in reactive granular materials},
    journal = {International Journal of Multiphase Flow},
    volume = {12},
    number = {6},
    pages = {861-889},
    year = {1986},
    issn = {0301-9322},
    author = {M. R. Baer and J. W. Nunziato},}

@article{kapila:2001,
    author = {Kapila, A. K. and Menikoff, R. and Bdzil, J. B. and Son, S. F. and Stewart, D. S.},
    title = "{Two-phase modeling of deflagration-to-detonation transition in granular materials: Reduced equations}",
    journal = {Physics of Fluids},
    volume = {13},
    number = {10},
    pages = {3002-3024},
    year = {2001},
    month = {10},
    issn = {1070-6631},}

@article{jain:2020,
    title = {A conservative diffuse-interface method for compressible two-phase flows},
    journal = {Journal of Computational Physics},
    volume = {418},
    pages = {109606},
    year = {2020},
    issn = {0021-9991},
    author = {Suhas S. Jain and Ali Mani and Parviz Moin},}

@article{allaire:2002,
    title = {A Five-Equation Model for the Simulation of Interfaces between Compressible Fluids},
    journal = {Journal of Computational Physics},
    volume = {181},
    number = {2},
    pages = {577-616},
    year = {2002},
    issn = {0021-9991},
    author = {Gr\'egoire Allaire and Sébastien Clerc and Samuel Kokh},}

@article{petitpas:2007,
    title = {A relaxation-projection method for compressible flows. Part II: Artificial heat exchanges for multiphase shocks},
    journal = {Journal of Computational Physics},
    volume = {225},
    number = {2},
    pages = {2214-2248},
    year = {2007},
    issn = {0021-9991},
    author = {Fabien Petitpas and Erwin Franquet and Richard Saurel and Olivier {Le M\'etayer}},}

@article{zein:2010,
    title = {Modeling phase transition for compressible two-phase flows applied to metastable liquids},
    journal = {Journal of Computational Physics},
    volume = {229},
    number = {8},
    pages = {2964-2998},
    year = {2010},
    issn = {0021-9991},
    author = {Ali Zein and Maren Hantke and Gerald Warnecke},}

@article{pelanti:2014,
    title = {A mixture-energy-consistent six-equation two-phase numerical model for fluids with interfaces, cavitation and evaporation waves},
    journal = {Journal of Computational Physics},
    volume = {259},
    pages = {331-357},
    year = {2014},
    issn = {0021-9991},
    author = {Marica Pelanti and Keh-Ming Shyue},}

@article{chiapolino:2017,
    title = {Sharpening diffuse interfaces with compressible fluids on unstructured meshes},
    journal = {Journal of Computational Physics},
    volume = {340},
    pages = {389-417},
    year = {2017},
    issn = {0021-9991},
    author = {Alexandre Chiapolino and Richard Saurel and Boniface Nkonga},}

@article{wong:2021,
    title = {A positivity-preserving high-order weighted compact nonlinear scheme for compressible gas-liquid flows},
    journal = {Journal of Computational Physics},
    volume = {444},
    pages = {110569},
    year = {2021},
    issn = {0021-9991},
    author = {Man Long Wong and Jordan B. Angel and Michael F. Barad and Cetin C. Kiris},}

@article{drew:1983,
    author = "Drew, D A",
    title = "Mathematical Modeling of Two-Phase Flow", 
    journal= "Annual Review of Fluid Mechanics",
    year = "1983",
    volume = "15",
    number = "Volume 15, 1983",
    pages = "261-291",
    publisher = "Annual Reviews",
    issn = "1545-4479",
    type = "Journal Article",}

@article{pelanti:2022,
    title = {Arbitrary-rate relaxation techniques for the numerical modeling of compressible two-phase flows with heat and mass transfer},
    journal = {International Journal of Multiphase Flow},
    volume = {153},
    pages = {104097},
    year = {2022},
    issn = {0301-9322},
    author = {Marica Pelanti},}

@article{saurel:2003, 
    title={A multiphase model with internal degrees of freedom: application to shock–bubble interaction}, 
    volume={495}, 
    journal={Journal of Fluid Mechanics}, 
    author={Saurel, Richard and Gavrilyuk, Sergey and Renaud, François}, 
    year={2003}, 
    pages={283–321}}

@article{jain:2022b,
    title = {Accurate conservative phase-field method for simulation of two-phase flows},
    journal = {Journal of Computational Physics},
    volume = {469},
    pages = {111529},
    year = {2022},
    issn = {0021-9991},
    author = {Suhas S. Jain},}

@article{kuya:2018,
    title = {Kinetic energy and entropy preserving schemes for compressible flows by split convective forms},
    journal = {Journal of Computational Physics},
    volume = {375},
    pages = {823-853},
    year = {2018},
    issn = {0021-9991},
    author = {Yuichi Kuya and Kosuke Totani and Soshi Kawai},}

@article{garrick:2017,
    title = {A finite-volume HLLC-based scheme for compressible interfacial flows with surface tension},
    journal = {Journal of Computational Physics},
    volume = {339},
    pages = {46-67},
    year = {2017},
    issn = {0021-9991},
    author = {Daniel P. Garrick and Mark Owkes and Jonathan D. Regele},}

@article{jain:2022a,
    title = {A kinetic energy–and entropy-preserving scheme for compressible two-phase flows},
    journal = {Journal of Computational Physics},
    volume = {464},
    pages = {111307},
    year = {2022},
    issn = {0021-9991},
    author = {S. S. Jain and P. Moin},}

@article{yoshida:2026,
    author = {S. Yoshida and S. Kawai and S. Kawai},
    title = {Strictly entropy-conserving seven-equation based KEEP scheme coupled with phase-ﬁeld method for compressible two-phase ﬂows},
    journal = {AIAA SCITECH 2026 Forum},
    year = {2026},}

@article{lemetayer:2004,
    title = {Élaboration des lois d'état d'un liquide et de sa vapeur pour les modèles d'écoulements diphasiques},
    journal = {International Journal of Thermal Sciences},
    volume = {43},
    number = {3},
    pages = {265-276},
    year = {2004},
    issn = {1290-0729},
    author = {O. {Le Métayer} and J. Massoni and R. Saurel},}

@article{dodd:2016, 
    title={On the interaction of Taylor length scale size droplets and isotropic turbulence}, 
    volume={806}, 
    journal={Journal of Fluid Mechanics}, 
    author={Dodd, Michael S. and Ferrante, Antonino}, 
    year={2016}, 
    pages={356–412}}

@article{battistella:2020,
    title = {On the terminal velocity of single bubbles rising in non-Newtonian power-law liquids},
    journal = {Journal of Non-Newtonian Fluid Mechanics},
    volume = {278},
    pages = {104249},
    year = {2020},
    issn = {0377-0257},
    author = {A. Battistella and S.J.G. {van Schijndel} and M.W. Baltussen and I. Roghair and M. {van Sint Annaland}},}

@article{crialesi:2022, 
    title={Modulation of homogeneous and isotropic turbulence in emulsions}, 
    volume={940}, 
    journal={Journal of Fluid Mechanics}, 
    author={Crialesi-Esposito, Marco and Rosti, Marco Edoardo and Chibbaro, Sergio and Brandt, Luca}, 
    year={2022}, 
    pages={A19}}

@article{hatashita:2026c,
    title = {Scalings and simulation requirements in two-phase flows},
    author = {Hatashita, Luis H. and Nathan, Pranav and Jain, Suhas S.},
    journal = {Phys. Rev. Fluids},
    volume = {11},
    issue = {7},
    pages = {074303},
    numpages = {30},
    year = {2026},
    month = {Jul},
    publisher = {American Physical Society},
    }

@article{hatashita:2025,
    author = {L. H. Hatashita and S. S. Jain},
    title = {Interface Preservation in Modeling Compressible Two-Phase Flows Using Six- and Seven-Equation Formulations},
    journal = {AIAA SCITECH 2025 Forum},
    year = {2025},}

@article{hatashita:2026a,
    author = {L. H. Hatashita and T. Samanta and S. S. Jain},
    title = {A Robust Seven-Equation Formulation for Compressible Two-Phase Flows for High-Speed Applications},
    journal = {AIAA SCITECH 2026 Forum},
    year = {2026},}

@inproceedings{hatashita:2026b,
    author = {L. H. Hatashita and S. S. Jain},
    title = {Advancements in the Seven-Equation Formulation for Compressible Two-Phase Flows for Modeling Leidenfrost Effect},
    booktitle = {Proceedings of ILASS-Americas 2026},
    year = {2026}}

@article{blaisdell:1996,
    title = {The effect of the formulation of nonlinear terms on aliasing errors in spectral methods},
    journal = {Applied Numerical Mathematics},
    volume = {21},
    number = {3},
    pages = {207-219},
    year = {1996},
    issn = {0168-9274},
    author = {G.A. Blaisdell and E.T. Spyropoulos and J.H. Qin},}

@article{perigaud:2005,
    title = {A compressible flow model with capillary effects},
    journal = {Journal of Computational Physics},
    volume = {209},
    number = {1},
    pages = {139-178},
    year = {2005},
    issn = {0021-9991},
    author = {Guillaume Perigaud and Richard Saurel},}

@article{murrone:2005,
    title = {A five equation reduced model for compressible two phase flow problems},
    journal = {Journal of Computational Physics},
    volume = {202},
    number = {2},
    pages = {664-698},
    year = {2005},
    issn = {0021-9991},
    author = {Angelo Murrone and Hervé Guillard},}

@article{saurel:2008, 
    title={Modelling phase transition in metastable liquids: application to cavitating and flashing flows}, 
    volume={607}, 
    journal={Journal of Fluid Mechanics}, 
    author={Saurel, Richard and Petitpas, Fabien and Abgrall, Remi}, 
    year={2008}, 
    pages={313–350}}

@article{abgrall:2014,
    title = {Discrete Equation Method (DEM) for the simulation of viscous, compressible, two-phase flows},
    journal = {Computers \& Fluids},
    volume = {91},
    pages = {164-181},
    year = {2014},
    issn = {0045-7930},
    author = {R. Abgrall and M.G. Rodio},}

@article{rodio:2015,
    title = {An innovative phase transition modeling for reproducing cavitation through a five-equation model and theoretical generalization to six and seven-equation models},
    journal = {International Journal of Heat and Mass Transfer},
    volume = {89},
    pages = {1386-1401},
    year = {2015},
    issn = {0017-9310},
    author = {M.G. Rodio and R. Abgrall},}

@article{sethian:2003,
   author = "Sethian, J. A. and Smereka, Peter",
   title = "LEVEL SET METHODS FOR FLUID INTERFACES", 
   journal= "Annual Review of Fluid Mechanics",
   year = "2003",
   volume = "35",
   number = "Volume 35, 2003",
   pages = "341-372",
   publisher = "Annual Reviews",
   issn = "1545-4479",
   type = "Journal Article",}

@article{saurel:2018,
   author = "Saurel, Richard and Pantano, Carlos",
   title = "Diffuse-Interface Capturing Methods for Compressible Two-Phase Flows", 
   journal= "Annual Review of Fluid Mechanics",
   year = "2018",
   volume = "50",
   number = "Volume 50, 2018",
   pages = "105-130",
   publisher = "Annual Reviews",
   issn = "1545-4479",
   type = "Journal Article",}

@article{abgrall:1996,
    title = {How to Prevent Pressure Oscillations in Multicomponent Flow Calculations: A Quasi Conservative Approach},
    journal = {Journal of Computational Physics},
    volume = {125},
    number = {1},
    pages = {150-160},
    year = {1996},
    issn = {0021-9991},
    author = {Rémi Abgrall},}

@article{saurel:1999b,
    author = {Saurel, Richard and Abgrall, R{\'e}mi},
    title = {A Simple Method for Compressible Multifluid Flows},
    journal = {SIAM Journal on Scientific Computing},
    volume = {21},
    number = {3},
    pages = {1115-1145},
    year = {1999},}

@article{johnsen:2012,
    title = {Preventing numerical errors generated by interface-capturing schemes in compressible multi-material flows},
    journal = {Journal of Computational Physics},
    volume = {231},
    number = {17},
    pages = {5705-5717},
    year = {2012},
    issn = {0021-9991},
    author = {Eric Johnsen and Frank Ham},}

@article{chiapolino:2017b,
    title = {A simple and fast phase transition relaxation solver for compressible multicomponent two-phase flows},
    journal = {Computers \& Fluids},
    volume = {150},
    pages = {31-45},
    year = {2017},
    issn = {0045-7930},
    author = {Alexandre Chiapolino and Pierre Boivin and Richard Saurel},}

@article{demou:2022,
    title = {A pressure-based diffuse interface method for low-Mach multiphase flows with mass transfer},
    journal = {Journal of Computational Physics},
    volume = {448},
    pages = {110730},
    year = {2022},
    issn = {0021-9991},
    author = {Andreas D. Demou and Nicolò Scapin and Marica Pelanti and Luca Brandt},}

@article{mirjalili:2017,
    title={Interface-capturing methods for two-phase flows: An overview and recent developments},
    author={Mirjalili, S. and Jain, S. S. and Dodd, M. S.},
    journal={Center for Turbulence Research Annual Research Briefs 2017},
    year={2017}}

@article{collis2022assessment,
  title={Assessment of WENO and TENO schemes for the four-equation compressible two-phase flow model with regularization terms},
  author={Collis, H and Mirjalili, S and Jain, SS and Mani, A and others},
  journal={Center for Turbulence Research Annual Research Briefs},
  pages={151--165},
  year={2022}
}

@article{clerc:2000,
    title = {Numerical Simulation of the Homogeneous Equilibrium Model for Two-Phase Flows},
    journal = {Journal of Computational Physics},
    volume = {161},
    number = {1},
    pages = {354-375},
    year = {2000},
    issn = {0021-9991},
    author = {S. Clerc},}

@article{panchal:2023,
    title = {A seven-equation diffused interface method for resolved multiphase flows},
    journal = {Journal of Computational Physics},
    volume = {475},
    pages = {111870},
    year = {2023},
    issn = {0021-9991},
    author = {Achyut Panchal and Spencer H. Bryngelson and Suresh Menon},}

@article{viqueira:2024,
    author = {Manuel Viqueira-Moreira and Christoph Brehm},
    title = {Implementation and Validation of a 7-Equation Model for High-Speed Droplet Impingement},
    journal = {AIAA AVIATION FORUM AND ASCEND 2024},
    year = {2024},
    }

@article{jain:2024,
    title = {Stable, entropy-consistent, and localized artificial-diffusivity method for capturing discontinuities},
    author = {Jain, Suhas S. and Agrawal, Rahul and Moin, Parviz},
    journal = {Phys. Rev. Fluids},
    volume = {9},
    issue = {2},
    pages = {024609},
    numpages = {27},
    year = {2024},
    month = {Feb},
    publisher = {American Physical Society},}

@article{honein:2004,
    title = {Higher entropy conservation and numerical stability of compressible turbulence simulations},
    journal = {Journal of Computational Physics},
    volume = {201},
    number = {2},
    pages = {531-545},
    year = {2004},
    issn = {0021-9991},
    author = {A. E. Honein and P. Moin},}

@article{chandrashekar:2013, 
    title={Kinetic Energy Preserving and Entropy Stable Finite Volume Schemes for Compressible Euler and Navier-Stokes Equations}, 
    volume={14}, 
    DOI={10.4208/cicp.170712.010313a}, 
    number={5}, 
    journal={Communications in Computational Physics}, 
    author={P. Chandrashekar}, 
    year={2013}, 
    pages={1252–1286}}

@article{tamaki:2022,
    title = {Comprehensive analysis of entropy conservation property of non-dissipative schemes for compressible flows: KEEP scheme redefined},
    journal = {Journal of Computational Physics},
    volume = {468},
    pages = {111494},
    year = {2022},
    issn = {0021-9991},
    author = {Y. Tamaki and Y. Kuya and S. Kawai},}

@article{furfaro:2015,
    title = {A simple HLLC-type Riemann solver for compressible non-equilibrium two-phase flows},
    journal = {Computers \& Fluids},
    volume = {111},
    pages = {159-178},
    year = {2015},
    issn = {0045-7930},
    author = {Damien Furfaro and Richard Saurel},}

@techreport{lallemand:2000,
    TITLE = {{Pressure Relaxation Procedures for Multiphase Compressible Flows}},
    AUTHOR = {Lallemand, Marie-H{\'e}l{\`e}ne and Saurel, Richard},
    NOTE = {Projet M3N},
    TYPE = {Research Report},
    NUMBER = {RR-4038},
    INSTITUTION = {{INRIA}},
    YEAR = {2000},
    HAL_ID = {inria-00072600},
    HAL_VERSION = {v1},}

@inproceedings{brill:2024,
    title={A Six-Equation Multimaterial Method for Compact Finite Differences},
    author={Brill, Steven and Olson, Britton},
    booktitle={APS Division of Fluid Dynamics Meeting Abstracts},
    pages={T20--007},
    year={2024}}

@article{huang2023consistent,
  title={A consistent and conservative Phase-Field method for compressible multiphase flows with shocks},
  author={Huang, Ziyang and Johnsen, Eric},
  journal={Journal of Computational Physics},
  volume={488},
  pages={112195},
  year={2023},
  publisher={Elsevier}
}
